\documentclass{ptephy_v1}

\preprintnumber{XXXX-XXXX} 

\usepackage{aas_macros}

\begin{document}

\title{Effects of nuclear matter and composition in core-collapse supernovae and long-term proto-neutron star cooling}


\author{Kohsuke Sumiyoshi}
\affil{Numazu College, National Institute of Technology,
Ooka 3600, Numazu, Shizuoka 410-8501, Japan
 \email{sumi@numazu-ct.ac.jp}}

\author{Shun Furusawa}
\affil{College of Science and Engineering, Kanto Gakuin University, 
1-50-1 Mutsuurahigashi, Kanazawa-ku, Yokohama, Kanagawa 236-8501, Japan}
\affil{Interdisciplinary Theoretical and Mathematical Sciences Program (iTHEMS), 
RIKEN, Wako, Saitama 351-0198, Japan}

\author{Hiroki Nagakura}
\affil{Division of Science, National Astronomical Observatory of Japan, 
2-21-1 Osawa, Mitaka, Tokyo 181-8588, Japan}

\author{Akira Harada} 
\affil{Interdisciplinary Theoretical and Mathematical Sciences Program (iTHEMS), RIKEN, Wako, Saitama 351-0198, Japan}

\author{Hajime Togashi} 
\affil{Department of Physics, Tohoku University, 
6-3 Aramaki Aza-Aoba, Aoba-ku, Sendai, Miyagi 980-8578,
Japan}

\author{Ken'ichiro Nakazato} 
\affil{Faculty of Arts and Science, Kyushu University, 
Fukuoka 819-0395, Japan}

\author{Hideyuki Suzuki} 
\affil{Faculty of Science and Technology, Tokyo University of Science, 
Noda, Chiba 278-8510, Japan}




\begin{abstract}%
We study the influence of hot and dense matter in core-collapse supernovae by adopting up-to-date nuclear equation of state (EOS) based on the microscopic nuclear many-body frameworks.  We explore effects of EOS based on the Dirac Br\"{u}ckner Hartree-Fock theory through comparisons with those based on the variational method.  We also examine effects of the differences in the composition of nuclei and nucleons by using the same EOS by the variational method but employing two different treatments in computations of nuclear abundances.  
We perform numerical simulations of core-collapse supernovae adopting the three EOSs.  
We also perform numerical simulations of the long-term evolution over 70 s of the proto-neutron star cooling.  
We show that impacts by different modeling of composition are remarkable as in those by different treatments of uniform matter in the gravitational collapse, bounce, and shock propagation.  
The cooling of proto-neutron star and the resulting neutrino emission are also affected by the compositional difference even if the same treatment in computing uniform matter of EOS.  

\end{abstract}

\subjectindex{xxxx, xxx}

\maketitle

\section{Introduction}

The equation of state (EOS) plays essential roles in the mechanism of core-collapse supernovae and the formation of neutron stars \cite{bet90,oer17}.  The stiffness of hot and dense matter determines the degree of compression of the central core to launch the shock wave for the supernova explosion and the composition affects the evolution of central core and the propagation of the shock wave via the weak interaction with neutrinos \cite{bar85,tak88,swe94,sum05,hem12,sum19}.  
These factors may be crucial for determining whether the explosion succeeds or fails \cite{mar09b,cou13,suw13,fis14,yas18,sch19b,har20} in the current scenario by the neutrino heating in multi-dimensional hydrodynamics \cite{jan12a,bur13,kot13,jan16,jan17b}.  The EOS is also influential in the thermal evolution of proto-neutron stars born in the supernova mechanism \cite{bur88,suz94,pon99,rob12,cam17,nak18,nak19,nak20}.  The supernova neutrinos emitted from the central object reflect the properties of EOS and carry the information of hot and dense matter \cite{jan17a,mul19} (See also \cite{suw19,nak19,nak20,war20,bax21,nag21a,nag21b,nag22a,nak22} for the neutrino signals at the detectors).  
It is, therefore, important to provide the sets of EOS for supernova simulations and explore the influence due to differences in the EOS.  
Careful studies on various factors in nuclear physics are necessary to connect the influence on the supernova phenomena with the properties of nuclear matter.  

Recently, there have been new developments of the EOS for supernova simulations based on the nuclear many-body frameworks.  
The variational method for nuclear matter has been applied to the construction of the table of EOS for numerical simulations in astrophysics \cite{tog17,fur17b}.  
It is a microscopic approach to describe nuclear many-body system based on the two- and three-nucleon potentials.  
The Dirac Br\"{u}ckner Hartree-Fock calculation has been utilized for the construction of the table of EOS more recently \cite{fur20}.  It describes the saturation of nuclear matter starting from the nucleon-nucleon interaction without the three-body interaction \cite{bro90}.  While the character of the former EOS has been studied by the numerical simulations of supernovae \cite{nag19c,iwa20,nak21}, the latter EOS has not been examined in the numerical simulations.  

It is the main purpose of the current study to clarify how the differences of the two sets of uniform matter bring the influence on the supernova mechanism, the birth of proto-neutron stars, and the associated emission of neutrinos.  
The properties of uniform nuclear matter and neutron matter obtained from the two frameworks are distinctively different from each other reflecting the characters of the relativistic and non-relativistic nuclear many-body frameworks.  The relativistic frameworks in general provide stiff EOS as compared with the counterpart in the non-relativistic frameworks \cite{oer17,sum21b}.  
The differences between the non-relativistic and relativistic frameworks have been discussed and their impact on supernova phenomena have been explored using the EOS tables constructed by the effective frameworks such as the relativistic mean field theory and the Skyrme Hartree-Fock theory \cite{sum04,sum05,hem12,ste13}.  
In the relativistic mean field theory, the repulsive interaction between the isovector-vector meson and nucleon tends to increase the symmetry energy at high densities \cite{ser86,sum94}.  
It is interesting to explore the differences in the microscopic many-body frameworks and their outcome in the supernova simulations.  

It is also the purpose of the current study to examine how the different models of composition of non-uniform matter at low densities make differences of the supernova dynamics adopting the same input of uniform matter.  
The construction of the EOS table requires the models to describe the non-uniform matter based on the EOS of uniform matter and to handle the mixture of nuclei.  
It is known that the main species of nuclei is different depending on the frameworks to handle the mixture of nuclei and may affect the dynamics of core-collapse supernovae \cite{hix03,hem10,hem12,fur17c}.  
The EOS table using the variational method is constructed by local density approximation to describe nuclei with a single nucleus approximation \cite{tog17} (hereafter denoted as VM-S EOS).  The elaborated version of the EOS table using the variational method is later constructed by the extended liquid drop model assuming the nuclear statistical equilibrium (NSE) \cite{fur17b} (hereafter denoted as VM EOS).  The EOS table using the Dirac Br\"{u}ckner Hartree-Fock calculation is recently constructed in the same NSE framework \cite{fur20} (hereafter denoted as DBHF EOS).  

We examine the influence of the hot and dense matter in two ways on the uniform matter and the composition by utilizing the three sets of EOS table.  We explore effects due to the different inputs of the uniform matter using the EOS tables: VM EOS and DBHF EOS.  We inspect effects due to the different models of the composition using the EOS tables: VM EOS and VM-S EOS.  
We perform the numerical simulations of the gravitational collapse and core bounce of the central core of massive stars using the three EOS tables to clarify the effects in the sets of comparison.  
We also perform the numerical simulations of the long-term evolution over 70 s of the proto-neutron star cooling.  

We demonstrate that the differences in the composition are important on the same footing as those in the uniform matter in core-collapse supernovae and proto-neutron stars.  
While we would like to reveal different features arising from the different EOS for uniform matter, we need to carefully assess the outcome from the different compositions and separate out the effect from uniform matter.  
It is known that the composition may affect the electron capture rates, neutrino scattering during the gravitational collapse and the cooling/heating rates by neutrinos during the propagation of the shock wave \cite{swe94,hix03,hem12,fur13b,fur17c}.  It is also pointed out that the composition may be influential as the trigger of the convective motion in multi-dimensional simulations \cite{har20}.  We examine these factors under the spherical symmetry to estimate the impact on simulations in multi-dimensions.  

This article is arranged as follows.  
In section \ref{section:EOS}, we explain the EOS tables adopted in the study.  We describe the EOS of uniform matter evaluated by the two nuclear many-body frameworks and the models to describe the composition of non-uniform matter.  
In section \ref{section:Simulation}, we describe the numerical simulations of the core-collapse supernovae and the cooling of proto-neutron stars.  
In section \ref{section:Results}, we report the numerical results of the core collapse and bounce in section \ref{section:Results_CCSNe} and the proto-neutron stars in section \ref{section:Results_PNSC}.  We separately compare the differences due to the uniform matter and the composition to examine their size of effects due to each factor.  We summarize the current study and discuss its implications in section 
\ref{section:Summary}.  

\section{Equation of state}\label{section:EOS}

The properties of uniform nuclear matter is the essential input for the construction of the EOS tables for supernova simulations.  They are used to evaluate the free energy of mixture of nucleons and nuclei by modelling the non-uniform matter.  While the behavior of uniform matter is directly related with the properties of neutron stars and the degree of core bounce in supernovae at high densities, it is also influential in determining the composition and main species of nuclei at low densities through the saturation properties and the symmetry energy for the description of nuclei.  

We adopt three sets of the data tables of EOS prepared for numerical simulations: the two sets based on the variational method \cite{tog17,fur17b} and the one based on the Dirac Br\"{u}ckner Hartree-Fock theory \cite{fur20}.  We describe the properties of the neutron matter and symmetric nuclear matter in the two frameworks below.  We also explain the models to treat the mixture of nuclei in non-uniform matter: the single nucleus approximation used in \cite{tog17} and the nuclear statistical equilibrium used in \cite{fur17b,fur20}.  

\subsection{EOS of uniform matter in many-body frameworks}\label{section:EOS_framework}

The calculations of uniform matter by the variational method \cite{tog13} (VM) have been made to provide the free energy for the construction of the two sets of EOS table by \cite{tog17,fur17b}.  It is a microscopic nuclear many-body calculation based on the variational principle method with realistic nuclear forces \cite{kan07,tog13}.  It starts from the non-relativistic nuclear Hamiltonian composed of the Argonne v18 two-nucleon potential \cite{wir95} and the Urbana IX three-nucleon potentials \cite{car83,pud95}.  The expectation value of the Hamiltonian for one- and two-body components is evaluated with the Jastrow wave function in the two-body cluster approximation.  The expectation value of the Hamiltonian for three-body component is evaluated by the Fermi-gas wave function.  
The parameters in the three-body term are determined so as to reproduce the empirical saturation properties. 
Furthermore, these are fine-tuned so as to reproduce the gross features of masses and radii of measured atomic nuclei in the Thomas-Fermi calculation.  
The evaluation of asymmetric nuclear matter is directly performed for arbitrary proton fraction under the same framework.  The resulting energies of symmetric nuclear matter and pure neutron matter are in close agreement with those in the APR EOS obtained by the Fermi hypernetted chain calculation \cite{akm98,muk09}.  The extension to finite temperature is made based on the variational method proposed by Schmidt and Pandharipande \cite{sch79,muk07}.  The data table of free energy for various density, proton fraction, and temperature is constructed and thus used for the construction of the EOS table.  

The evaluation by the Dirac Br\"{u}ckner Hartree-Fock approach \cite{kat13} (DBHF) is utilized to provide the energy contributions for the construction of the EOS table \cite{fur20}.  It is based on a microscopic nuclear many-body calculation in a relativistic formulation \cite{bro90,kat13}.  It starts from the Bonn nucleon-nucleon interaction which is determined by the nucleon-nucleon scattering experiments  \cite{bro90,gro99}.  The relativistic formulation of the Dirac Br\"{u}ckner Hartree-Fock approach provides the nuclear saturation point with only the two-body nucleon interaction without introducing the three-body nucleon interaction, being different from the non-relativistic many-body frameworks such as the variational method.  The Dirac Br\"{u}ckner Hartree-Fock calculation provided the fitting formulae as function of Fermi momentum for the energy, vector and scalar potentials for symmetric nuclear matter and pure neutron matter \cite{kat13}.  The asymmetric nuclear matter with arbitrary proton fractions is described by the interpolation with a parabolic expression using the formulae.  The extension to the finite temperature is made using the expression of the kinetic energy at finite temperatures \cite{fur20}.  Applications to the table of EOS have been made through these expressions.  

\begin{figure*}[ht]
\centering
\includegraphics[width=0.45\textwidth]{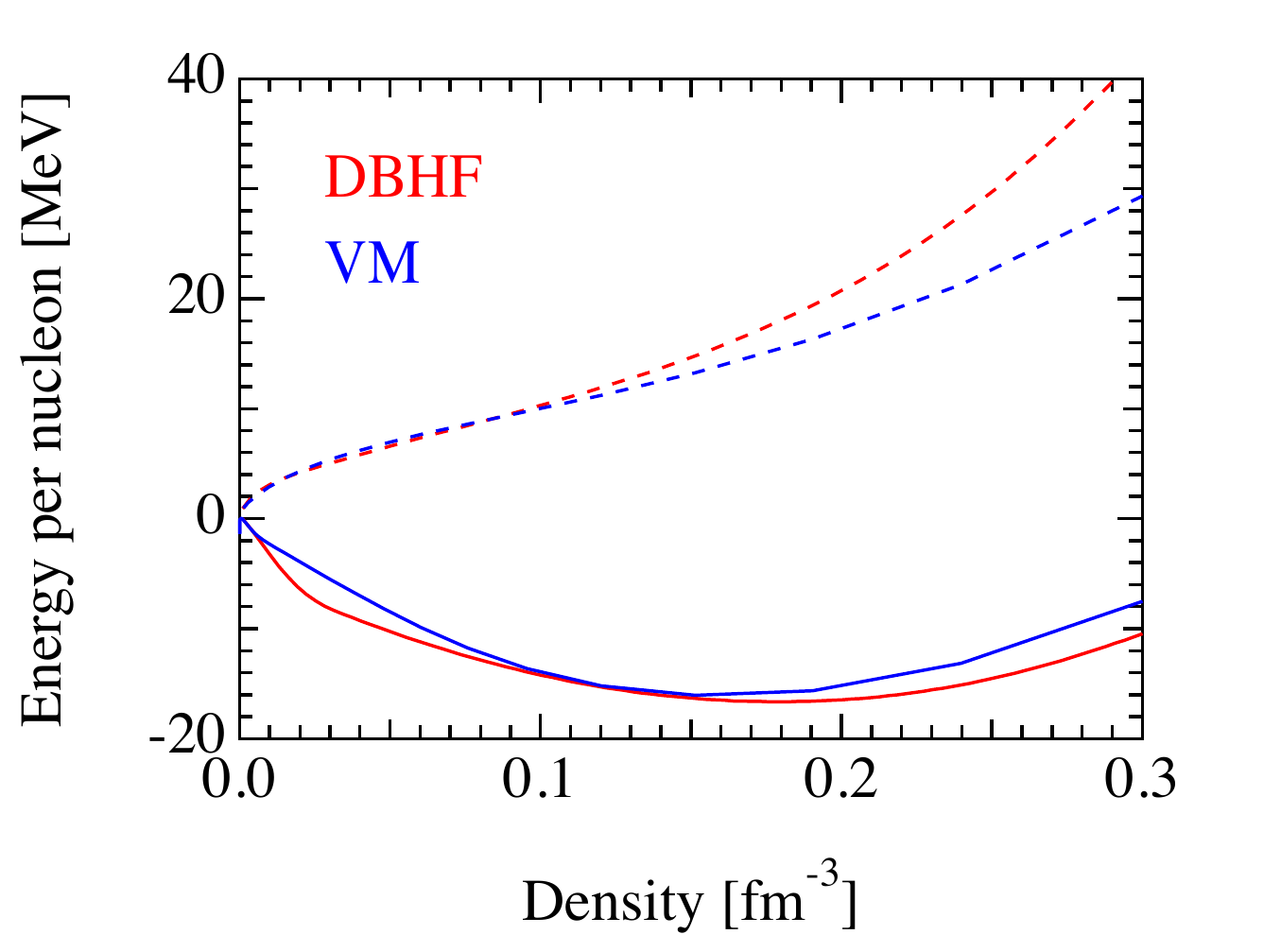}
\includegraphics[width=0.45\textwidth]{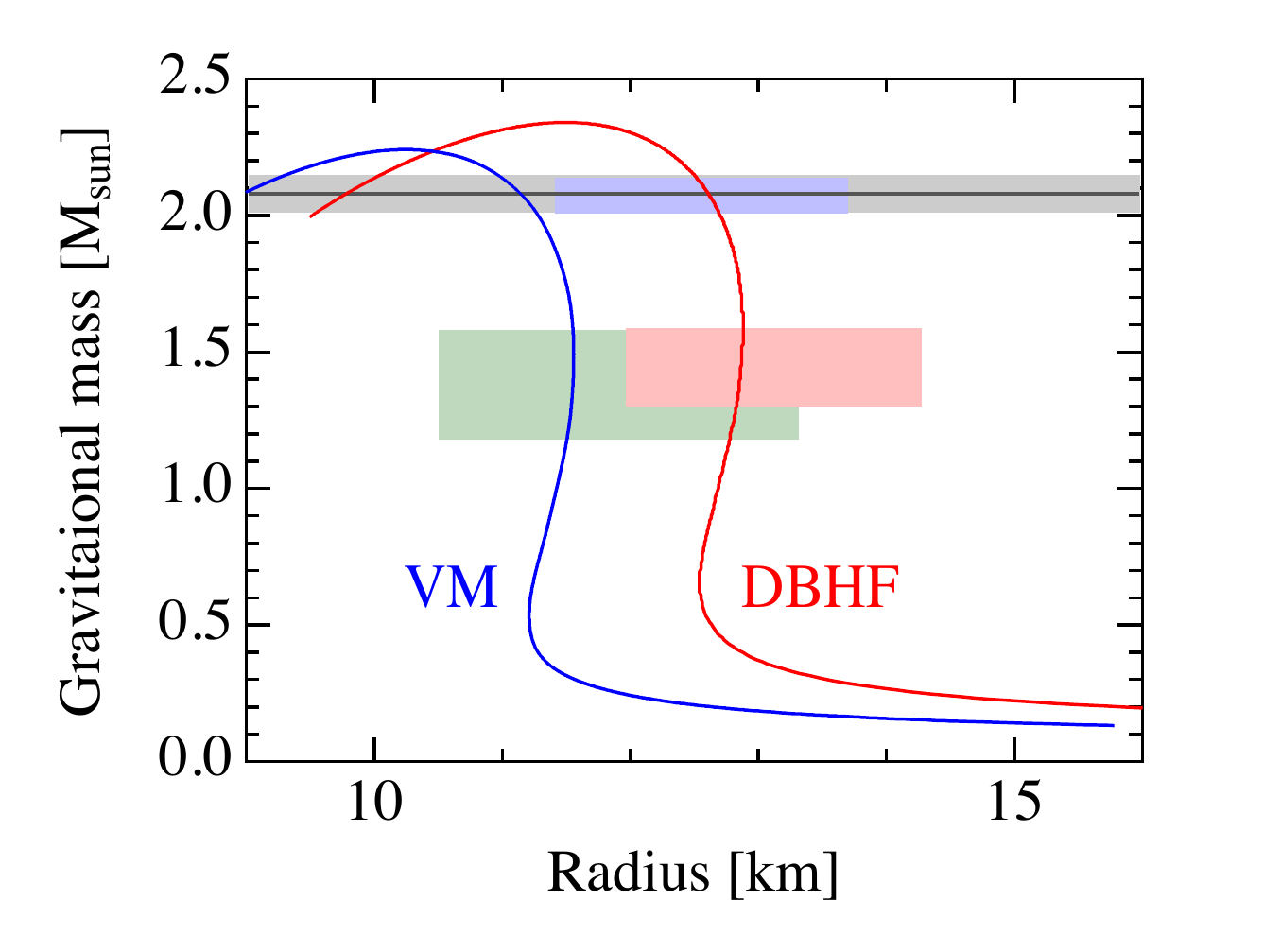}
\caption{Energy per nucleon of nuclear matter and neutron matter for DBHF EOS and VM EOS is shown by solid and dashed lines, respectively, as a function of nucleon density in the left panel.  
The relation between gravitational mass and radius of neutron stars constructed by DBHF EOS and VM EOS is shown in the right panel.  
The quantities for DBHF and VM EOSs are shown by red and blue lines, respectively.  
The maximum mass derived by the observation of the pulsar PSR J0740+6620 \cite{fon21} is shown by gray band.  
The ranges of mass and radius derived by the detection of the gravitational wave, GW170817, \cite{abb18}, the X-ray observations of the pulsars PSR J0030+0451 \cite{ril21}, and PSR J0740+6620 \cite{mcmil21} are shown by light green, light red, and light blue shaded areas, respectively.  
\label{fig:ns_M-R_dbhf}}
\end{figure*}

The energies per nucleon of symmetric nuclear matter and pure neutron matter at zero temperature for VM and DBHF EOSs are compared in the left panel of Fig. \ref{fig:ns_M-R_dbhf}.  The bulk properties at the saturation density is listed in Table \ref{tab:EOS}.  The energies of symmetric nuclear matter in the two EOSs fulfill the properties of nuclear saturation, however, the saturation point is slightly deeper in DBHF EOS.  The energies of pure neutron matter in the two EOSs are similar to each other at low densities.  The increase of the energy at high densities is faster in DBHF EOS than that in VM EOS.  Accordingly, the symmetry energy in DBHF EOS is larger.  In the right panel of Fig. \ref{fig:ns_M-R_dbhf}, the mass and radius of neutron stars constructed by the two EOSs are shown \cite{tog17,fur20}.  The maximum masses in both models are larger than 2.0$M_{\odot}$ and fulfills the observational constraints \cite{dem10,ant13,cro20,fon21}.  The radii for the two EOSs are consistent with the range derived by the observation by gravitational wave \cite{abb18}, however, the radius of DBHF EOS is larger than that of VM EOS and consistent with the range derived by the NICER observation \cite{mcmil19,ril19,mcmil21,ril21}.  Judging from the properties of uniform matter and neutron stars, DBHF EOS is stiffer than VM EOS.  

\begin{table}[htbp]
\caption{Properties of uniform matter of the VM and DBHF EOSs.  The saturation density, $n_0$, and energy, $E_0$, incompressibility, $K$, symmetry energy, $E_{\mathrm{sym}}$, and its slope parameter, $L$, are listed for two models.  The values of the symmetry energy are defined here as coefficients of the quadratic term of the asymmetry parameter in the expansion.  }
\centering
\begin{tabular}{lccccc}
\hline
     & $n_0$ [fm$^{-3}$] & $E_0$ [MeV] & $K$ [MeV]  & $E_{\mathrm{sym}}$ [MeV] & $L$ [MeV] \\
\hline
VM   & 0.160 & $-$16.1 & 245 & 29.1  & 38.7 \\ 
DBHF & 0.179 & $-$16.6  & 232 & 34.5  & 66.8 \\ 
\hline
\end{tabular}
\label{tab:EOS}
\end{table}

These differences between the two EOSs are similar to the features seen in the comparison of the effective many-body theories in a non-relativistic and relativistic frameworks \cite{sum93,sum21b}.  The relativistic mean field theory, which is an effective theory of microscopic relativistic many-body theory such as the Dirac Br\"{u}ckner Hartree-Fock  theory, has the tendency of large symmetry energy due to the increase of neutron matter as compared with that in the Skyrme Hartree-Fock theory as an example of non-relativistic effective many-body theory.  The neutron star radii in the relativistic mean field EOS are larger than those in the Skyrme Hartree-Fock EOS in many cases.  It is to be noted that the EOS in relativistic many-body frameworks (both Dirac Br\"{u}ckner Hartree-Fock and relativistic mean field theories) automatically fulfills the causality condition while the EOS in the non-relativistic frameworks at high densities often leads to the sound speed larger than the speed of light, breaking the causality condition.  

\subsection{Description of composition of non-uniform matter}\label{section:EOS_composition}

For the construction of EOS tables for supernova simulations, it is mandatory to describe various phases of hot and dense matter for a wide range of density, proton fraction, and temperature.  In order to evaluate the thermodynamical quantities in a consistent manner, it is necessary to set the model to evaluate the free energy of the non-uniform matter composed of nucleons and nuclei.  There are differences of the models to describe the profile of nuclei and to determine the mixture of nuclei in the EOS tables constructed so far \cite{oer17}.  

In the VM-S EOS table, the non-uniform matter is modeled by the Thomas-Fermi approximation as in the case of the Shen EOS \cite{she98a,she98b,she11,she20}.  The phase of nuclei is described by using the functional form of distributions of neutrons, protons, and alpha particles in the Wigner-Seitz cell.  The non-uniform matter is assumed to be the mixture of free neutrons, free protons, alpha particles, and a single species of heavy nuclei.  The free energy of matter is calculated by the sum of the bulk energy, the gradient term, and the Coulomb energy through the volume integration using the input of energy of uniform matter, which are provided by the nuclear many-body calculations explained above.  In this model, single species of a representative nuclei is assumed as an approximation to handle the composition of nuclei.  

In the VM EOS table, the non-uniform matter is modeled by the mixture of various nuclei and nucleons assuming the nuclear statistical equilibrium \cite{fur11,fur13a,fur17a}.  The masses of heavy nuclei are evaluated by the compressible liquid drop models as the sum of bulk, surface, shell, and Coulomb energies.  The variations of the saturation density, energy, surface coefficient, and shell energies are considered with the dependence on the density, proton fraction, and temperature.  The light nuclei including deuterons, tritons, helions, and alpha particles are treated with the energy shift and Pauli effect in the matter.  The distributions in the mixture of nuclei are obtained by the minimization of the free energy of the multi-species of nuclei covering the wide ranges of neutron and proton numbers.   The ensemble average of nuclei is listed in the EOS table and used in the numerical simulations.  The DBHF EOS table is constructed in the same model of the non-uniform matter but with different input of the uniform matter.  

\begin{figure*}[ht]
\centering
\includegraphics[width=0.9\textwidth]{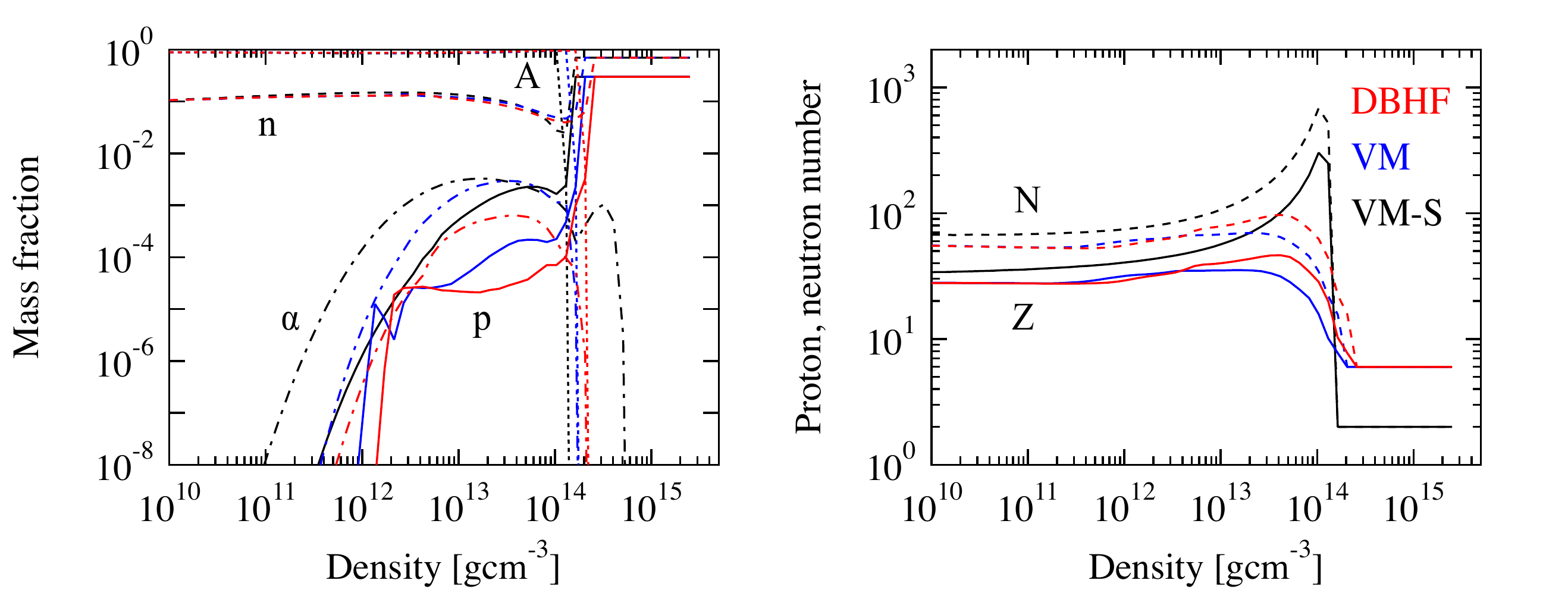}
\caption{Composition of the matter with the fixed electron fraction, $Y_e=0.3$ and entropy per baryon, $S=1$~$k_B$.  The mass fraction and the proton and neutron number of nuclei are shown as functions of the density in left and right panels, respectively.  The quantities of DBHF, VM-S, and VM EOSs are shown by red, black, and blue lines, respectively.  
The mass fraction of nuclei (dotted), alpha particles (dot-dashed), neutrons (dashed), and protons (solid) are shown in the left panel.  The neutron (dashed) and proton (solid) numbers are shown in the right panel.  
\label{fig:eos_compo_Rhodep}}
\end{figure*}

We examine the behavior of composition in the hot and dense matter of the three sets of EOS to demonstrate importance of the treatment of mixture.  We utilize the EOS tables with electrons, positrons, and photons for simulations.  
We investigate the matter at the fixed electron fraction, $Y_e=0.3$, and entropy per baryon, $S=1$~$k_B$ as a typical condition in supernova cores.  In Fig. \ref{fig:eos_compo_Rhodep}, we show the composition of the matter in the three sets of EOS as a function of density.  We see differences in the mass fractions of alpha particles and free protons among the three EOSs with slight differences in the mass fractions of nuclei ($\sim0.9$) and free neutrons ($\sim0.1$).  The species of nuclei in VM-S EOS behaves in a different way as compared to those in VM EOS and DBHF EOS.  The proton and neutron numbers of nuclei in VM-S EOS are larger than the others.  They increase rapidly around 10$^{14}$~g\,cm$^{-3}$ being different from the behavior in VM and DBHF EOSs, which are similar to each other.  The treatment of mixture has an impact on the composition whereas that of the uniform matter has a minor effect.  These differences may have influence on the supernova dynamics through the reaction rates of electron captures and neutrino scattering as we will see in section \ref{section:Results_CCSNe}.  

\begin{figure*}[ht]
\centering
\includegraphics[width=0.9\textwidth]{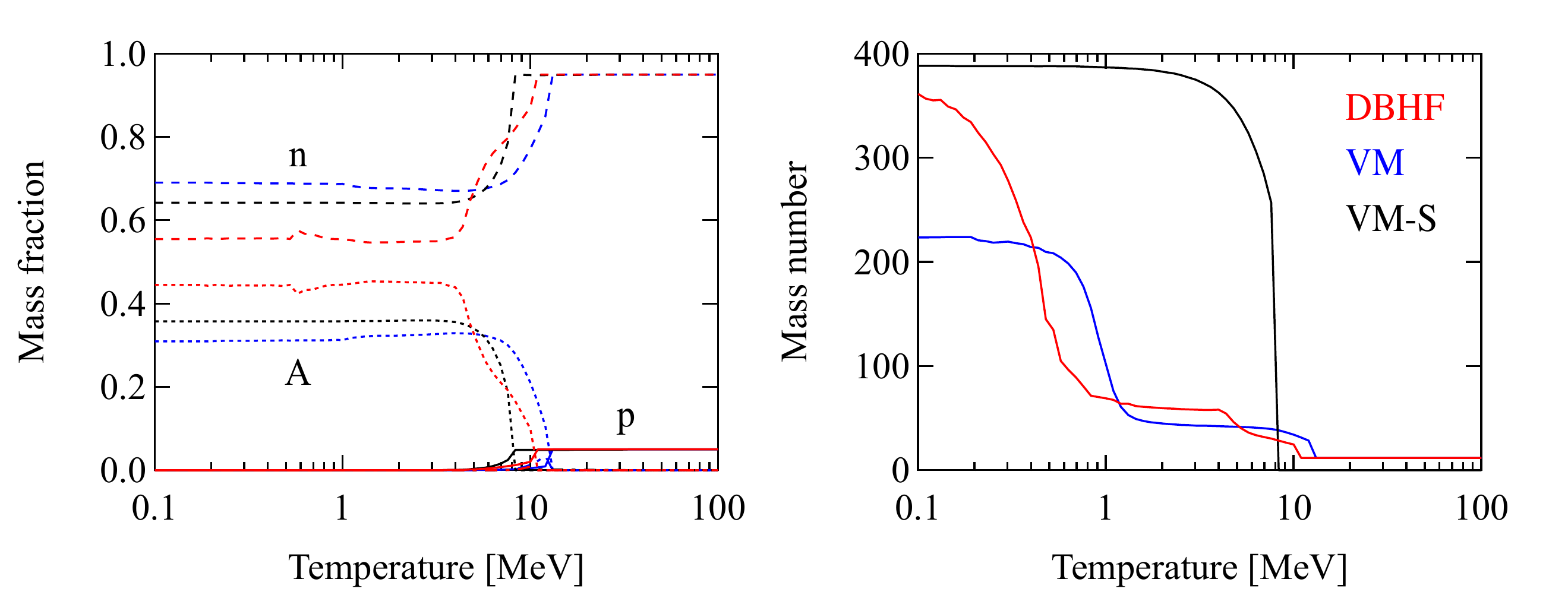}
\caption{Composition of the matter at the fixed density, $5\times10^{13}$~g\,cm$^{-3}$, and electron fraction, $Y_e=0.3$.  The mass fraction and the mass number of nuclei of DBHF, VM-S, and VM EOSs are shown as functions of temperature in left and right panels, respectively.  
The quantities of DBHF, VM-S, and VM EOSs are shown by red, black, and blue lines, respectively.  
The mass fraction of nuclei (dotted), neutrons (dashed), and protons (solid) are shown in the left panel.  
\label{fig:eos_compo_Tdep}}
\end{figure*}

It is interesting to see changes in the composition at the conditions in the proto-neutron star cooling.  In Fig. \ref{fig:eos_compo_Tdep}, we show the composition of the matter in the three sets of EOS as a function of temperature.  We examine variations at the fixed density, $5\times10^{13}$~g\,cm$^{-3}$, and electron fraction, $Y_e=0.3$.  
It is remarkable that the mass number of nuclei (right panel) becomes large $\sim400$ for VM-S EOS as the temperature goes below 5~MeV.  They are rather small $< 100$ for DBHF and VM EOS down to temperature $\sim1$~MeV.  The difference of the mass number persists during the cooling of the proto-neutron star in our simulations as we will see in section \ref{section:Results_PNSC}.  
The mass fraction (left panel) is different among the three sets of EOS.  The mass fraction of nuclei becomes large below $\sim10$~MeV and remains constant.  It is larger for DBHF EOS than those for VM-S EOS and VM EOS.  

There is a minor difference in the description of translational motion of nuclei.  The kinetic energy of nuclei is considered in the energy evaluation by the expression of the Boltzmann gas with an excluded volume correction in the VM and DBHF model.  In the VM-S model (also in the series of Shen EOS), the translational energy of nuclei is not included in the evaluation of the free energy.  This difference is small in the low density and temperature inside the phase of nuclei, but it may bring some differences in the thermodynamical quantities such as the entropy.  We examine this issue by performing additional simulations and report the results in the Appendix.  We show that the influence due to the translational motion of nuclei is negligible in the dynamics of core bounce except for overall shifts of the entropy per baryon.  

\section{Numerical simulations}\label{section:Simulation}

\subsection{Core-collapse of massive stars}\label{section:Simulation_CCSNe}

Numerical simulations of the gravitational collapse and bounce of the core of massive stars are done by the numerical code for the general relativistic neutrino-radiation hydrodynamics under the spherical symmetry \cite{sum05,sum06,sum07,sum19,nak21}.  The code solves a set of the equations for hydrodynamics and neutrino transfer in general relativity simultaneously in a time implicit manner.  The general relativistic Boltzmann equations for neutrino distribution are directly solved by the S$_n$ method, retaining the information of angular distributions.  
The neutrino distributions for four species of neutrinos ($\nu_e$, $\bar{\nu}_e$, $\nu_{\mu}$ and $\bar{\nu}_{\mu}$) are solved by assuming the same reactions for $\mu$-type and $\tau$-type (anti-)neutrinos.  
The neutrino distributions are descretized by the mesh with 6 angle grids and 14 energy grids.  The number of grids for radial mass coordinate is 511.  
A basic set of reactions via the weak interaction are implemented in the collision term based on the Bruenn's prescription with the extension for the nucleon-nucleon bremsstrahlung.  The same setting as in \cite{sum05} is adopted for comparisons with the numerical results for VM EOS \cite{nak21} and other conventional sets of EOS \cite{sum19}.  

The initial model is taken from the central Fe core of the massive stars of 11.2M$_{\odot}$ by \cite{woo02} and 15M$_{\odot}$ by \cite{woo95}.  
We take the central part of 1.41M$_{\odot}$ and 1.81M$_{\odot}$ of the 11.2M$_{\odot}$ and 15M$_{\odot}$ stars, respectively.  We adopt the profile of density, electron fraction, and temperature as functions of the baryon mass coordinate.  
We set up the initial condition by using each set of EOS to obtain other quantities such as the entropy per baryon.  The resulting configuration is unstable against the gravitational collapse.  
We follow the time evolution of the gravitational collapse, core bounce, and the shock propagation for 0.3 s after the core bounce.  We mainly report comparison for 11.2M$_{\odot}$ to extract difference and similarity and supplement generality by the comparison for 15M$_{\odot}$.  

\subsection{Cooling of proto-neutron stars}\label{section:Simulation_PNSC}

Numerical simulations of the thermal evolution of the proto-neutron stars just born in the core-collapse supernovae are done by the numerical code for the time evolution of  quasi-hydrostatic structure with an approximated neutrino transfer in general relativity under the spherical symmetry \cite{suz94,nak13a,nak18}.  
The neutrino transport is handled by the flux-limited diffusion approximation.  The equation for energy density and flux of neutrinos are solved adopting the diffusion coefficient with the flux limiter.  
Three species of neutrinos ($\nu_e$, $\bar{\nu}_e$, $\nu_{x}$) are handled by collectively treating the $\mu$-type and $\tau$-type (anti-)neutrinos into one species ($\nu_{x}$).  
The number of grids for radial mass coordinate and energy are 99 and 25, respectively.  
The set of reactions via the weak interaction corresponding to the one in the core-collapse simulations is implemented based on the Bruenn's rate and its extension \cite{sum95c}.  

The initial model is set up by adopting the profile of central object born in a supernova core after the core bounce in the same way as in \cite{nak18,sum19} (See section \ref{section:Results_PNSC} for details).  
The initial profile as a reference for three sets of EOS is taken from the supernova core of the 15M$_{\odot}$ star at 0.3 s after the bounce with the Shen EOS.  
We adopt the distributions of electron fraction and entropy per baryon as functions of the radial mass coordinate.  
The baryon mass of the chosen profile for proto-neutron stars is 1.47M$_{\odot}$.  
The initial configuration of hydrostatic structure with the stationary state of neutrino flow is constructed in the following manner.  
We first obtain the mass density profile by solving the Tolman-Oppenheimer-Volkoff equation under the fixed distributions of electron fraction and entropy per baryon in the baryon mass coordinate.  We search for the central mass density for the profile so as to match with the total baryon mass at the surface.  
We then solve the neutrino transfer under the obtained profile of density, electron fraction, and entropy per baryon until it reaches the stationary state.  
%
%

We set the common profile of the electron fraction and entropy per baryon for the three EOSs in the current study to mainly focus on the long-term evolution over 70 s.  We note that, in principle, the initial profiles may have differences due to the EOSs during the core-collapse simulations.  However, there are uncertainties in setting the initial model for the proto-neutron star cooling due to accretion since the spherical simulations do not provide any successful explosions for the ordinary Fe cores in massive stars (See \cite{nak13a} to implement the effect of accretion).  Hence, we adopt the same setting for the birth of proto-neutron star and explore the influence afterward.  This is also beneficial for the comparison with the previous simulations by adopting other EOSs \cite{nak18,sum19}.  Ultimately, it is ideal to perform numerical simulations in two or three dimensions and take out the central object in successful explosions to examine the EOS effects in a consistent manner.  

\section{Numerical results}\label{section:Results}

\subsection{Core-collapse and bounce}\label{section:Results_CCSNe}

\subsubsection{Profiles of DBHF and VM models}\label{section:Results_CCSNe_DBHF}

\begin{figure*}[ht]
\centering
\includegraphics[width=0.45\textwidth]{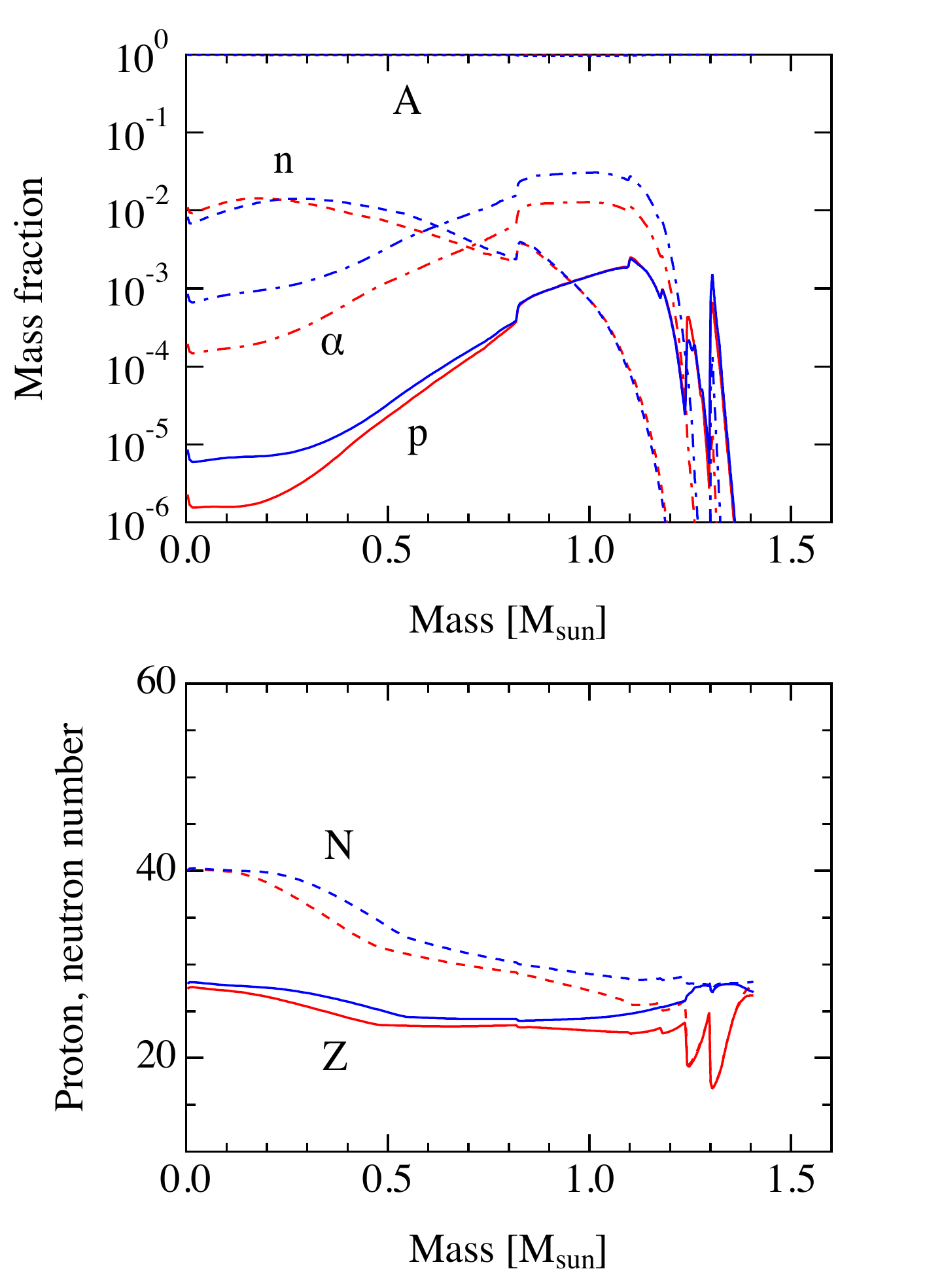}
\includegraphics[width=0.45\textwidth]{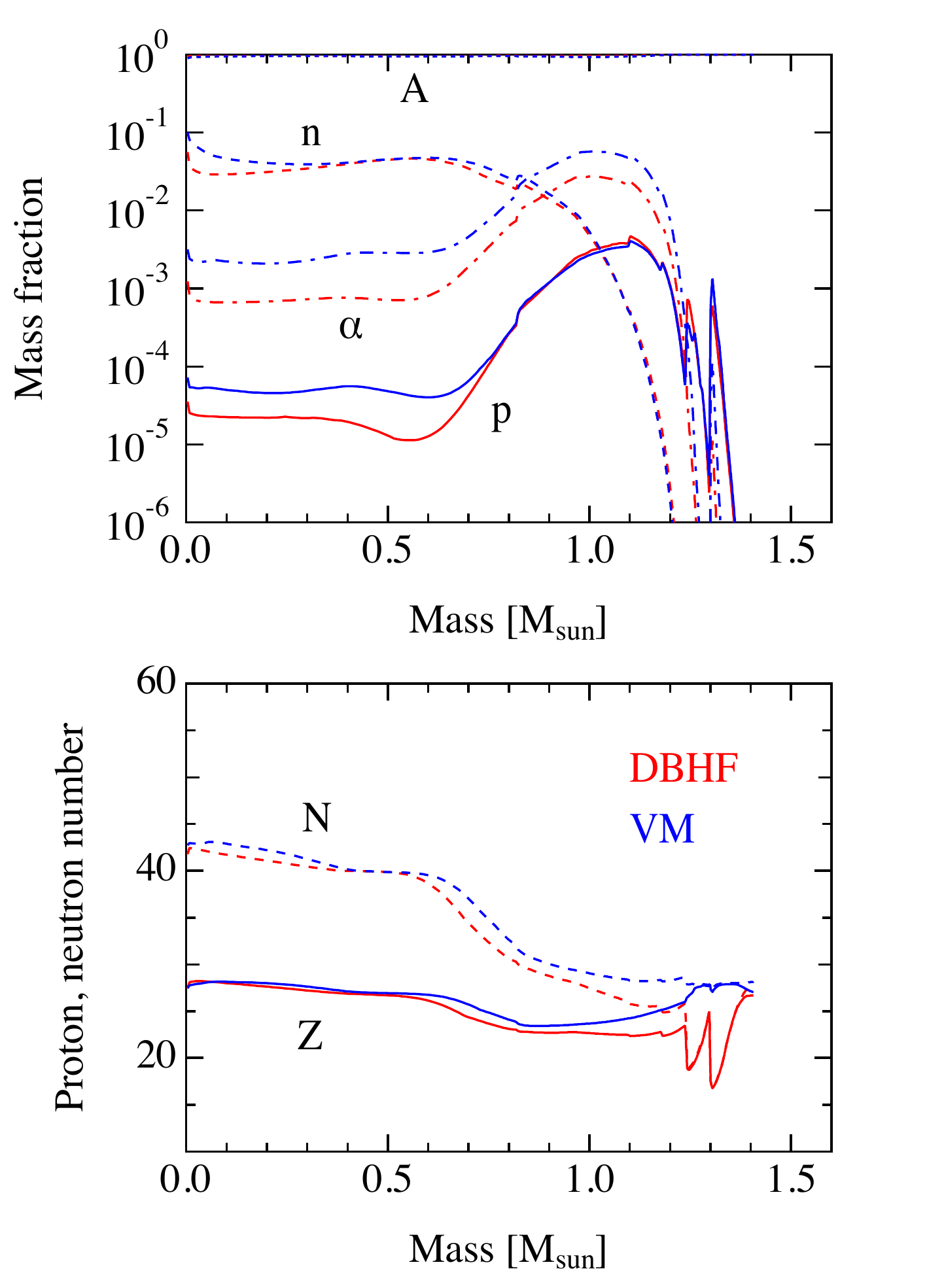}
\caption{The composition of matter is compared for the two models of 11.2M$_{\odot}$ star using DBHF EOS and VM EOS.  
The DBHF and VM models are shown by red and blue colors, respectively.  
The snapshots when the central density reaches $10^{11}$ and $10^{12}$ g\,cm$^{-3}$ are shown as functions of the baryon mass coordinate in the left and right panels, respectively.  The mass fraction of proton, neutron, alpha particles, and nuclei is shown by solid, dashed, dot-dashed, and dotted lines in the upper panels.  The proton and neutron number of nuclei is shown by solid and dashed lines, respectively, in the lower panels.  
\label{fig:radhyd_rhoc11_DBHF}}
\end{figure*}

We discuss the profiles during the collapse and bounce of the central core of 11.2M$_{\odot}$ star using DBHF EOS and VM EOS.  
We show the composition of matter during the collapse for the two models 
in Fig. \ref{fig:radhyd_rhoc11_DBHF}.  
The composition of the matter at the central densities, $\sim 10^{11}-10^{12}$ g\,cm$^{-3}$, during the gravitational collapse is important to determine the lepton fraction through neutrino emission and trapping by the weak reactions.  
The nuclear species in the two models are similar to each other and the effect of neutrino trapping becomes similar.  
The nuclei are not so neutron-rich ($N < 40$) at the central density $10^{11}$ g\,cm$^{-3}$ (left panel) and becomes more neutron-rich ($N > 40$) at $10^{12}$ g\,cm$^{-3}$ (right panel).  
The zigzag feature in the DBHF model occurs due to transitions of the main species of nuclei.  
The mass fraction of alpha particles is smaller in the DBHF model.  

It is noticeable to see the mass fraction of free protons of the DBHF model is smaller than that of the VM model in the central region.  
This is due to the larger symmetry energy of DBHF EOS than that of the VM EOS at high densities.  
For the neutron-rich matter, the isovector part of the nuclear interaction, which contributes to the symmetry energy, provides attraction to protons and repulsion to neutrons.  
The large symmetry energy leads to a large difference between the neutron and proton chemical potentials.  
The proton chemical potential becomes low, leading to a small fraction of free protons, while the neutron chemical potential becomes high.  
This difference in the mass fraction of free protons may lead to a difference of the weak reaction rate of electron capture on free protons and the resulting amount of trapped leptons \cite{bru89a}.  
The electron capture on protons proceeds when the electron capture on nuclei is hindered for $N > 40$ due to the blocking factor in the prescription of the Bruenn's rate.  

\begin{figure*}[ht]
\centering
\includegraphics[width=0.9\textwidth]{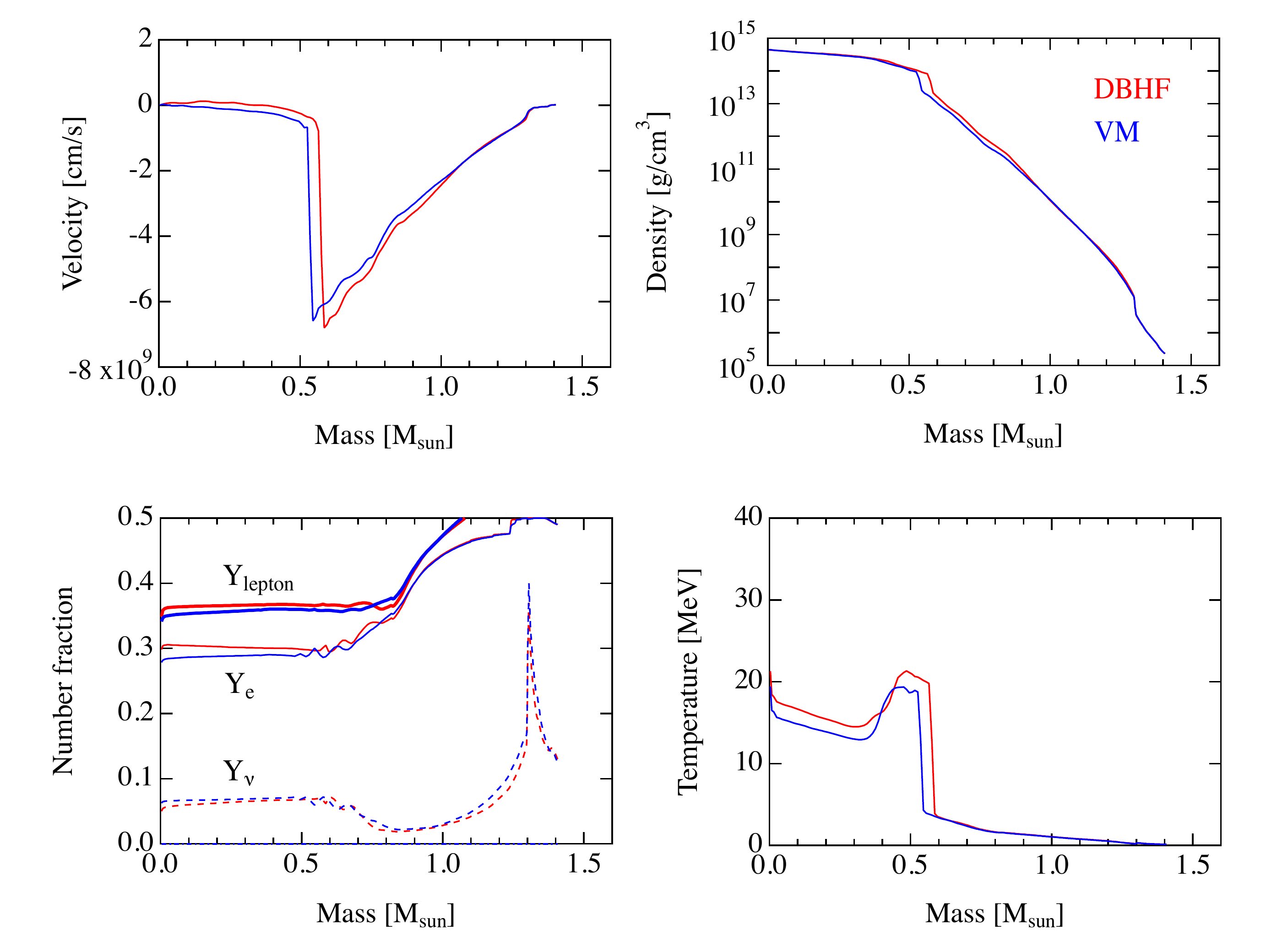}
\caption{The profiles at the core bounce are compared for the two models of 11.2M$_{\odot}$ star using DBHF EOS and VM EOS.  
The velocity, density, lepton fractions, and temperature are shown as functions of the baryon mass coordinate in the upper-left, upper-right, lower-left, and lower-right panels.  
The lepton, electron, and neutrino fractions are shown by thick-solid, solid, and dashed lines, respectively, in the lower-left panel.  
\label{fig:radhyd_tpb0_DBHF}}
\end{figure*}

\begin{figure*}[ht]
\centering
\includegraphics[width=0.45\textwidth]{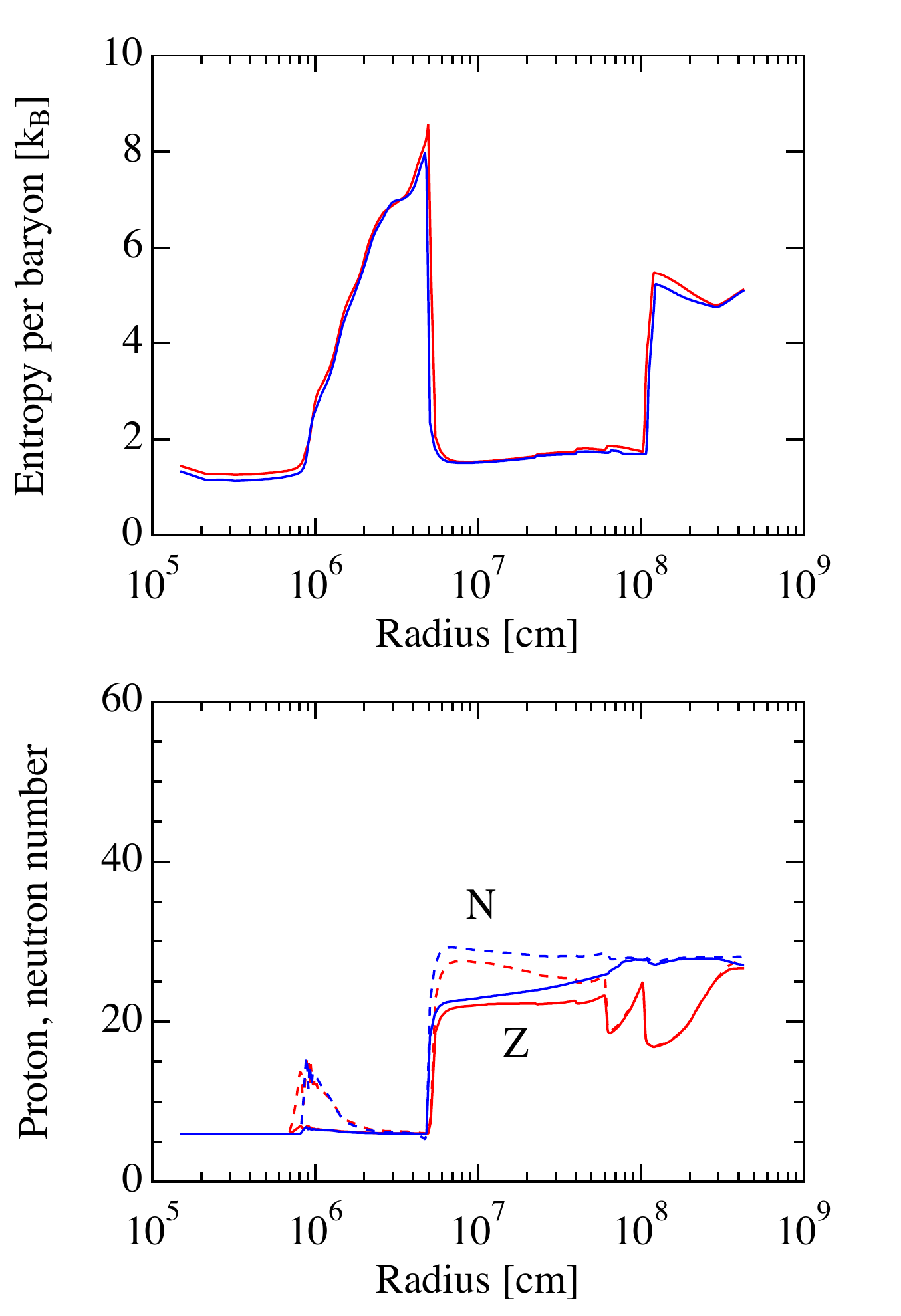}
\includegraphics[width=0.45\textwidth]{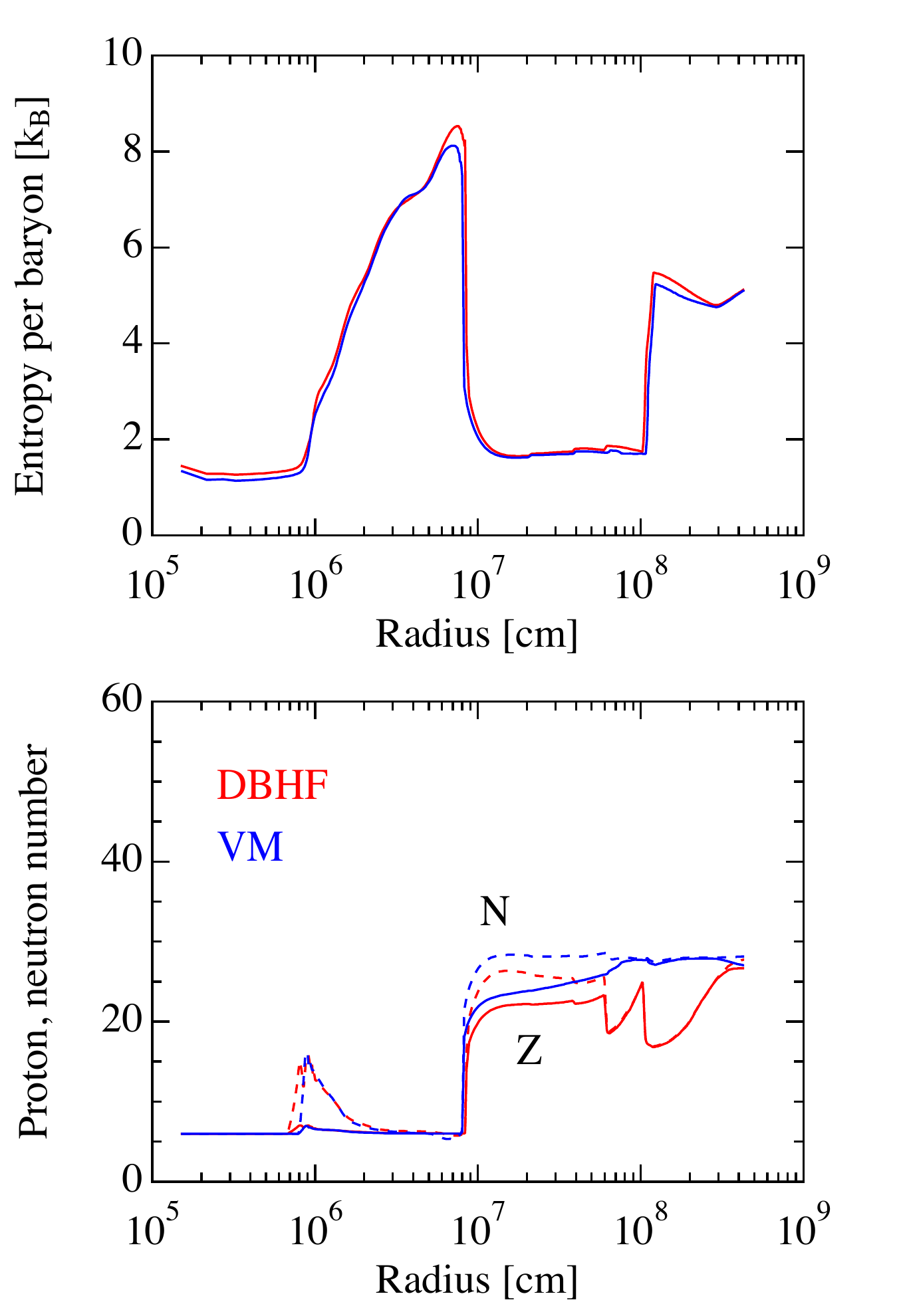}
\caption{Radial profiles of entropy per baryon (top) and proton, neutron number of nuclei (bottom) for the two models of 11.2M$_{\odot}$ star using DBHF EOS and VM EOS are shown for the snapshots at 1 and 2 ms after the bounce in the left and right panels, respectively.  
The proton and neutron number of nuclei is shown by solid and dashed lines, respectively, in the lower panels.  
Note that the species of nuclei becomes $^{12}$C (other than light nuclei such as $^{4}$He) in the central region although the mass fraction is very small.  
\label{fig:radhyd_tpb1_DBHF}}
\end{figure*}

We see the consequence of the different uniform matter through composition in the profiles at the core bounce in Fig. \ref{fig:radhyd_tpb0_DBHF}.  
The size of bounce core is slightly larger in the DBHF model than in the VM model, while the central densities are similar.  
The position of the shock wave is located at 0.575M$_{\odot}$ and 0.535M$_{\odot}$ in the mass coordinate for the DBHF and VM models, respectively.  
This is consistent with the fact that the lepton and electron fractions are larger in the DBHF model, reflecting the difference in the mass fraction of free protons.  
Note that the neutrino fraction in the DBHF model is slightly smaller than that of the VM model because the larger symmetry energy leads to a smaller neutrino chemical potential ($\mu_{\nu}=\mu_e+\mu_p-\mu_n$) under the beta equilibrium condition.  
This makes the difference of the lepton fraction smaller than that of the electron fraction.  
The temperature in the DBHF model is slightly higher than that in the VM model.  
The entropy per baryon in the DBHF model is slightly larger at the core bounce (not shown here, but see Fig. \ref{fig:radhyd_tpb1_DBHF}) due to a small difference in the initial model with the same temperature 
and the following increase through the neutronization during the collapse.  

In order to explore the influence of the EOS after the bounce, we examine the composition of the central core right after the core bounce for the two models in Fig. \ref{fig:radhyd_tpb1_DBHF}.  
Different composition of nuclei may lead to a different size of energy loss due to nuclear dissociation at the passage of shock wave and result in different development of convective motion through negative gradient of entropy distribution as pointed out in the analysis of EOS effects \cite{har20}.  
In the comparison between the DBHF and VM models, there is a slight difference of the proton and neutron numbers of nuclei in unshocked matter of outer layers at $> 50$~km and $>80$~km at 1 and 2 ms, respectively, after the bounce (bottom panels).  
Resulting profiles of the entropy per baryon are similar (upper panels), but the peak profile of the DBHF model is more sharp with the smaller mass number as compared with those of VM model, which have wider extension of negative gradient.  
As we will see below, the difference of composition is more clearly seen in the different treatments of the non-uniform matter.  

\begin{figure*}[ht]
\centering
\includegraphics[width=0.9\textwidth]{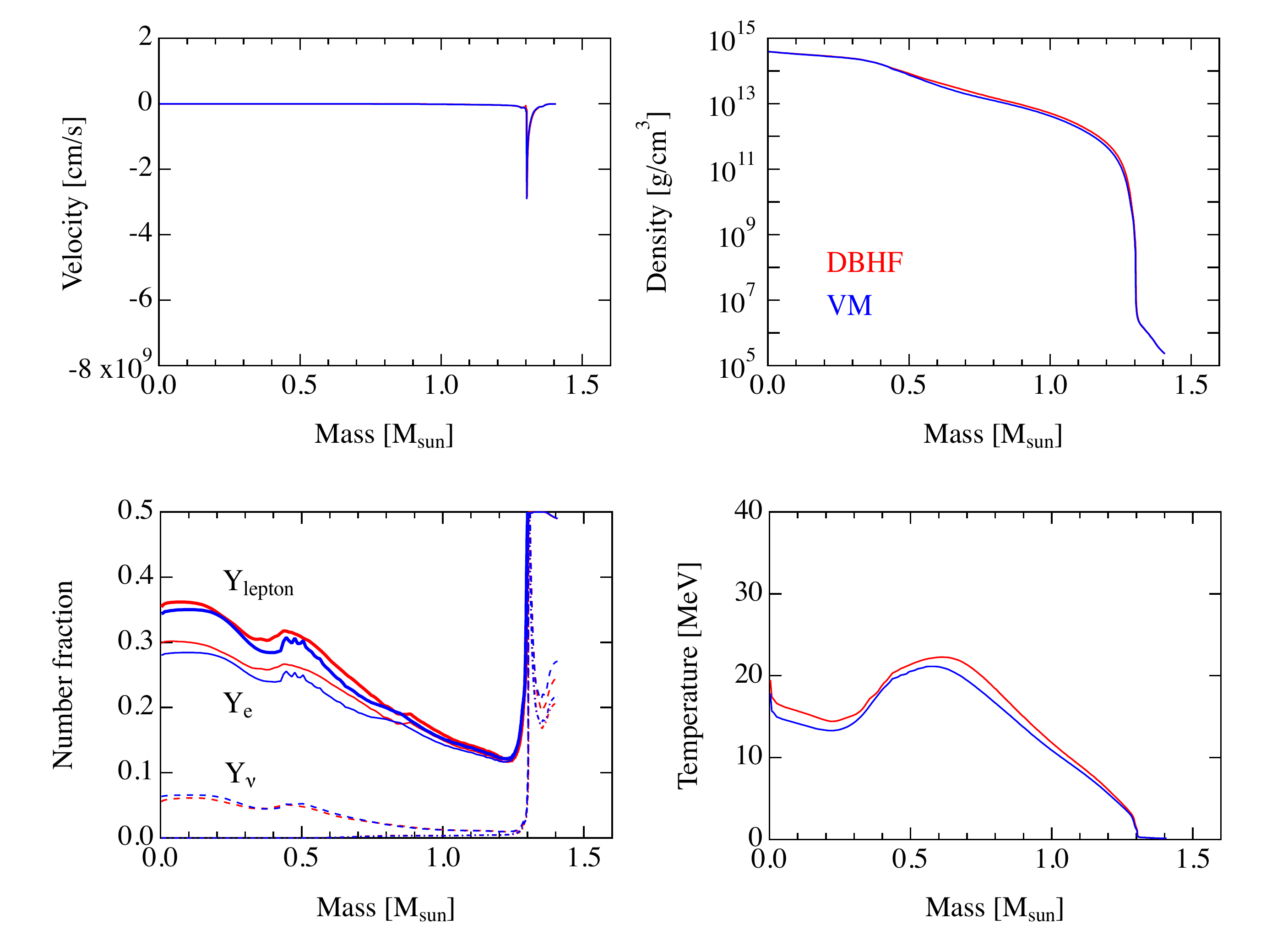}
\caption{The profiles at 100 ms after the bounce are compared for the two models of 11.2M$_{\odot}$ star using DBHF EOS and VM EOS in the same way as in Fig. \ref{fig:radhyd_tpb0_DBHF}.  
\label{fig:radhyd_tpb100_DBHF}}
\end{figure*}

In Fig. \ref{fig:radhyd_tpb100_DBHF}, we compare the profiles of the two models at 100 ms after the bounce.  The shock propagation up to this point is similar to each other.  The central density is almost the same while the density in the DBHF model is slightly higher in the outer part ($>0.5M_{\odot}$).  The temperature in the DBHF model is higher than that in the VM model, however, the difference is small within $\sim$2~MeV.  The differences in lepton, electron, and neutrino fractions remain similar to the situation at the core bounce.  

\subsubsection{Profiles of VM-S and VM models}\label{section:Results_CCSNe_VMS}

\begin{figure*}[ht]
\centering
\includegraphics[width=0.45\textwidth]{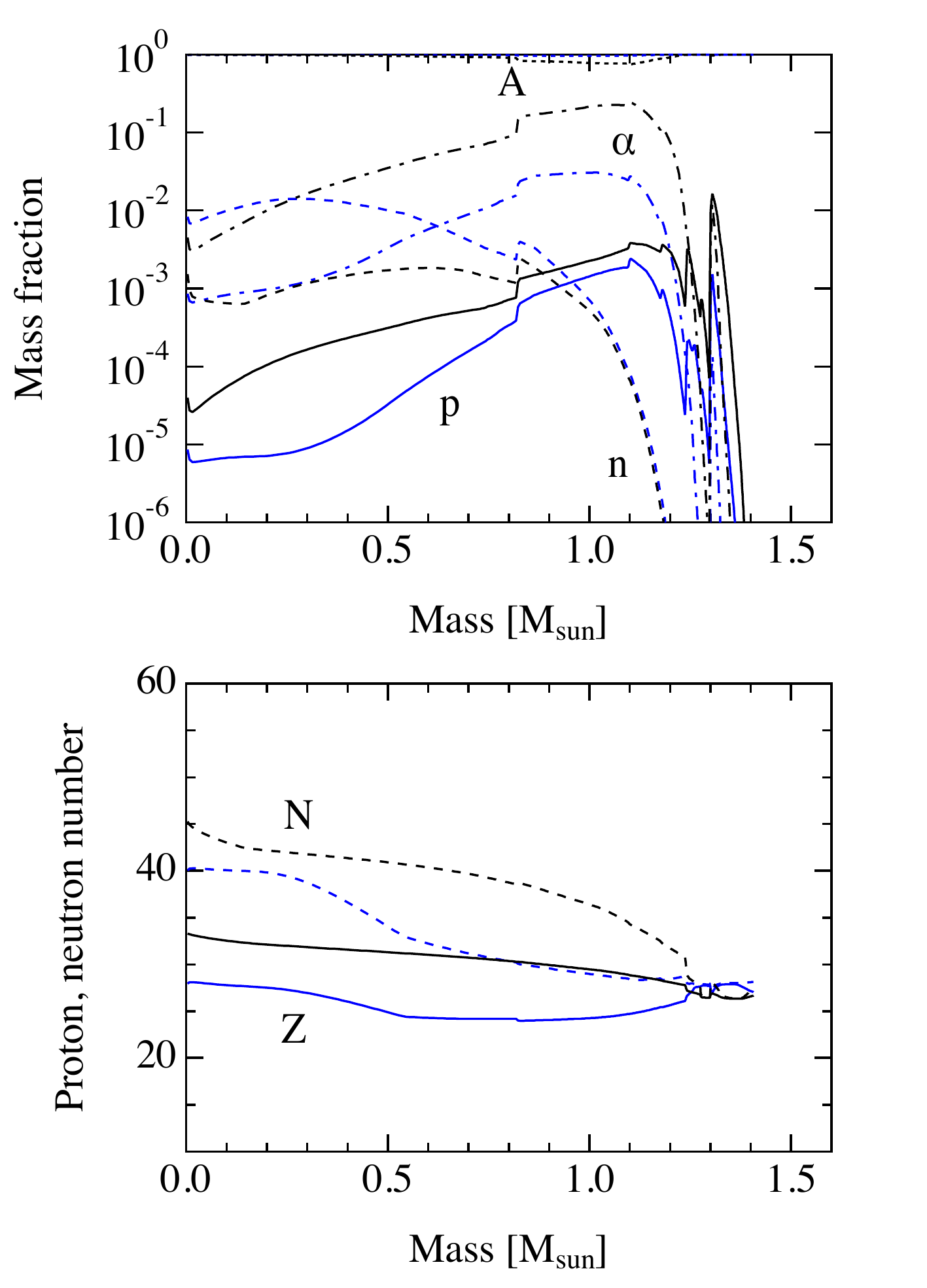}
\includegraphics[width=0.45\textwidth]{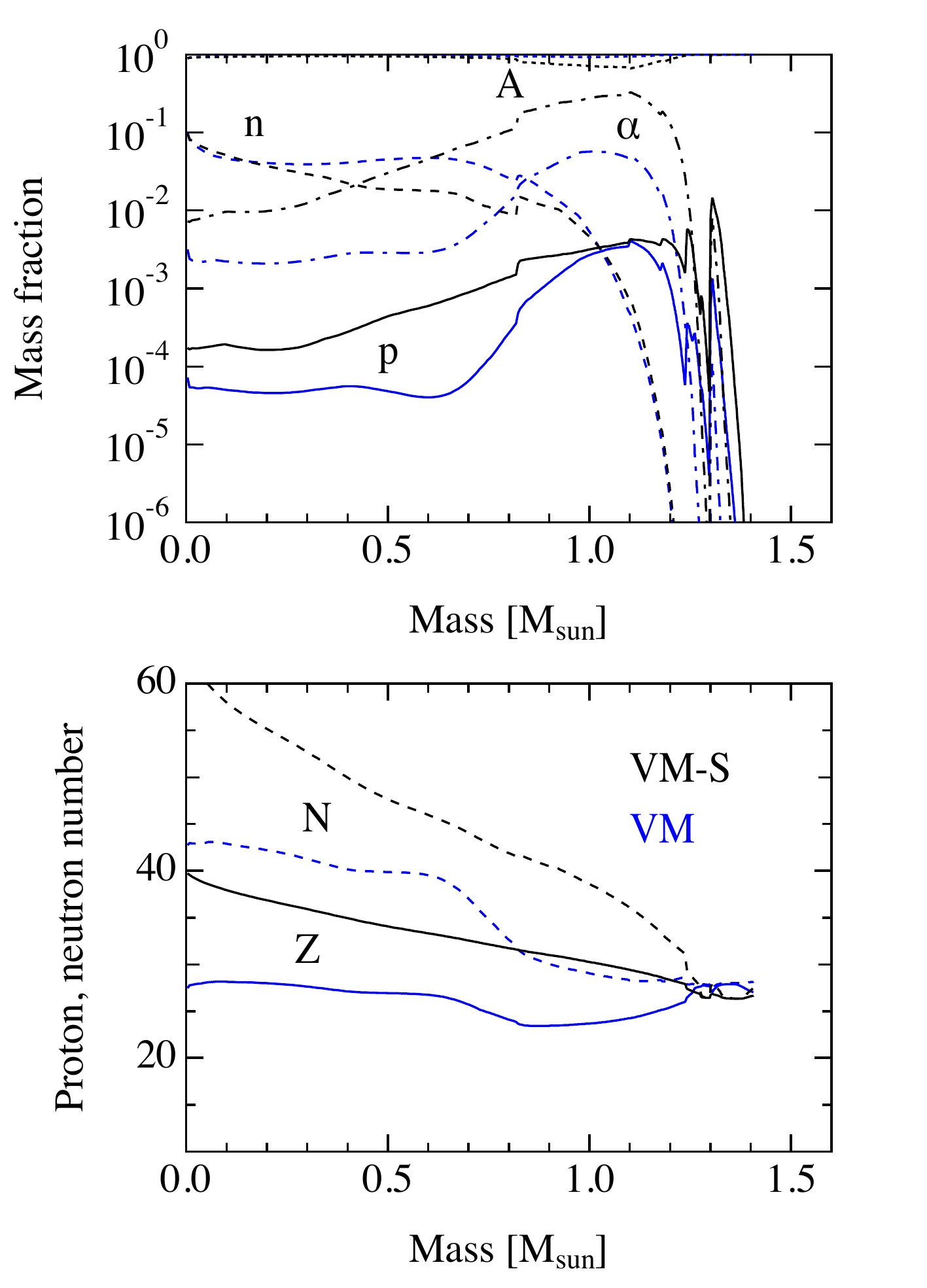}
\caption{The composition of matter is compared for the two models of 11.2M$_{\odot}$ star using VM-S EOS and VM EOS.  
The VM-S and VM models are shown by black and blue colors, respectively.  
The snapshots when the central density reaches $10^{11}$ and $10^{12}$ g\,cm$^{-3}$ are shown as a function of the baryon mass coordinate in the left and right panels, respectively.  The mass fraction of proton, neutron, alpha particles, and nuclei is shown by solid, dashed, dot-dashed, and dotted lines, respectively, in the upper panels.  The proton and neutron number of nuclei is shown by solid and dashed lines, respectively, in the lower panels.  
\label{fig:radhyd_rhoc11_VMS}}
\end{figure*}

We compare the profiles during the collapse and bounce of the central core of 11.2M$_{\odot}$ star using VM-S EOS and VM EOS.  
We show the composition of matter during the collapse for the two models in Fig. \ref{fig:radhyd_rhoc11_VMS}.  
The mass fractions are clearly different between the two models reflecting the different frameworks to describe the non-uniform matter (upper panels).  
The mass fractions of alpha particles and free protons in the VM-S model are larger than those in the VM model.  
The amount of free protons may affect the distribution of electron fraction through the neutrino emission via electron captures.  
The mass fraction of nuclei in the two models is similar in the inner part ($\lesssim 0.5$M$_{\odot}$ in the baryon mass coordinate) although it becomes small for the VM-S model in the outer layer ($\sim1.0$M$_{\odot}$) due to the appearance of alpha particles.  
The proton and neutron number of nuclei in the VM-S model are generally larger than in the VM model.  
Larger mass numbers contribute to a larger opacity for neutrinos through the coherent scattering on nuclei and affect the neutrino trapping.  
It is beneficial to provide detailed distribution of nuclei in the NSE to precisely determine the opacity \cite{bur84,lan03,fur11}.  

The main channel of electron captures changes during the collapse in the two models depending on the composition.  
The neutron number is larger in the VM-S model and has tendency of suppression of the electron capture on nuclei using the blocking factor for $N > 40$ in the Bruenn's prescription.  
The electron capture on protons proceeds in the VM-S model at the central density $10^{11}$ g\,cm$^{-3}$ while the electron capture on nuclei proceeds in the VM model.  The electron capture on nuclei in the two models is hindered afterward and the electron capture on protons becomes the main source at the central density $10^{12}$ g\,cm$^{-3}$.  
This behavior may be altered when different rates of the electron capture on nuclei are adopted.  The thermal unblocking in nuclei can enhance the electron capture rates \cite{lan03prl,joh22} and may weaken the effects of free protons.  
It is to be noted that the weak reactions using the mixture of nuclei in the NSE may affect the outcome of the collapse as studied in \cite{hix03}.  
The numerical studies by solving the Boltzmann equation using the electron capture rates averaged over the mixture of nuclei \cite{nag18} have been made to clarify the influence of various treatments with VM EOS \cite{nag19a}.  
Further studies on the electron captures are necessary and numerical studies using the electron capture rates consistent with DBHF EOS are going on and will be reported elsewhere \cite{nag22x}.  

\begin{figure*}[ht]
\centering
\includegraphics[width=0.9\textwidth]{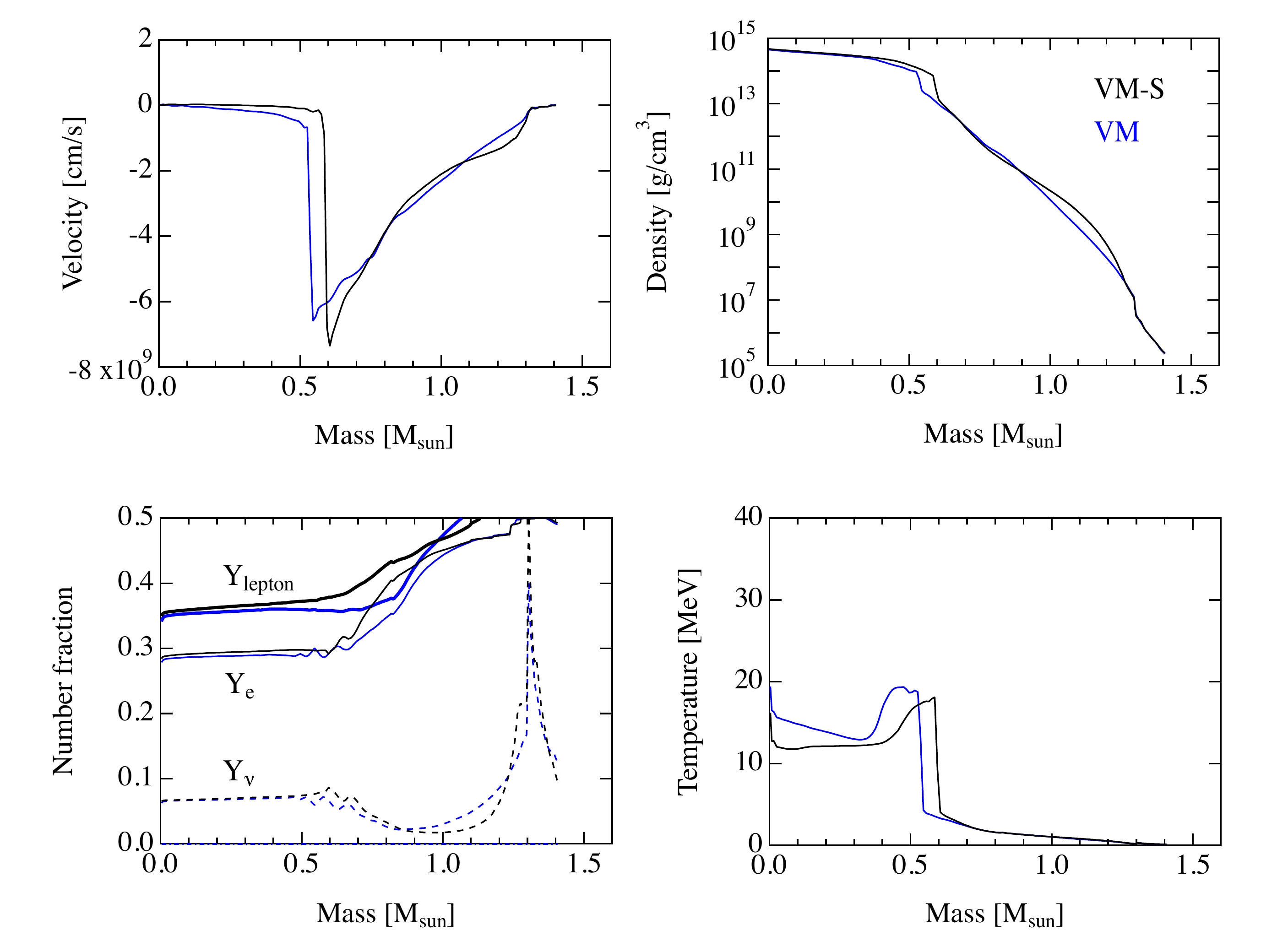}
\caption{The profiles at the core bounce are compared for the two models of 11.2M$_{\odot}$ star using VM-S EOS and VM EOS.  
The velocity, density, lepton fractions, and temperature are shown as functions of the baryon mass coordinate in the upper-left, upper-right, lower-left, and lower-right panels, respectively.  
The lepton, electron, and neutrino fractions are shown by thick-solid, solid, and dashed lines, respectively, in the lower-left panel.  
\label{fig:radhyd_tpb0_VMS}}
\end{figure*}

Figure \ref{fig:radhyd_tpb0_VMS} shows the profiles at the core bounce in the VM-S and VM models.  
The size of bounce core in the VM-S model is larger than that of the VM model.  
The position of the shock wave is located at 0.595M$_{\odot}$ and 0.535M$_{\odot}$ in the mass coordinate for the VM-S and VM models, respectively.  
The different size of bounce core is consistent with the fact that the lepton fraction in the VM-S model is larger than that in the VM model.  
The large lepton fraction is attained due to the enhanced neutrino trapping with the larger mass nuclei in the VM-S model.  
It is also attained through the difference in the electron fractions.  
The electron fraction in the VM-S model is found to be larger despite there is competition among the channels of electron capture rates.  
The neutrino fraction is similar at the center and even larger at $\sim$0.6M$_{\odot}$ due to strong neutrino trapping.  
The lepton fraction, which is the sum of electron fraction and neutrino fraction, in the VM-S model is larger accordingly.  Note that the size of the bounce core is affected by the lepton fraction and other factors such as finite temperature and composition \cite{bar90,tim96}.  
It is interesting to see that the difference of the bounce core due to the composition (VM-S vs VM) is larger than that due to the uniform matter (DBHF vs VM).  The different treatments of the composition can be more influential through the amount of neutrino trapping than the different properties of uniform matter.  

The density in the VM-S model is slightly higher than in the VM model while the temperature is lower.  The entropy per baryon in the VM-S model is lower throughout the collapse keeping the initial difference for the same initial temperature.  The differences are more clear than those in the case of the comparison between DBHF and VM models.  

\begin{figure*}[ht]
\centering
\includegraphics[width=0.45\textwidth]{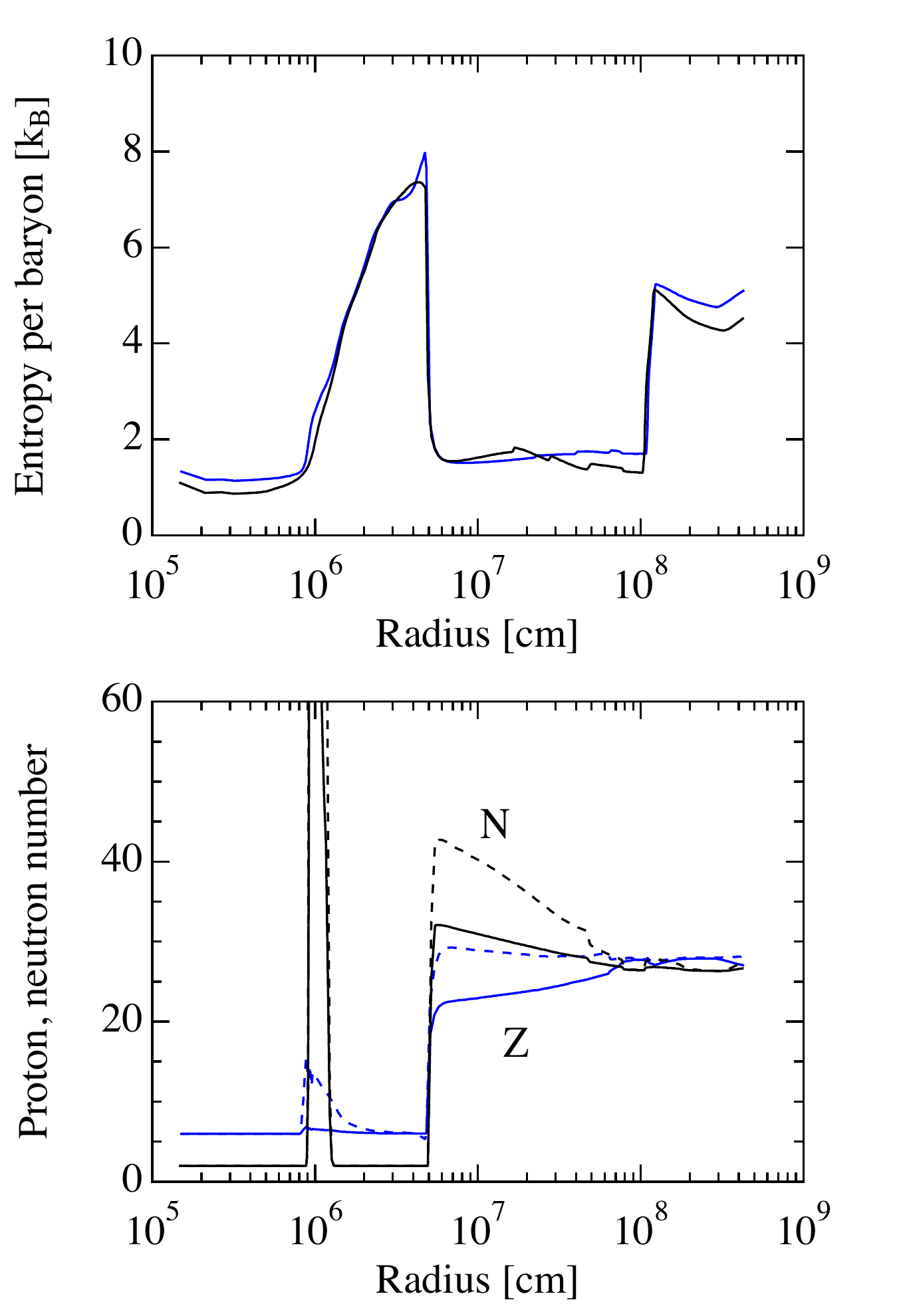}
\includegraphics[width=0.45\textwidth]{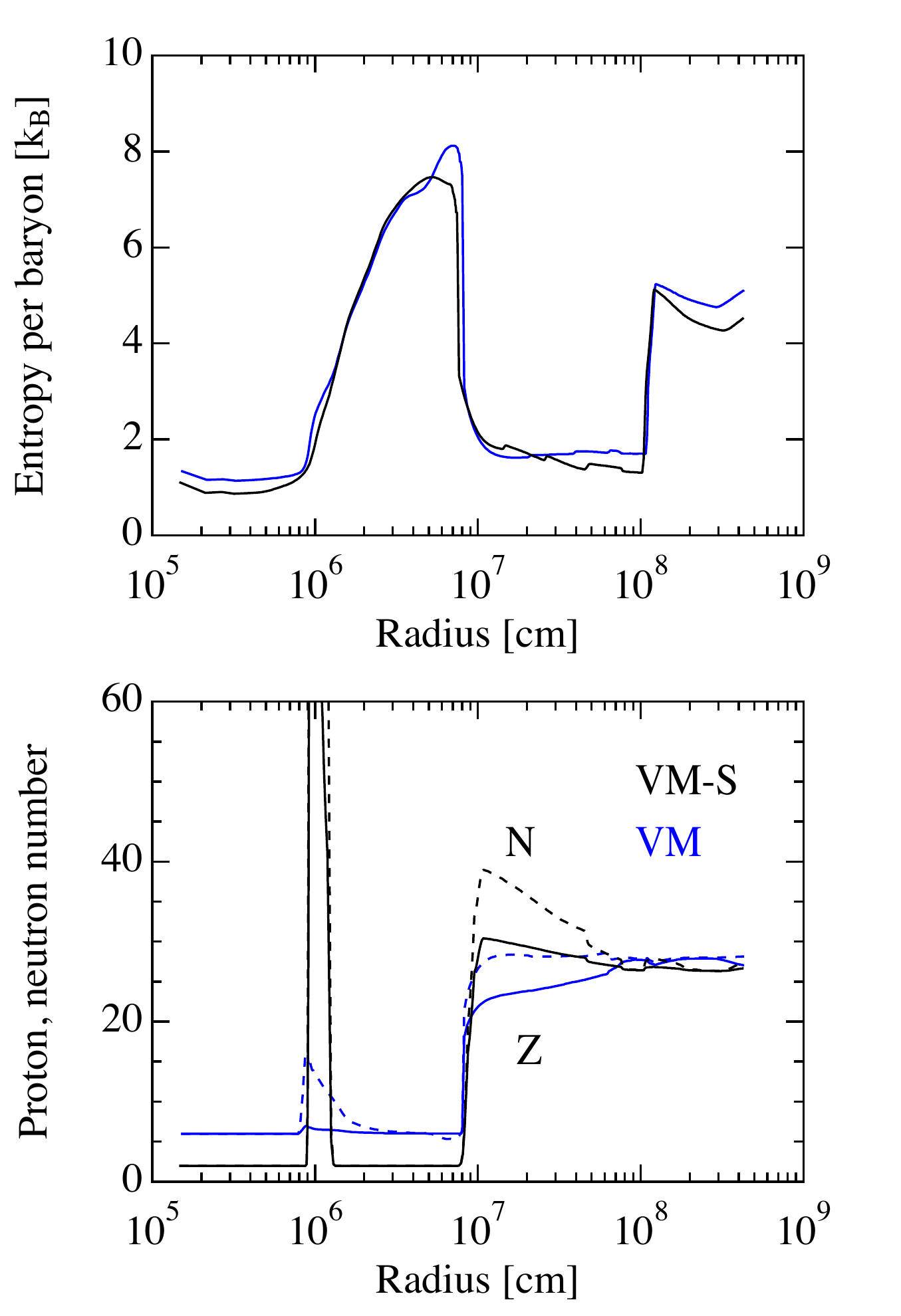}
\caption{Radial profiles of entropy per baryon (top) and proton, neutron number of nuclei (bottom) for the two models of 11.2M$_{\odot}$ star using VM-S EOS and VM EOS are shown for the snapshots at 1 and 2 ms after the bounce in the left and right panels, respectively.  
The proton and neutron number of nuclei is shown by solid and dashed lines, respectively, in the lower panels.  
The peak values of proton and neutron numbers at the radius $10^6$~cm are $\sim$370 and $\sim$910 at 1~ms and $\sim$410 and $\sim$1060 at 2~ms, respectively, for the VM-S model.  
\label{fig:radhyd_tpb1_VMS}}
\end{figure*}

It is interesting to find the clear differences in the composition right after the core bounce between the VM-S and VM models due to the different treatments of composition.  
We compare in Fig. \ref{fig:radhyd_tpb1_VMS} the proton and neutron number of nuclei as a function of radius for the two models.  
The proton and neutron number in the unshocked matter in the layers above the shock wave at $\sim50$~km is larger in the VM-S model than in the VM model at 1 ms after the bounce (left panels).  
This difference remains during the propagation of the shock wave up to $\sim80$~km at 2 ms after the bounce (right panels).  
The large mass number of nuclei in the VM-S model leads to the large energy loss due to dissociation of nuclei when the falling matter hits the shock wave.  
This leads to the decrease of entropy through the neutrino emission by weak reactions with dissociated neutrons and protons.  
It contributes to a more blunt peak in the entropy profile in the VM-S model than in the VM model.  

This change of entropy profile may affect the occurrence of the convection due to different degree of its negative gradient.  
The effect of composition on the convective condition is seen in the numerical study of 2D calculation of core-collapse supernova \cite{har20} by adopting the Furusawa EOS \cite{fur17a} using the NSE treatment and the Lattimer-Swesty EOS \cite{lat91} using the single nucleus approximation.  
We stress that the current study shows the influence of composition using the common input of uniform matter while those two sets of EOS table adopted the different frameworks of non-uniform matter as well as uniform matter.  

\begin{figure*}[ht]
\centering
\includegraphics[width=0.9\textwidth]{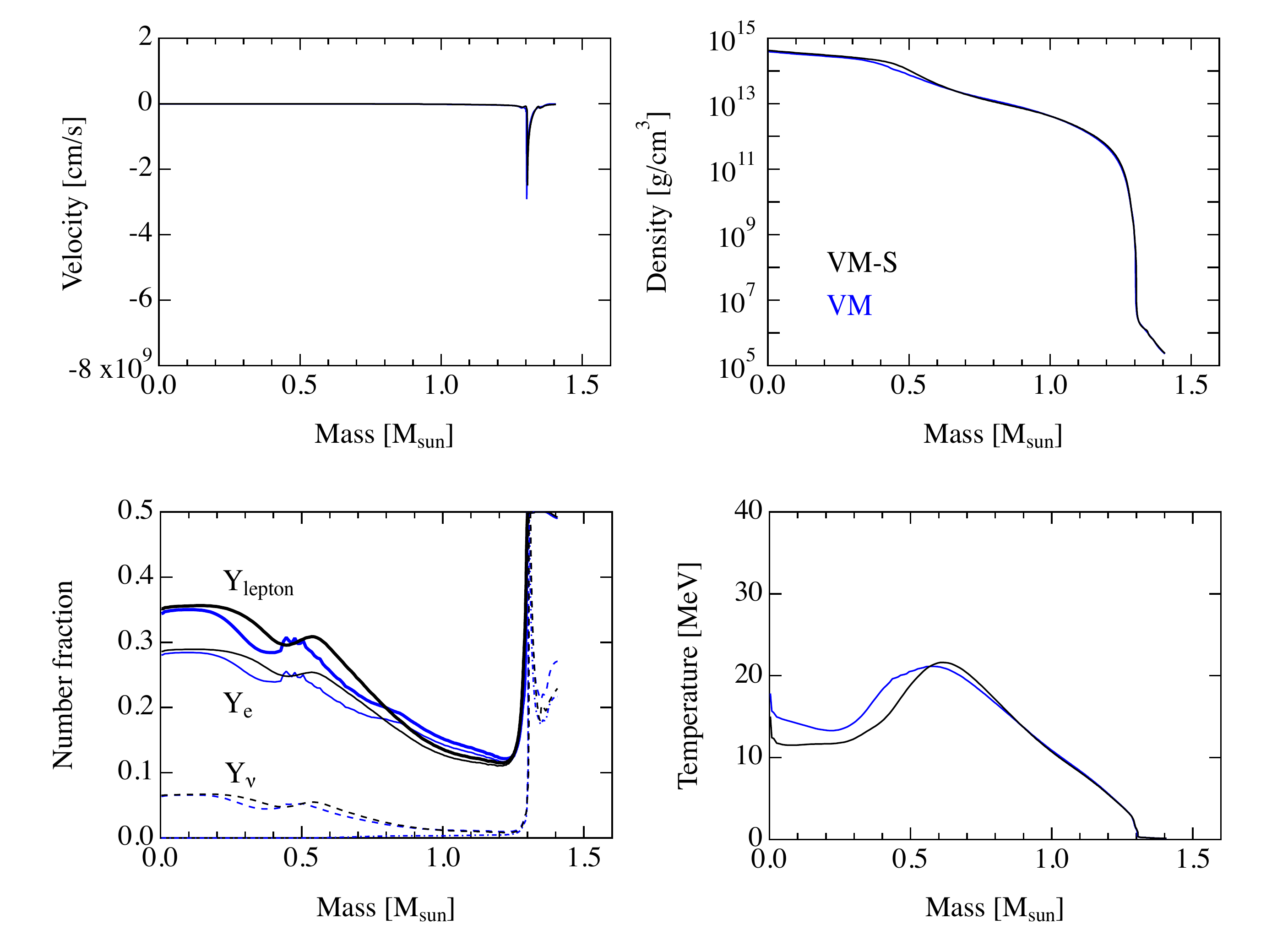}
\caption{The profiles at 100 ms after the bounce are compared for the two models of 11.2M$_{\odot}$ star using VM-S EOS and VM EOS in the same way as in Fig. \ref{fig:radhyd_tpb0_VMS}.  
\label{fig:radhyd_tpb100_VMS}}
\end{figure*}

We show in Fig. \ref{fig:radhyd_tpb100_VMS} the profiles at 100 ms after the bounce for the two models.  
The difference is found to be rather mild with similar shock positions and small differences of density, temperature, and lepton fractions.  
The peak temperatures are almost the same and the temperature at the center is slightly lower in the VM-S model.  
The size of differences between the VM-S and VM models is comparable or even a little larger than that between the DBHF and VM models.  
%


\subsubsection{Neutrino emission}\label{section:Results_CCSNe_Neutrino}

\begin{figure*}[ht]
\centering
\includegraphics[width=0.45\textwidth]{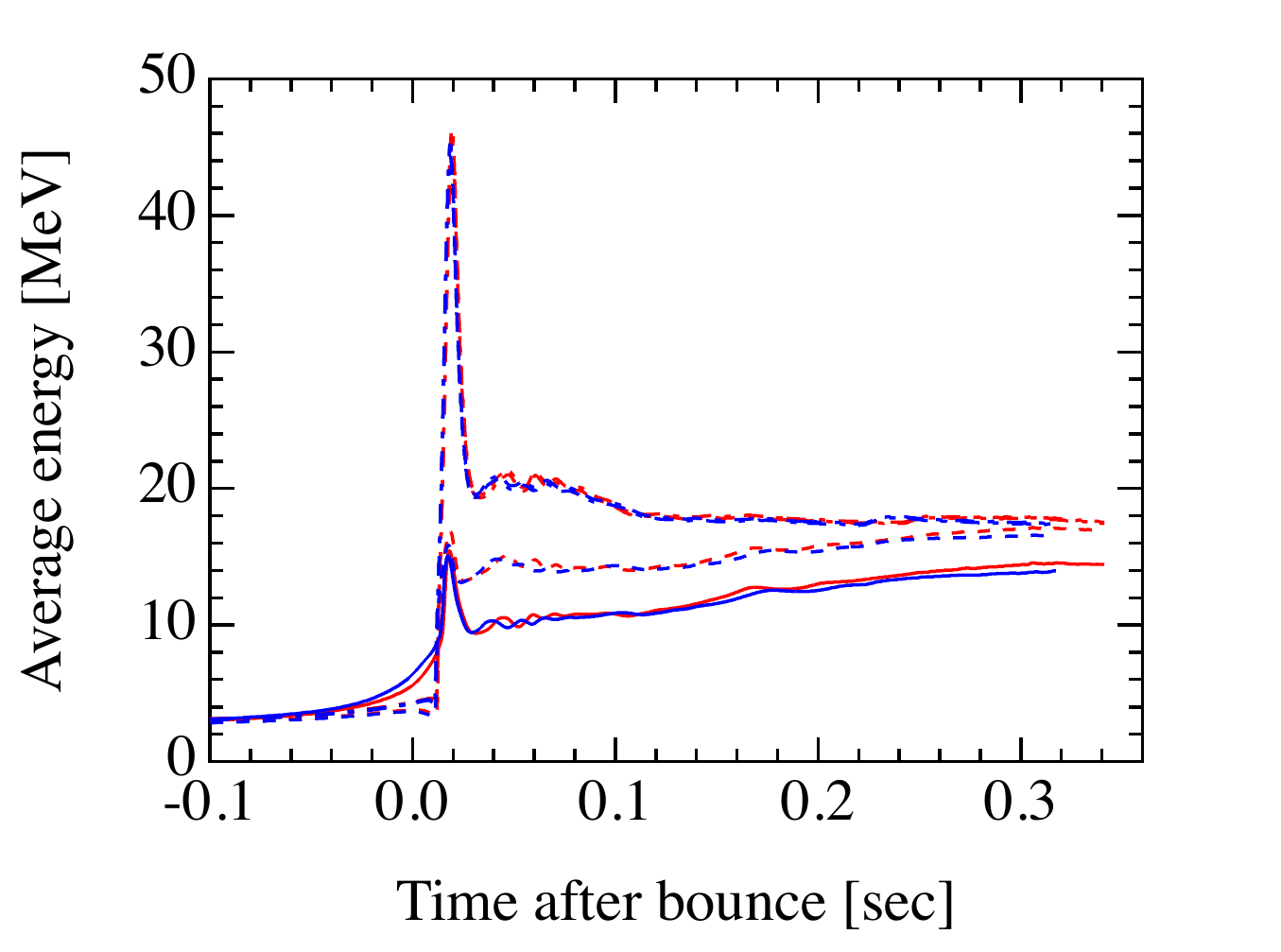}
\includegraphics[width=0.48\textwidth]{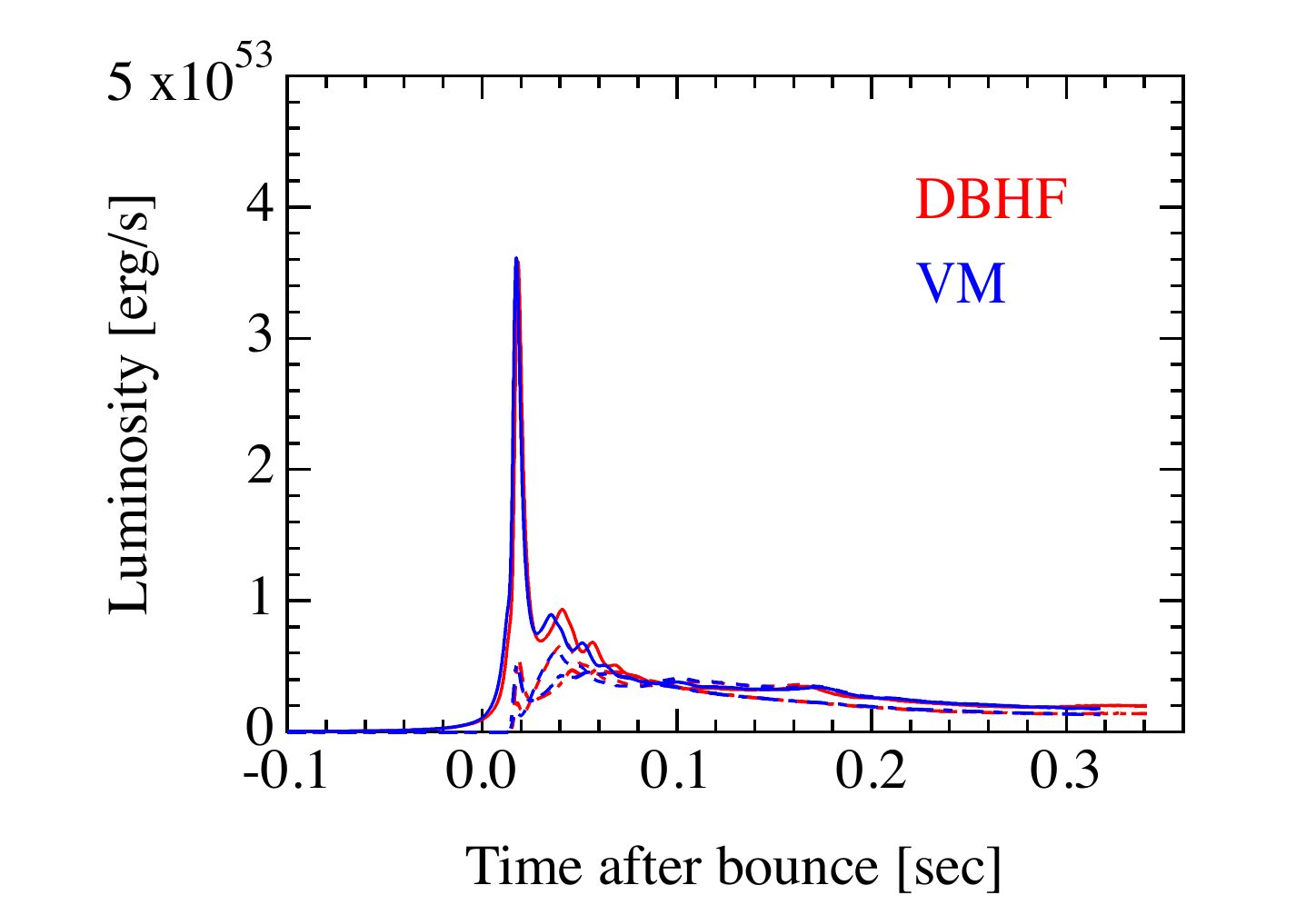}
\includegraphics[width=0.45\textwidth]{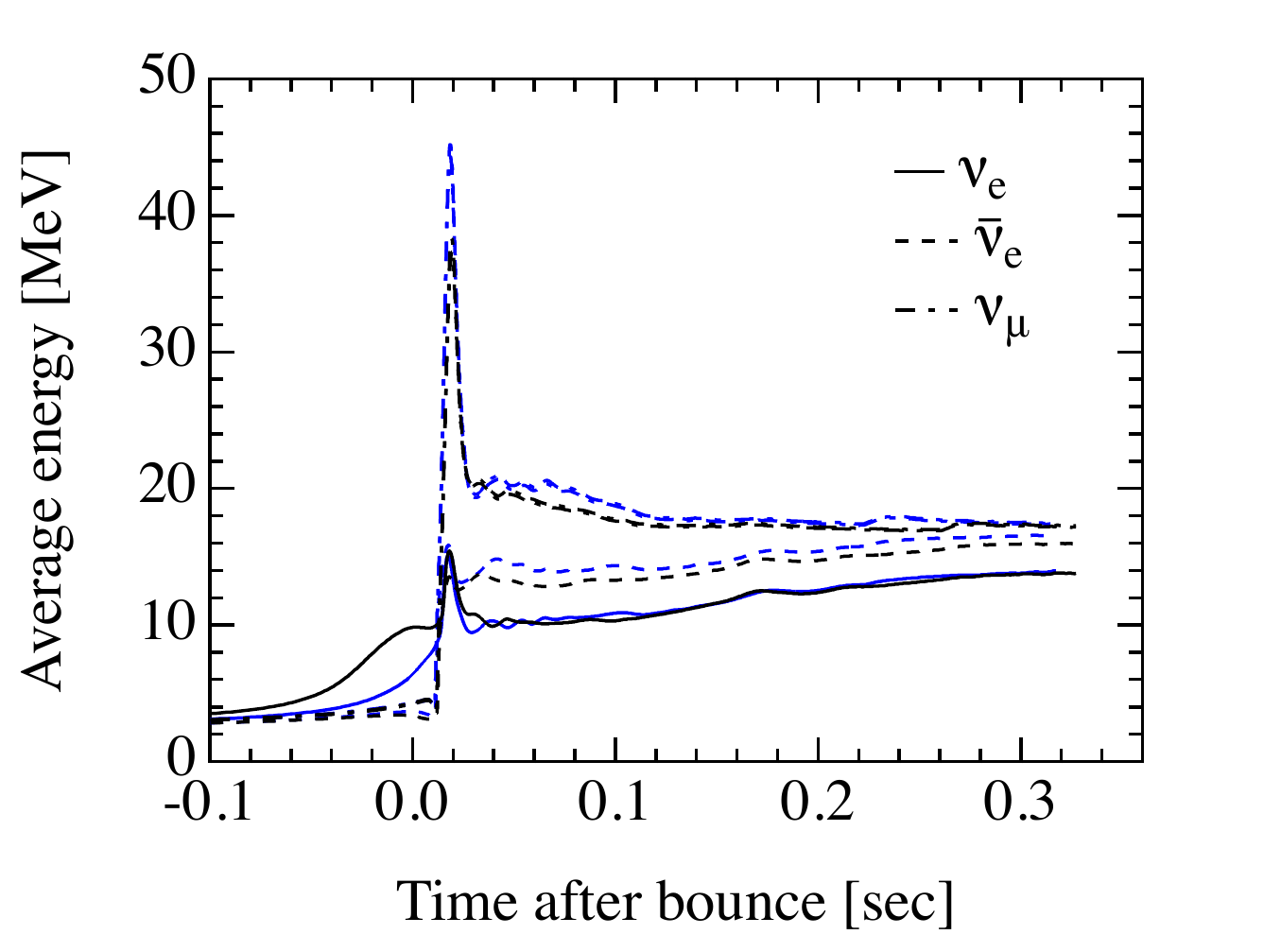}
\includegraphics[width=0.48\textwidth]{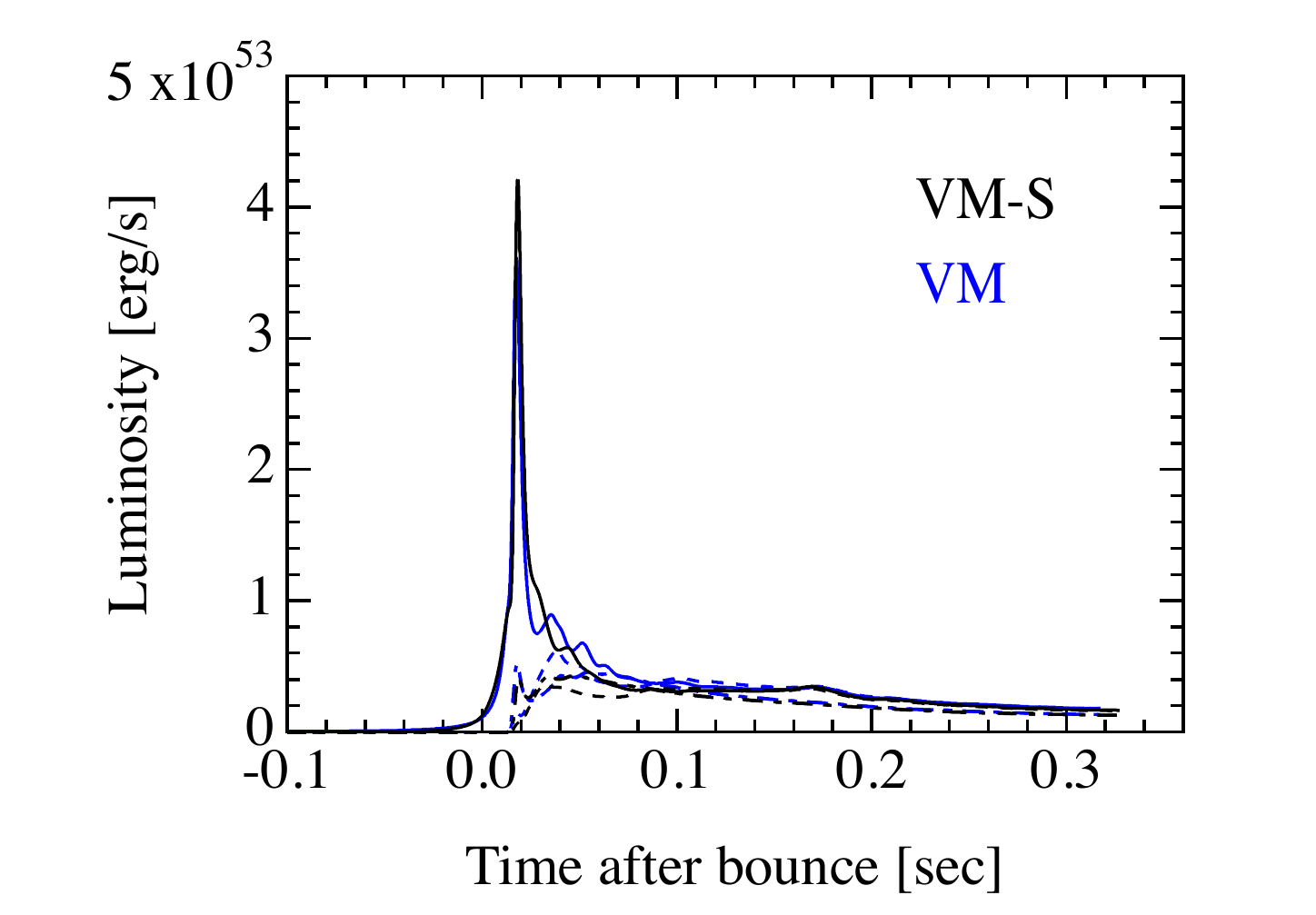}
\caption{Average energies (left) and luminosities (right) of neutrinos emitted from the core-collapse simulation of 11.2M$_{\odot}$ star using the three sets of EOS.  
The DBHF and VM models are shown by red and blue colors,  respectively, in the top panels.  
The VM-S and VM models are shown by black and blue colors, respectively, in the bottom panels.  
The neutrino species, $\nu_{e}$, $\bar{\nu}_e$, and $\nu_{\mu}$ are denoted by solid, dashed, and dot-dashed lines, respectively.  
\label{fig:radhyd_Nu_3eos}}
\end{figure*}

We show in Fig. \ref{fig:radhyd_Nu_3eos} the average energies and luminosities of neutrinos as a function of the time after the bounce for three species, $\nu_{e}$, $\bar{\nu}_e$, and $\nu_{\mu}$.  Comparison of the DBHF and VM-S models with respect to the VM model is shown in the upper and lower panels, respectively.  
The neutrino signal of the DBHF model is quite similar to the one of the VM model, reflecting the small difference of the profiles seen above.  In the case of the VM-S model, differences appear in the average energies and luminosities around the core bounce.  The difference of average energy amounts to a few MeV in some cases between the VM-S and VM models.  The different treatments of composition has influence on the neutrino signal more clearly than those of the uniform matter.  

\subsubsection{different progenitor}\label{section:Results_CCSNe_Progenitor}

\begin{figure*}[ht]
\centering
\includegraphics[width=0.9\textwidth]{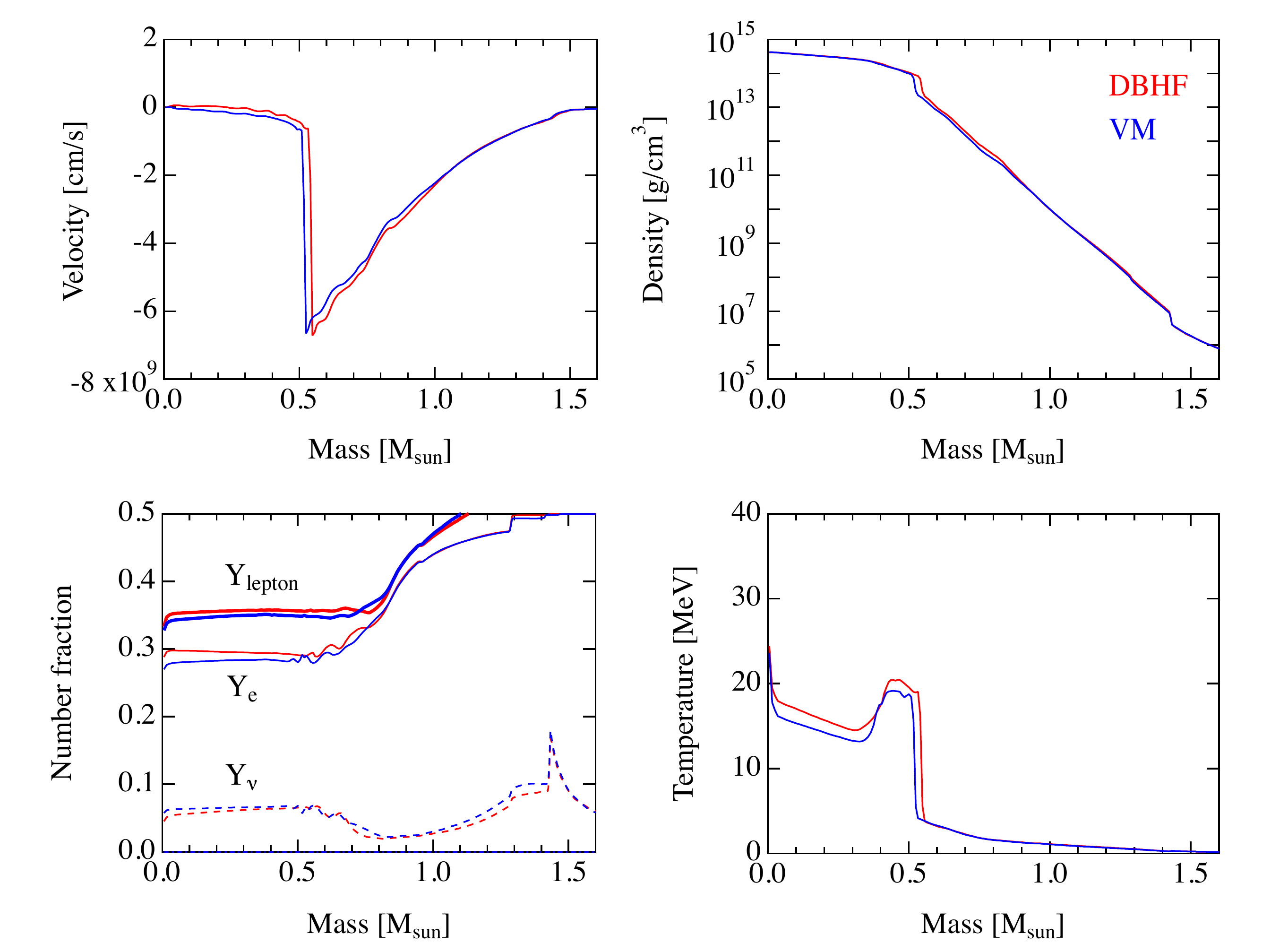}
\caption{The profiles at the core bounce are compared for the two models of 15M$_{\odot}$ star using DBHF EOS and VM EOS.  
The DBHF and VM models are shown by red and blue colors, respectively.  
The velocity, density, lepton fractions, and temperature are shown as functions of the baryon mass coordinate in the upper-left, upper-right, lower-left, and lower-right panels, respectively.  
The lepton, electron, and neutrino fractions are shown by thick-solid, solid, and dashed lines, respectively, in the lower-left panel.  
\label{fig:radhyd_15M_tpb0_DBHF}}
\end{figure*}

\begin{figure*}[ht]
\centering
\includegraphics[width=0.9\textwidth]{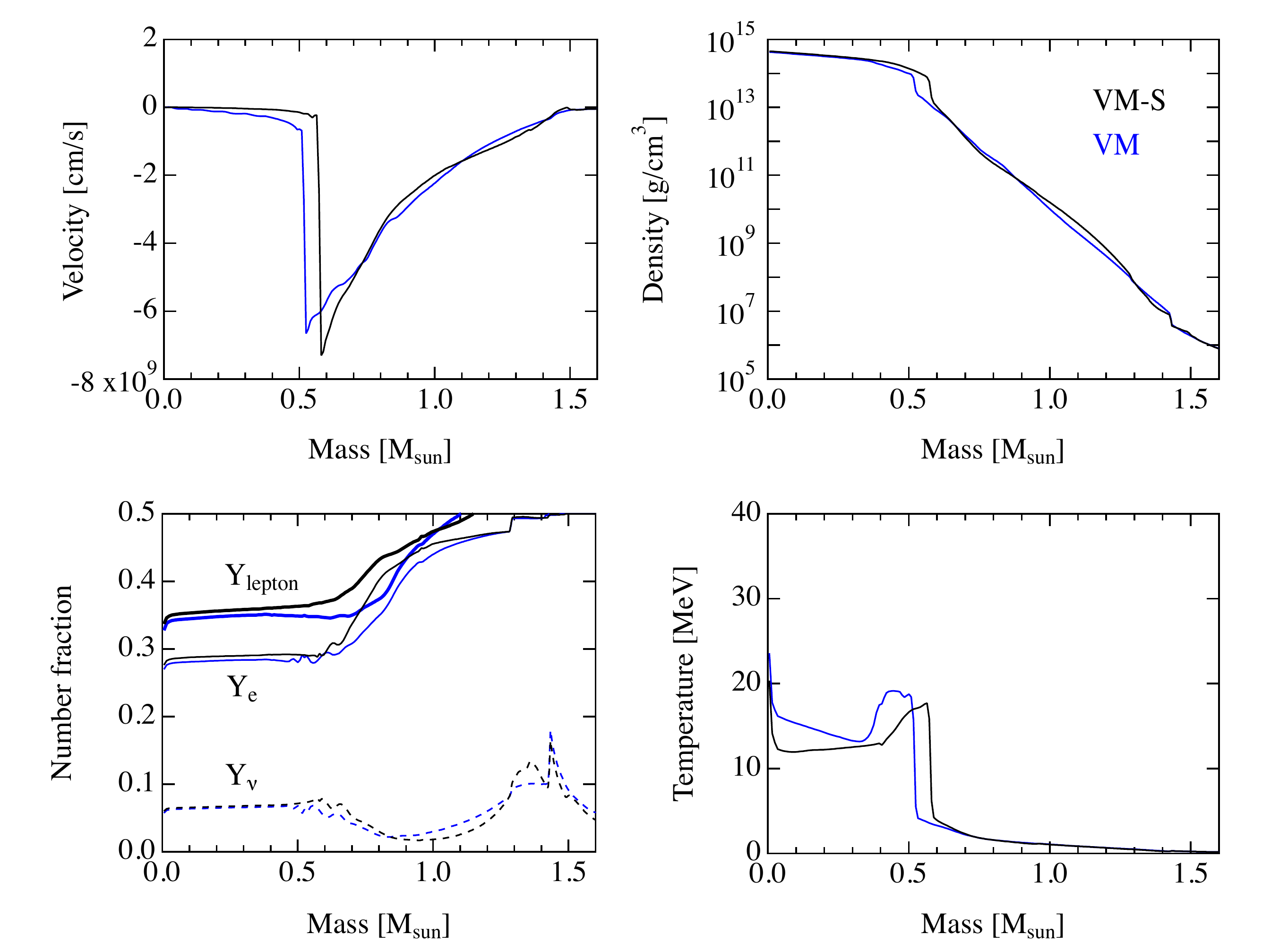}
\caption{The profiles at the core bounce are compared for the two models of 15M$_{\odot}$ star using VM-S EOS and VM EOS.  
The VM-S and VM models are shown by black and blue colors, respectively.  
The velocity, density, lepton fractions, and temperature are shown as functions of the baryon mass coordinate in the upper-left, upper-right, lower-left, and lower-right panels, respectively.  
The lepton, electron, and neutrino fractions are shown by thick-solid, solid, and dashed lines, respectively, in the lower-left panel.  
\label{fig:radhyd_15M_tpb0_VMS}}
\end{figure*}

We additionally show the model using the 15M$_{\odot}$ star to strengthen generality of the features we have seen above.  
In Fig, \ref{fig:radhyd_15M_tpb0_DBHF}, comparison of the profiles at the core bounce is shown for the two models of 15M$_{\odot}$ using DBHF EOS and VM EOS.  The feature of difference is similar to the case of 11.2M$_{\odot}$ as shown in Fig. \ref{fig:radhyd_tpb0_DBHF}.  The size of bounce core, lepton fraction, and temperature in the DBHF model is slightly larger than those in the VM model.  In Fig. \ref{fig:radhyd_15M_tpb0_VMS}, comparison at the core bounce is shown for the two models of 15M$_{\odot}$ using VM-S EOS and VM EOS.  As in the case of 11.2M$_{\odot}$ as shown in Fig. \ref{fig:radhyd_tpb0_VMS}, the size of the bounce core and the leptron fraction in the VM-S model are larger than those in the VM model.  The temperature in the VM-S model is lower than that of the VM model, similarly to Fig. \ref{fig:radhyd_tpb0_VMS}.  The differences of the profiles between the VM-S and VM models are more clear than the differences between the DBHF and VM models.  This is consistent with the comparison in the case of 11.2M$_{\odot}$.  

\begin{figure*}[ht]
\centering
\includegraphics[width=0.45\textwidth]{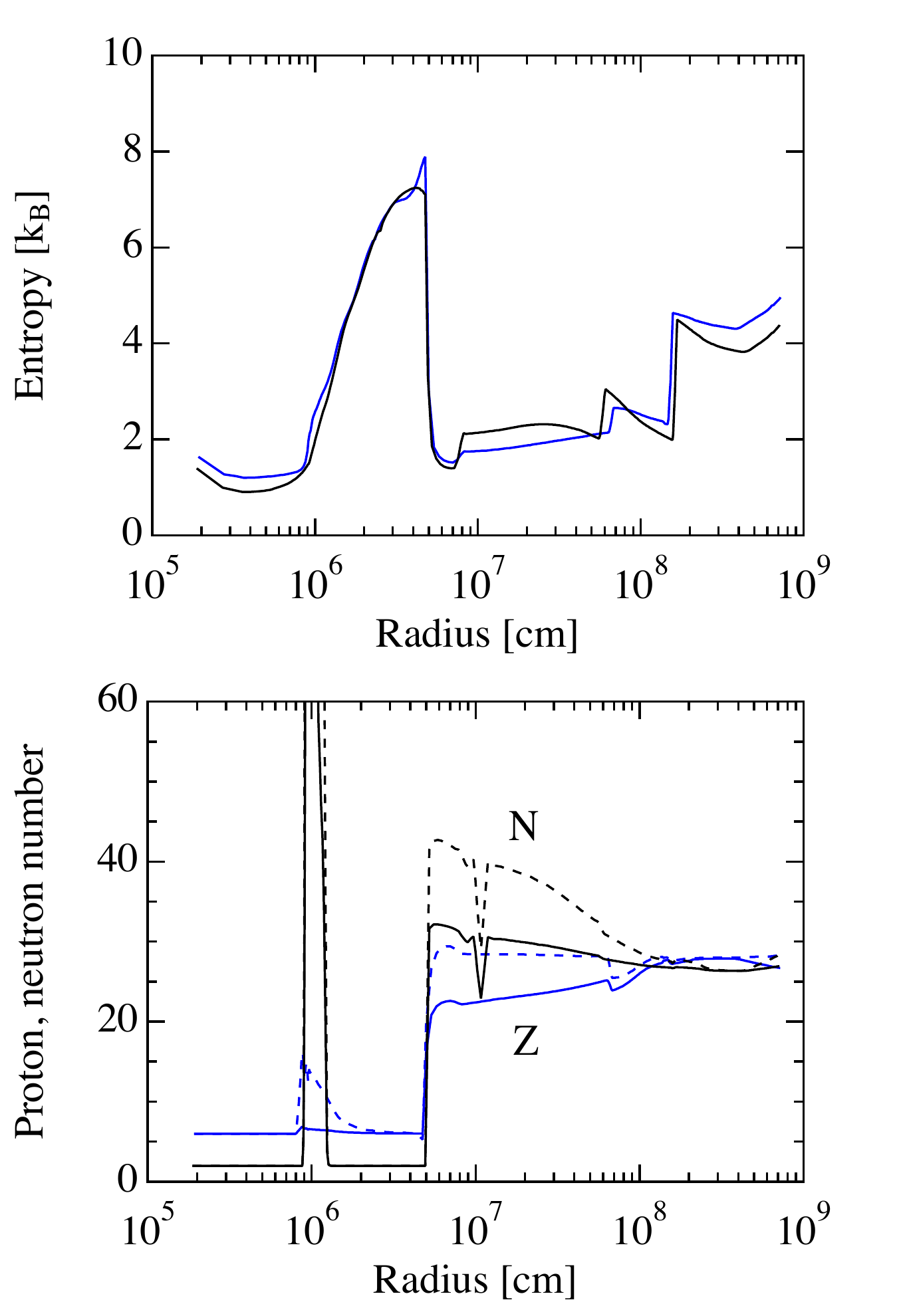}
\includegraphics[width=0.45\textwidth]{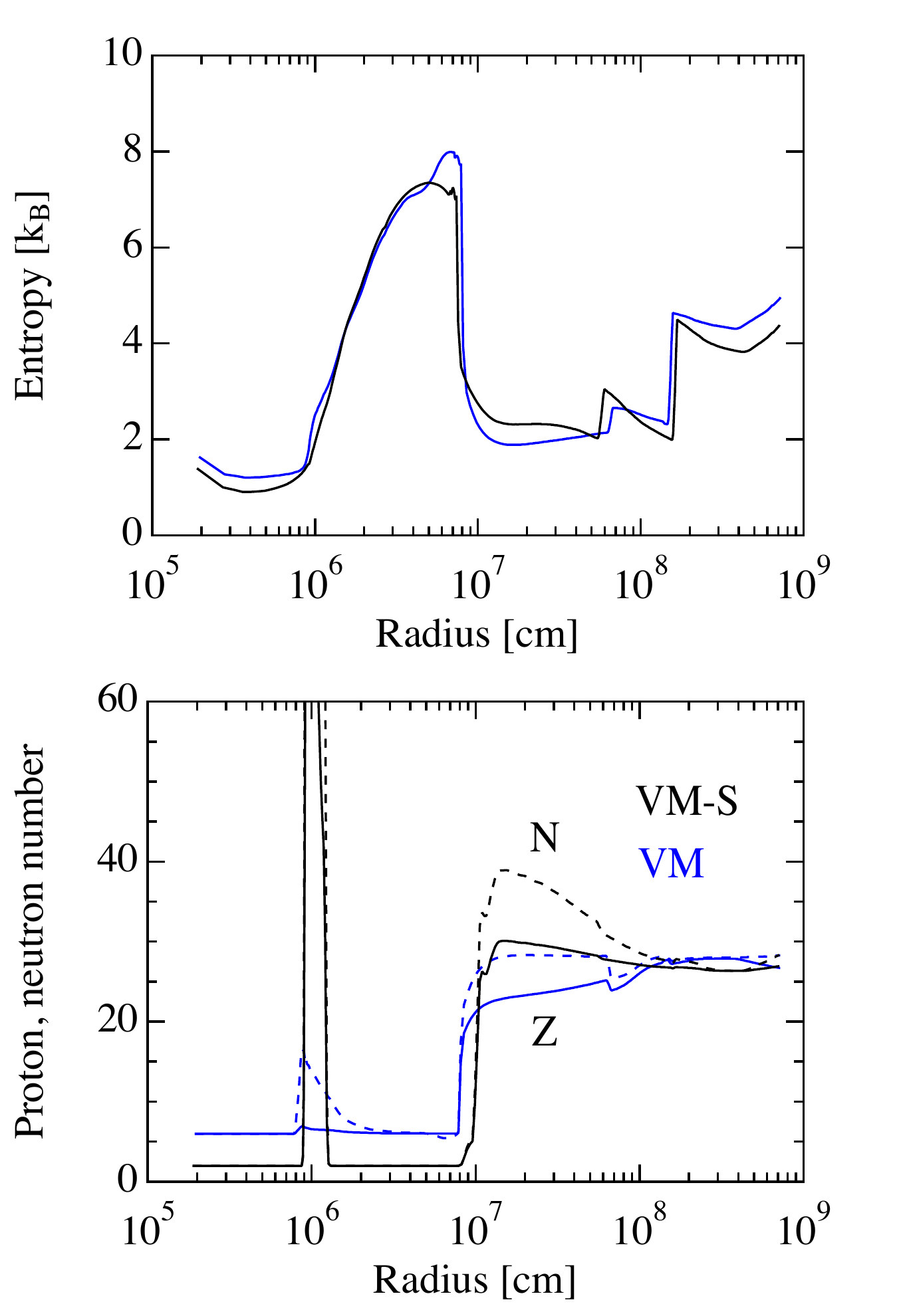}
\caption{Radial profiles of entropy per baryon (top) and proton, neutron number of nuclei (bottom) for the two models of 15M$_{\odot}$ star using VM-S EOS and VM EOS are shown for the snapshots at 1 and 2 ms after the bounce in the left and right panels, respectively.  
The proton and neutron number of nuclei is shown by solid and dashed lines, respectively, in the lower panels.  
The peak values of proton and neutron numbers at the radius $10^6$~cm are $\sim$380 and $\sim$970 at 1~ms and $\sim$420 and $\sim$1120 at 2~ms, respectively, for the VM-S model.  
\label{fig:radhyd_15M_tpb1_VMS}}
\end{figure*}

We found that 
differences of the composition in the collapse and bounce of the 15M$_{\odot}$ star appear more clearly in the comparison of the VM-S and VM models than in the comparison of the DBHF and VM models.  This tendency is consistent with the case of 11.2M$_{\odot}$.  The difference of nuclear species and its influence on the entropy right after the bounce can be also seen in the case of 15M$_{\odot}$.  Figure \ref{fig:radhyd_15M_tpb1_VMS} shows the comparison of the profiles in the case of 15M$_{\odot}$.  In the VM-S model, the proton and neutron numbers of nuclei are larger than that in the VM model.  Accordingly the entropy distribution in the VM-S model does not have the sharp peak at the shock wave suggesting the loss of energy due to the dissociation of heavy nuclei.  The negative gradient of the entropy in the VM-S model may trigger the convective motion also in this case.  


\subsection{Proto-neutron star cooling}\label{section:Results_PNSC}

\begin{figure*}[ht]
\centering
\includegraphics[width=0.9\textwidth]{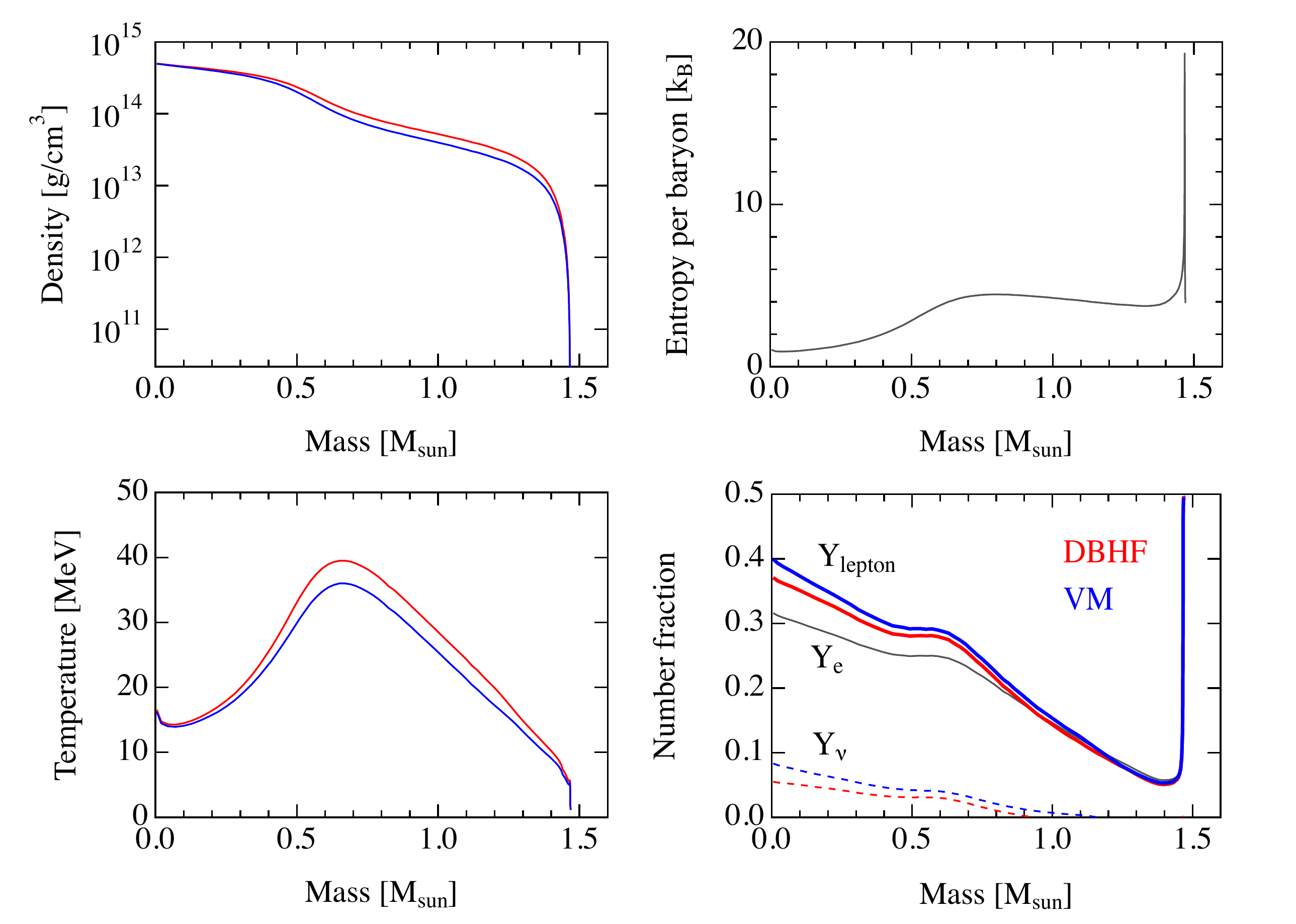}
\caption{Initial profiles of proto-neutron stars 
using DBHF EOS and VM EOS are shown by red and blue lines, respectively.  
Profiles of the density, entropy per baryon, temperature, and number fractions (leptons, electrons, and neutrinos) are shown in upper-left, upper-right, lower-left, and lower-right panels, respectively.  
The profiles of entropy per baryon and electron fraction (gray lines) are given to construct the initial model for both EOSs.  
\label{fig:pnsc_profile0_DBHF}}
\end{figure*}

In Fig. \ref{fig:pnsc_profile0_DBHF}, we show the initial models for the DBHF and VM models for the cooling simulations.  
We adopt the initial profiles of electron fraction and entropy per baryon as functions of the radial mass coordinate to construct the stationary configuration as described in section \ref{section:Simulation_PNSC}.  
We set the common initial profiles for the three EOSs to focus on the long-term evolution without uncertain factors of accretion and to separate effects from those due to the core-collapse and bounce in the current study (See discussion in section \ref{section:Simulation_PNSC}).  
The profiles of the VM-S model (not shown) are almost identical to those of the VM model.  The density and temperature in the DBHF model are slightly higher than those in the VM model although the central values are almost the same.  
The lepton fraction in the DBHF model is smaller than that in the VM model since the neutrino fraction is smaller due to the larger symmetry energy.  
The large symmetry energy affects the difference of neutron and proton chemical potentials and reduces the neutrino chemical potential under the chemical equilibrium.  

\subsubsection{Profiles of DBHF and VM models}\label{section:Results_PNSC_DBHF}

\begin{figure*}[ht]
\centering
\includegraphics[width=0.32\textwidth]{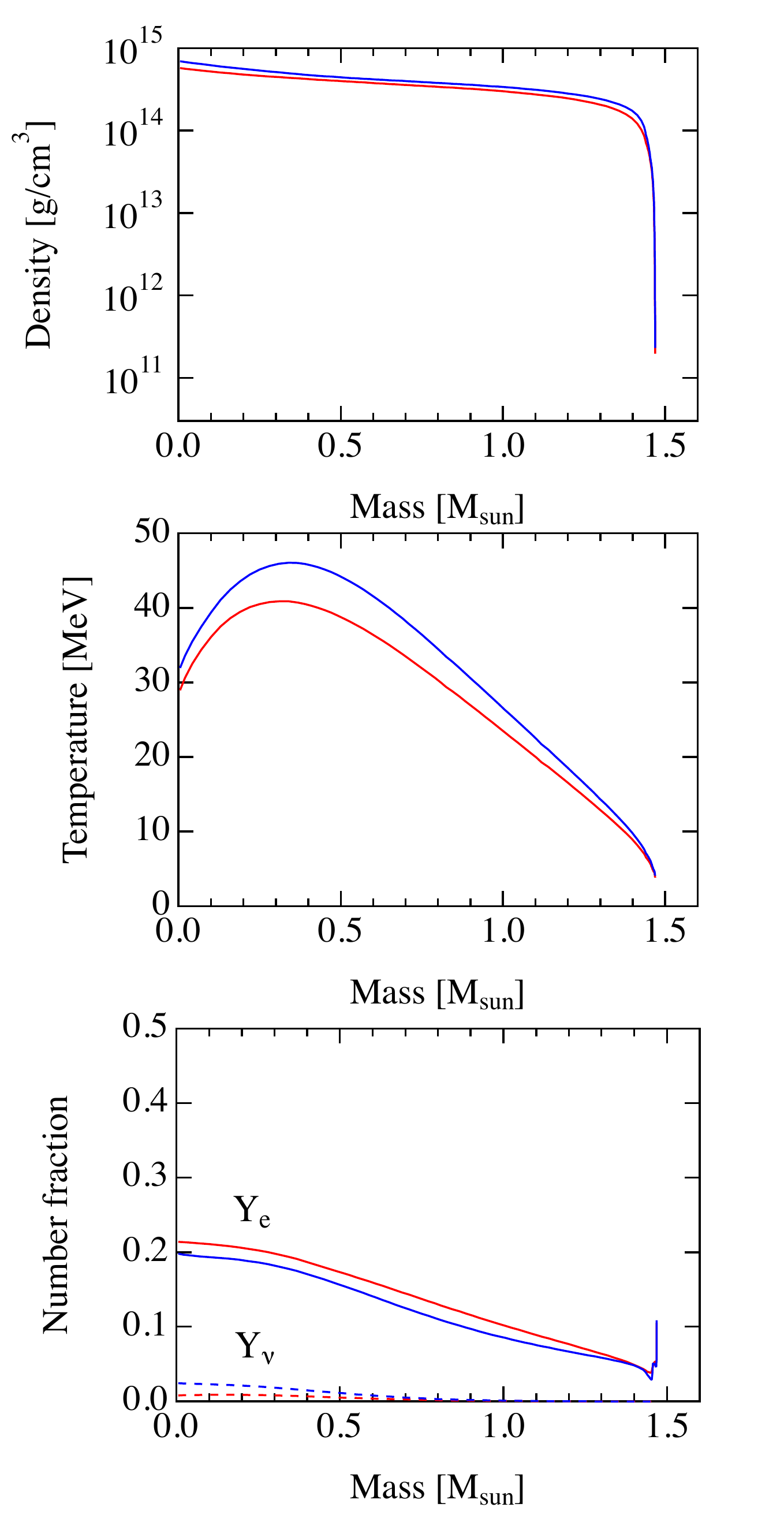}
\includegraphics[width=0.32\textwidth]{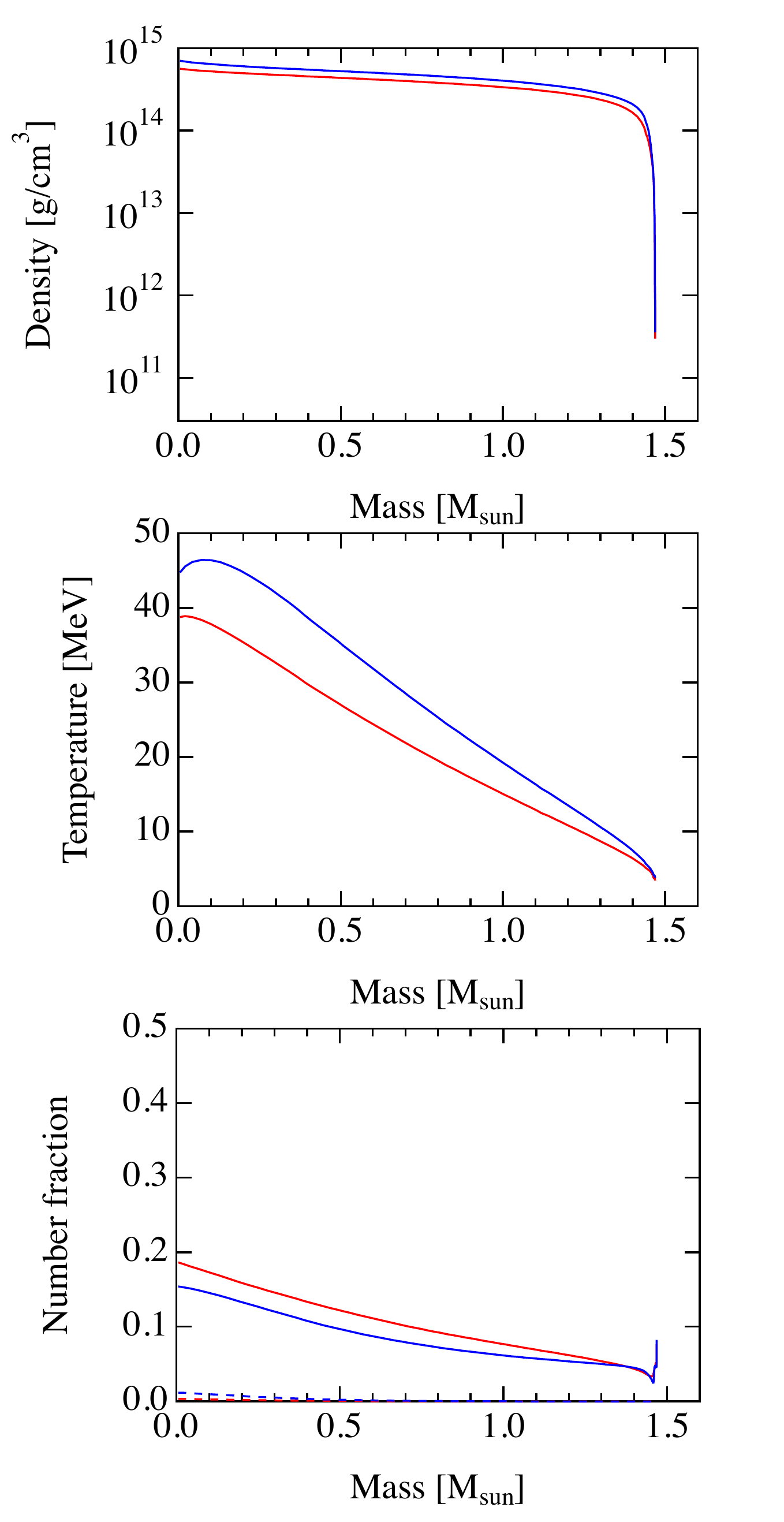}
\includegraphics[width=0.32\textwidth]{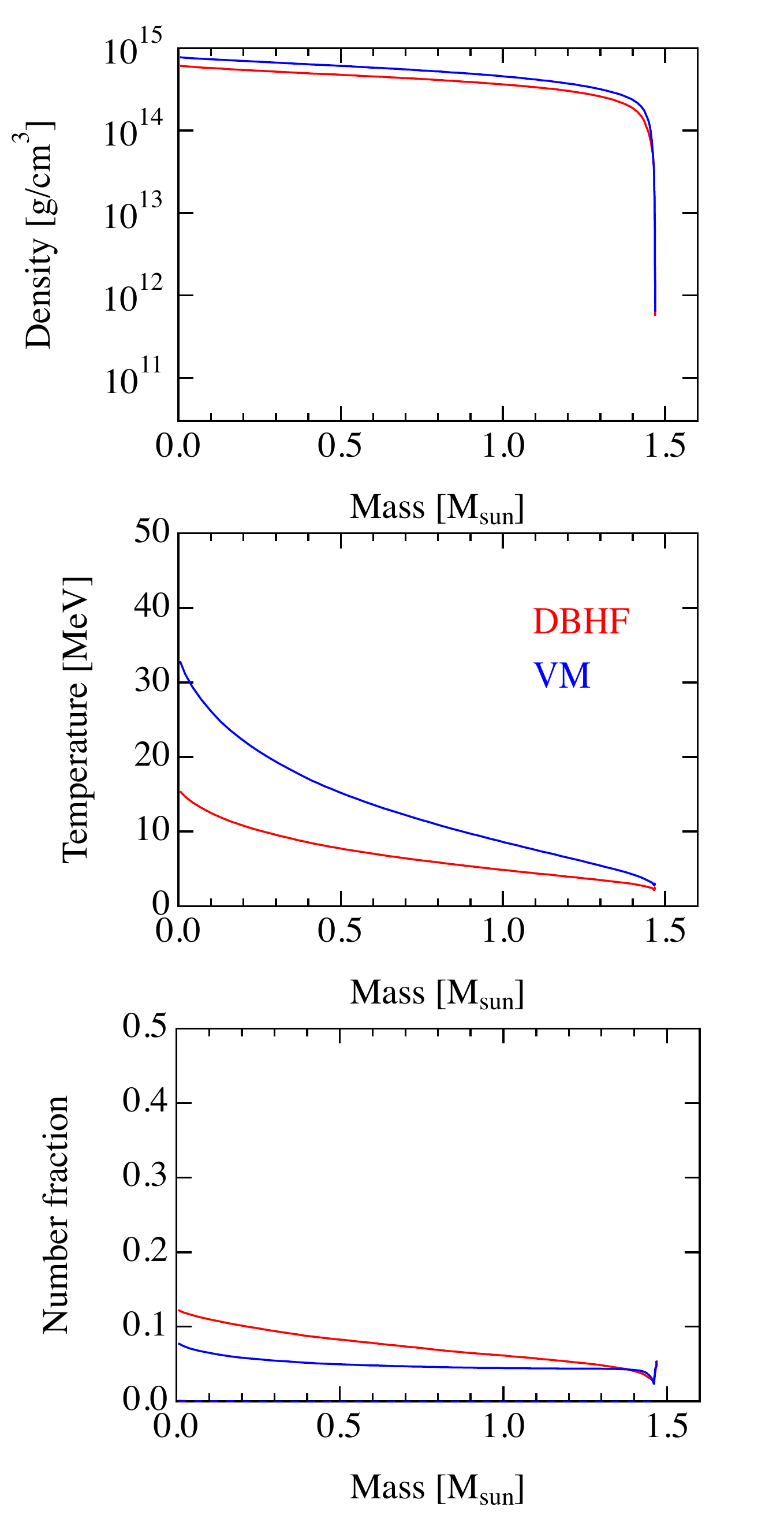}
\caption{Profiles of proto-neutron stars at 10, 20, and 50 s using DBHF EOS and VM EOS are shown in left, middle, and right panels, respectively.  
The DBHF and VM models are shown by red and blue colors, respectively.  
The density, temperature, and lepton fractions are shown in top, middle, and bottom panels, respectively.  The electron and neutrino fractions are drawn by the solid and dashed lines, respectively, in the bottom panels.  
\label{fig:pnsc_profile_DBHF}}
\end{figure*}

We show the thermal evolution of proto-neutron stars in the models using VM EOS and DBHF EOS in Fig. \ref{fig:pnsc_profile_DBHF}.  
The profiles at 10, 20 and 50 s after the start of the cooling simulations are shown.  
%
The thermal evolution proceeds through the neutrino transport and the proto-neutron stars become more compact with high densities.  
At 10 s (left panels), the peak temperature is shifted toward the center due to the neutrino flow and the temperature becomes higher due to the contraction as a whole.  The lepton fractions decrease through deleptonization as can be seen by the decrease of neutrino fraction.  
The central peaks of temperature are formed at 20 s and the outward flow of energy and lepton number proceeds through neutrino diffusion (middle panels).  
At 50 s (right panels), the neutrino fractions becomes almost zero and the deleptonization process is nearly finished.  
The values of electron fraction are approaching to the beta equilibrium values for cold neutron stars.  

\begin{figure*}[ht]
\centering
\includegraphics[width=0.32\textwidth]{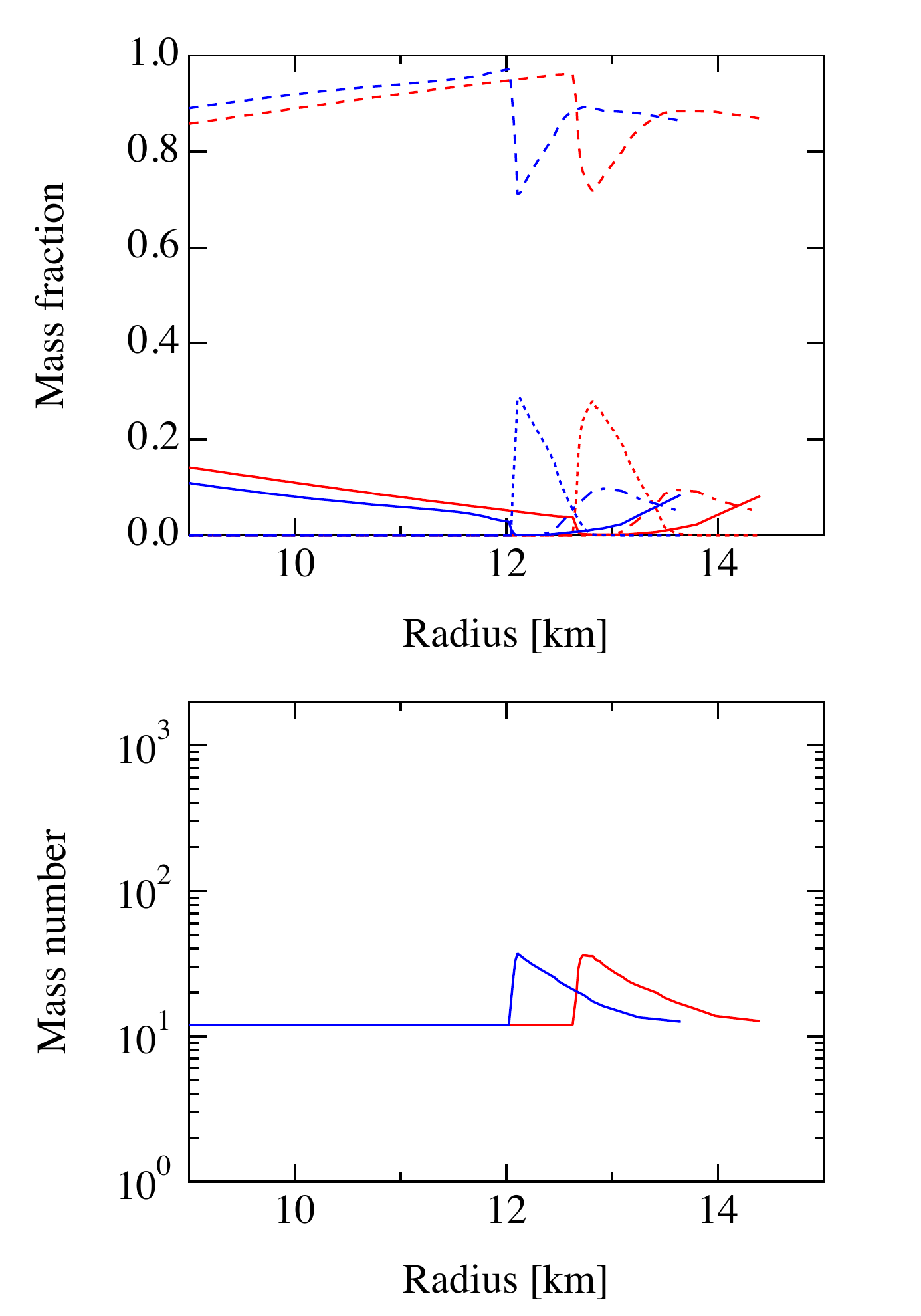}
\includegraphics[width=0.32\textwidth]{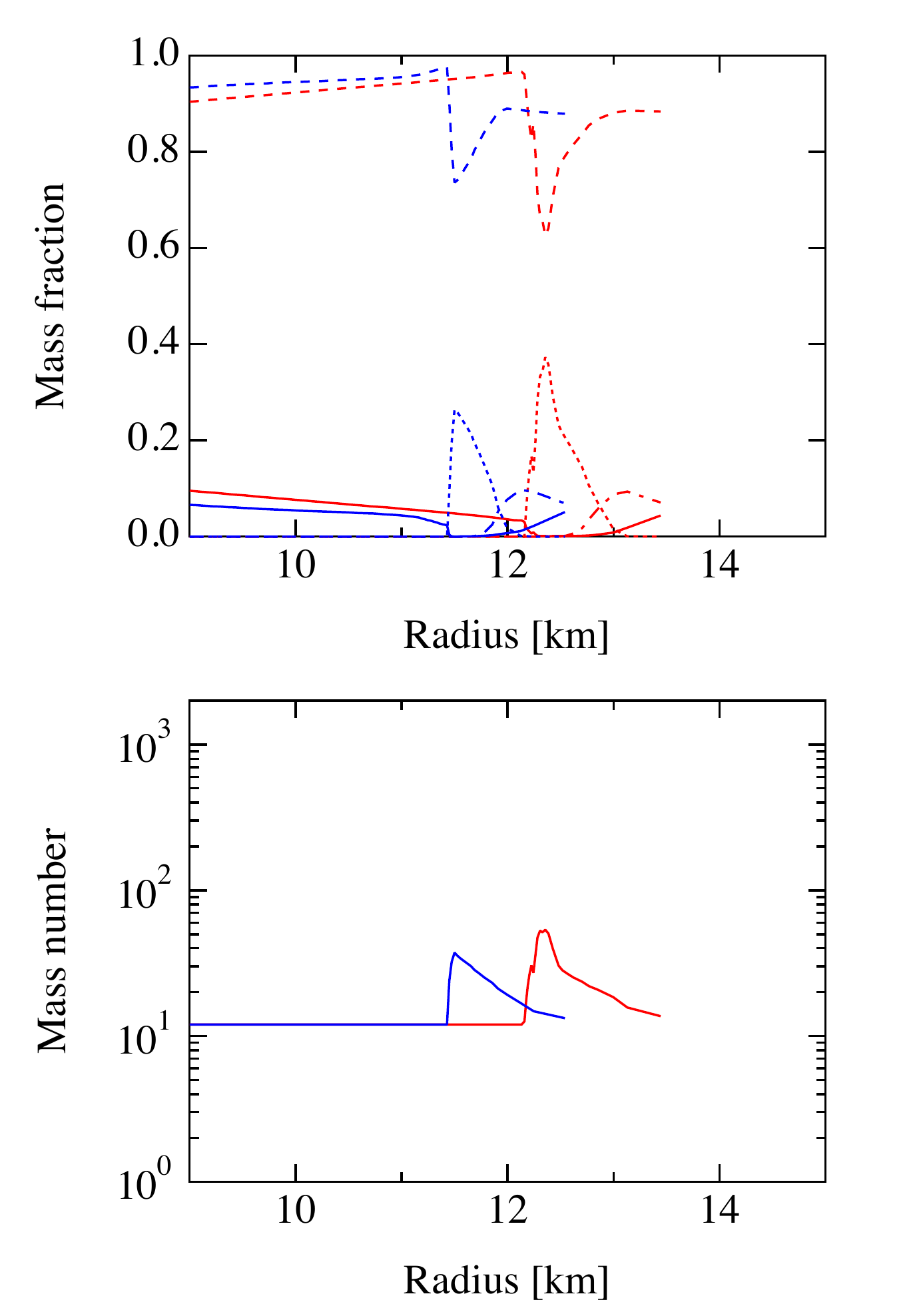}
\includegraphics[width=0.32\textwidth]{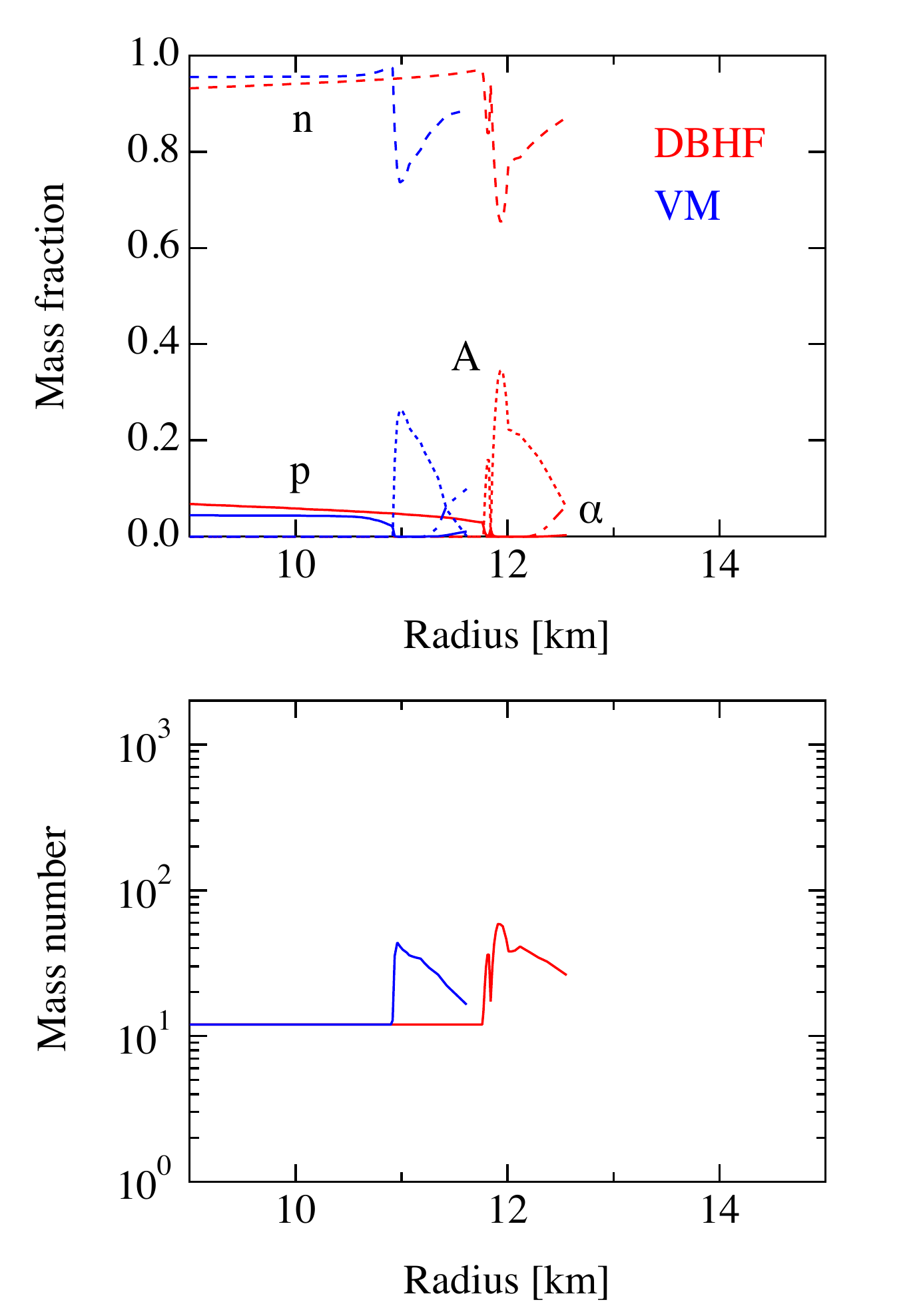}
\caption{Composition inside the proto-neutron stars at 10, 20 and 50 s using DBHF EOS and VM EOS are shown in the left, middle and right panels, respectively.  
The DBHF and VM models are shown by red and blue colors, respectively.  
The mass fraction of nuclei and mass number are shown in top and bottom panels, respectively.  
Note that the species of nuclei becomes $^{12}$C in the inner region although the mass fraction is very small.  
\label{fig:pnsc_composition_DBHF}}
\end{figure*}

The influence of EOS appears in interesting ways in the evolution of quantities.  
The density inside the evolving proto-neutron star of the VM model is higher than that of the DBHF model due to the difference of the stiffness of the two EOSs.  The radius of the VM model becomes smaller accordingly (See Fig. \ref{fig:pnsc_composition_DBHF}) and the contraction is more drastic.  
The temperature in the VM model becomes higher than that in the DBHF model at 10 s with the influence of high degree of the contraction.  
The difference of the temperature persists after 10 s and becomes larger at 50 s.  
The difference in cooling is enhanced due to a larger amount of the thermal energy contained in the VM model due to the softness of the EOS than in the DBHF model \cite{nak19,nak20}.  
The electron fraction of the VM model becomes smaller than that in the DBHF model at 10 s, starting with the same profile.  
The decrease of electron fraction proceeds through the escape of the net lepton number through the different emissions of $\nu_{e}$ and $\bar{\nu}_e$.  
It proceeds faster in the VM model due to a larger gradient of number densities under higher neutrino chemical potential than in the DBHF model \cite{sum95c}.  
This is due to the effect of symmetry energy, which provides the difference of chemical potentials between neutrons and protons, as we have discussed in section \ref{section:Results_CCSNe_DBHF}.  
The difference is enhanced toward the final value of the electron fraction determined by the neutrino-less beta equilibrium.  
Note that the final value of electron fraction in the VM model is smaller than that of the DBHF model due to the smaller symmetry energy.  
The deleptonization toward the state of neutrino-less neutron star takes more time in the VM model since the initial amount of the trapped neutrinos is large and the final electron fraction is small.  
%

%
%

We compare the composition of hot and dense matter in the proto-neutron stars for the two models in Fig. \ref{fig:pnsc_composition_DBHF}.  
The main composition is protons and neutrons in the central region and nuclei appear only in the outer region near the surface (upper panel).  
The mass number is moderate, $\sim40-60$, and similar for both the two models using the same framework to describe the composition of non-uniform matter (lower panel).  
Those nuclei do not have any significant consequence in the comparison of the DBHF and VM models.  
We will see differences in the composition using the different treatment of non-uniform matter and its impact in section \ref{section:Results_PNSC_VMS}.  
We remark that the radii of the proto-neutron star in the DBHF model is larger than that in the VM model reflecting the different stiffness.  

\subsubsection{Profiles of VM-S and VM models}\label{section:Results_PNSC_VMS}

\begin{figure*}[ht]
\centering
\includegraphics[width=0.32\textwidth]{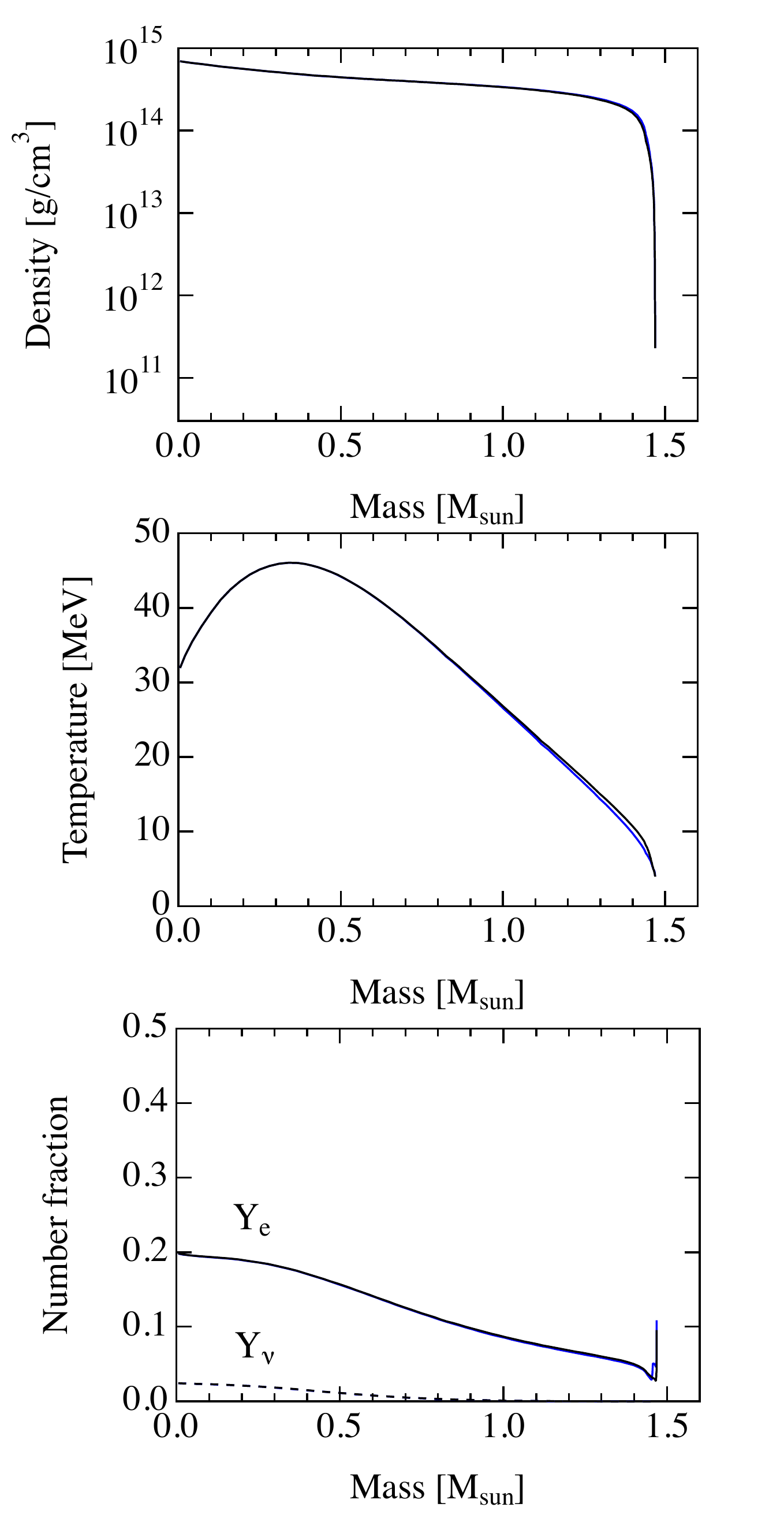}
\includegraphics[width=0.32\textwidth]{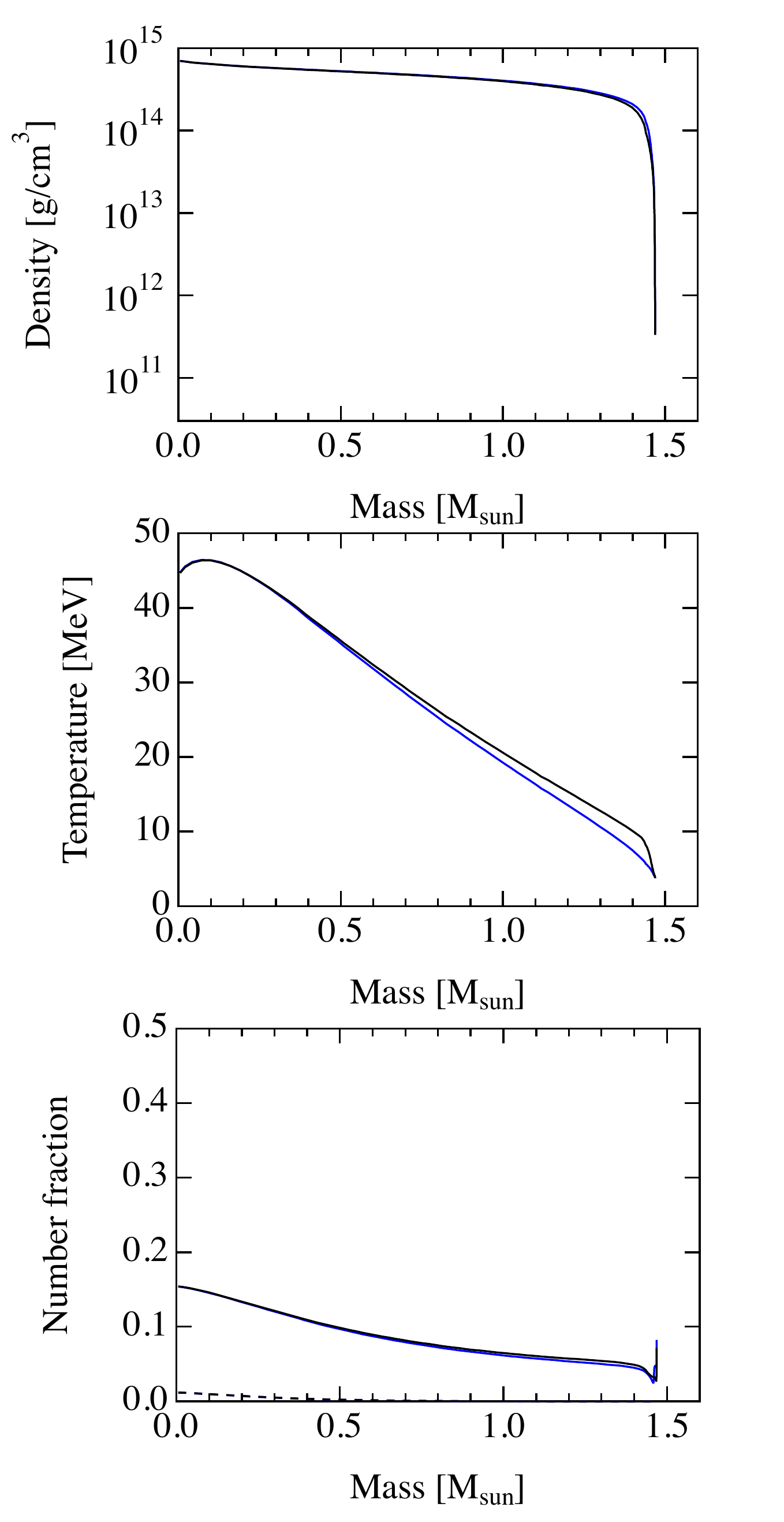}
\includegraphics[width=0.32\textwidth]{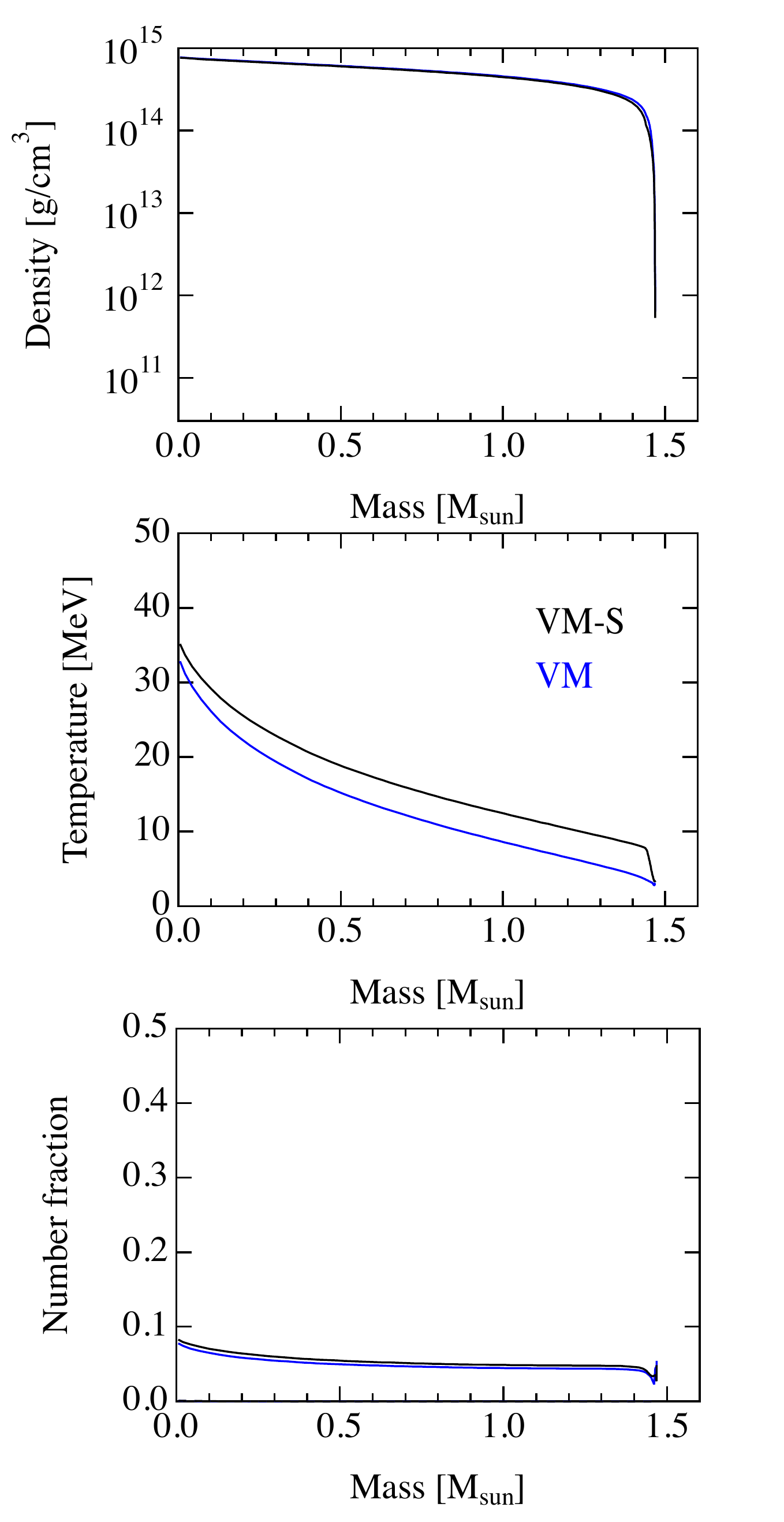}
\caption{Profiles of proto-neutron stars at 10, 20, and 50 s using VM-S EOS and VM EOS are shown in the left, middle, and right panels, respectively.  
The VM-S and VM models are shown by black and blue colors, respectively.  
The density, temperature, and lepton fractions are shown in the top, middle, and bottom panels, respectively.  
The electron and neutrino fractions are drawn by the solid and dashed lines, respectively, in the bottom panels.  
\label{fig:pnsc_profile_VMS}}
\end{figure*}

We compare the thermal evolution of proto-neutron stars in the models using VM-S EOS and VM EOS in Fig. \ref{fig:pnsc_profile_VMS} to discuss the influence of different composition.  
The two models start from almost the same configurations with a slight difference of density and temperature in the outer region ($\geq 1$M$_{\odot}$) due to the different frameworks of non-uniform matter.  
The thermal structure of the VM-S model is getting different from that of the VM model as it evolves over 20 s.  
The temperature of the VM-S model becomes higher in the outer region at 20 s and in the whole region at 50 s.  
This is because the different composition of nuclei in the surface region as we will see below.  
The transport of thermal energy via neutrinos is hindered by the large opacity of the neutrino-nucleus scattering.  

\begin{figure*}[ht]
\centering
\includegraphics[width=0.32\textwidth]{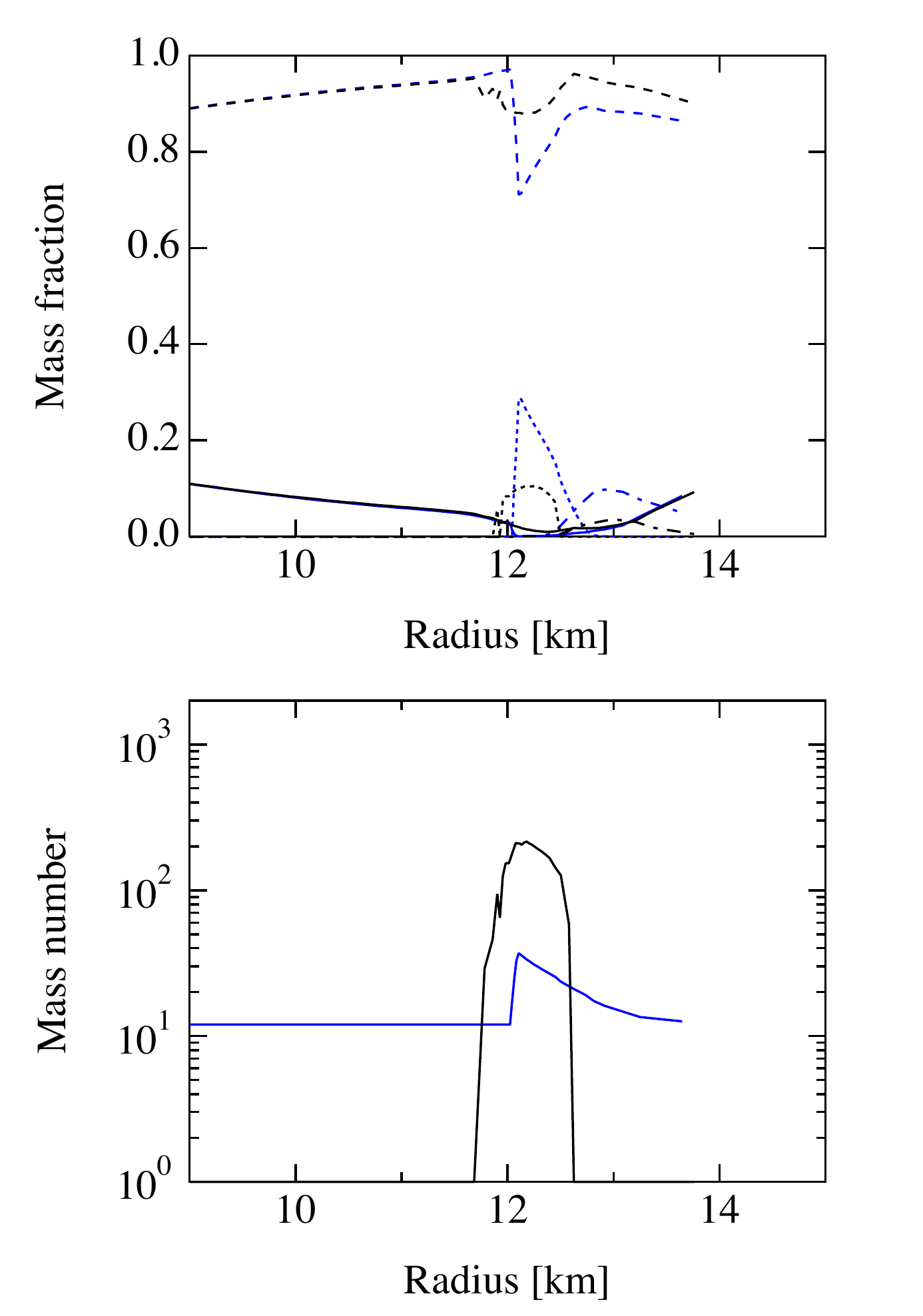}
\includegraphics[width=0.32\textwidth]{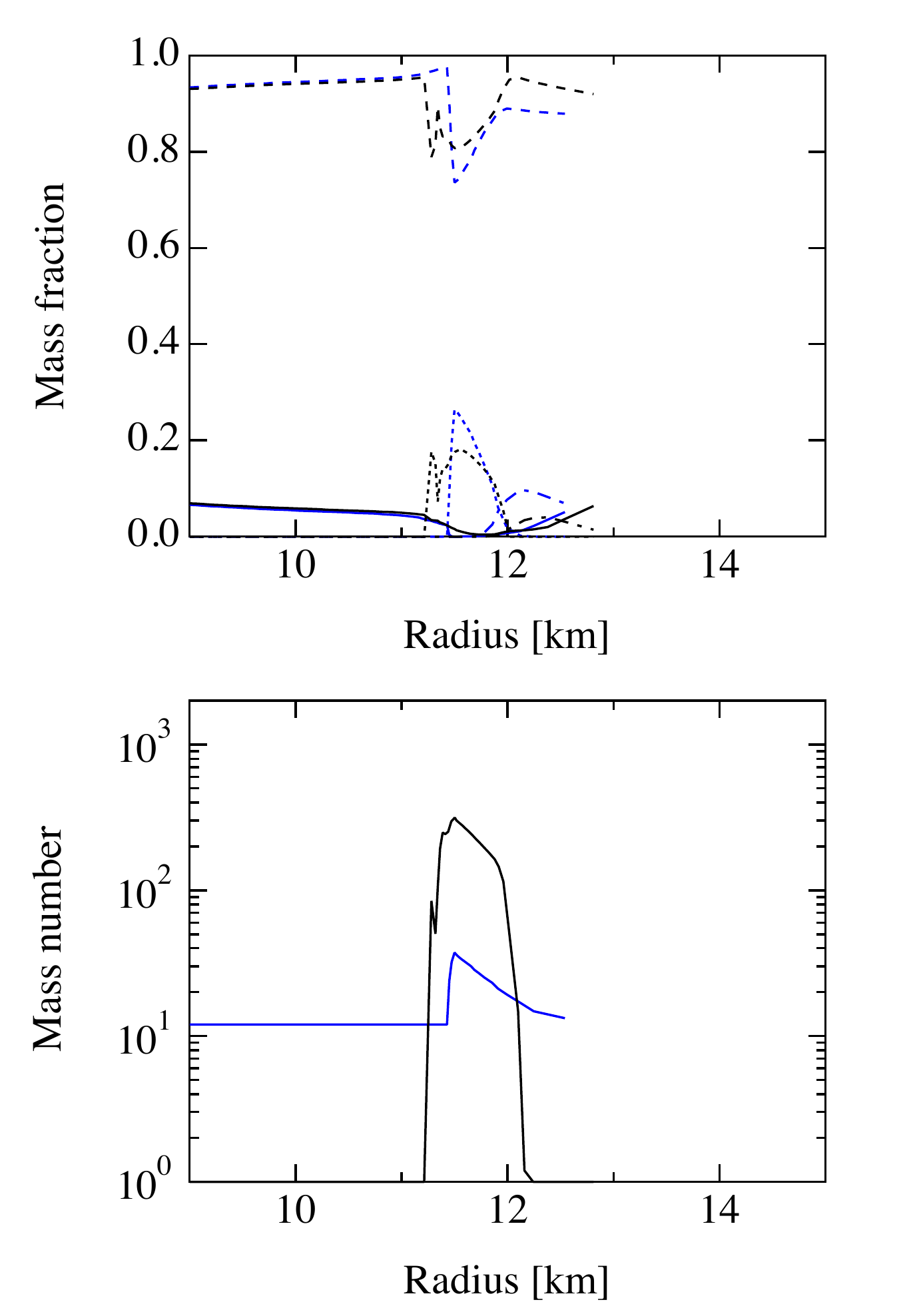}
\includegraphics[width=0.32\textwidth]{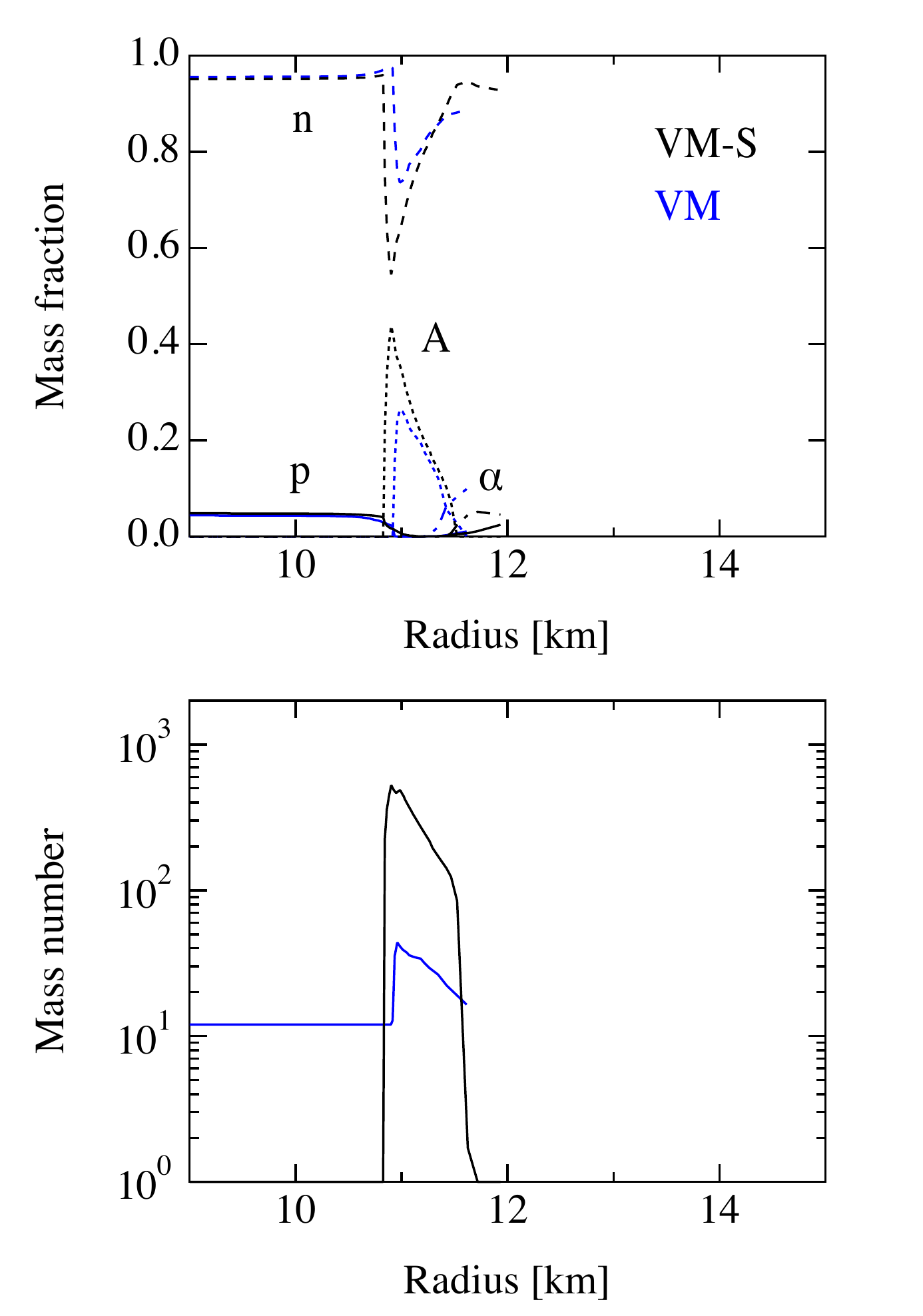}
\caption{Composition inside the proto-neutron stars at 10, 20, and 50 s using VM-S EOS and VM EOS are shown in the left, middle, and right panels, respectively.  
The VM-S and VM models are shown by black and blue colors, respectively.  
The mass fraction of nuclei and mass number are shown in top and bottom panels, respectively.  
\label{fig:pnsc_composition_VMS}}
\end{figure*}

We show the different composition of the two models in Fig. \ref{fig:pnsc_composition_VMS}.  
The mass number of nuclei in the VM-S model is large $\sim300-700$ at $\sim12$~km and distinctively different from the other models.  
The mass number becomes large as the outcome of favorable species in the single nucleus approximation of VM-S EOS whereas those of medium mass number coexist in the NSE treatment of VM EOS to lower the total energy.  
Note that the VM EOS takes into account the bubble phase around the transitional region at around 10$^{14}$~g\,cm$^{-3}$.  
The nuclei with large mass numbers contribute to the major source of opacities, which block the neutrino transport \cite{nak18}.  
They make energy transport slow in this region and the thermal energy remains inside at higher temperature than the VM model.  
This tendency becomes more clear at 50 s with large size of the mass number and mass fraction.  
This feature leads to influence on the properties of neutrino emission as we will see below.  


\subsubsection{Neutrino emission}\label{section:Results_PNSC_Neutrino}

\begin{figure*}[ht]
\centering
\includegraphics[width=0.45\textwidth]{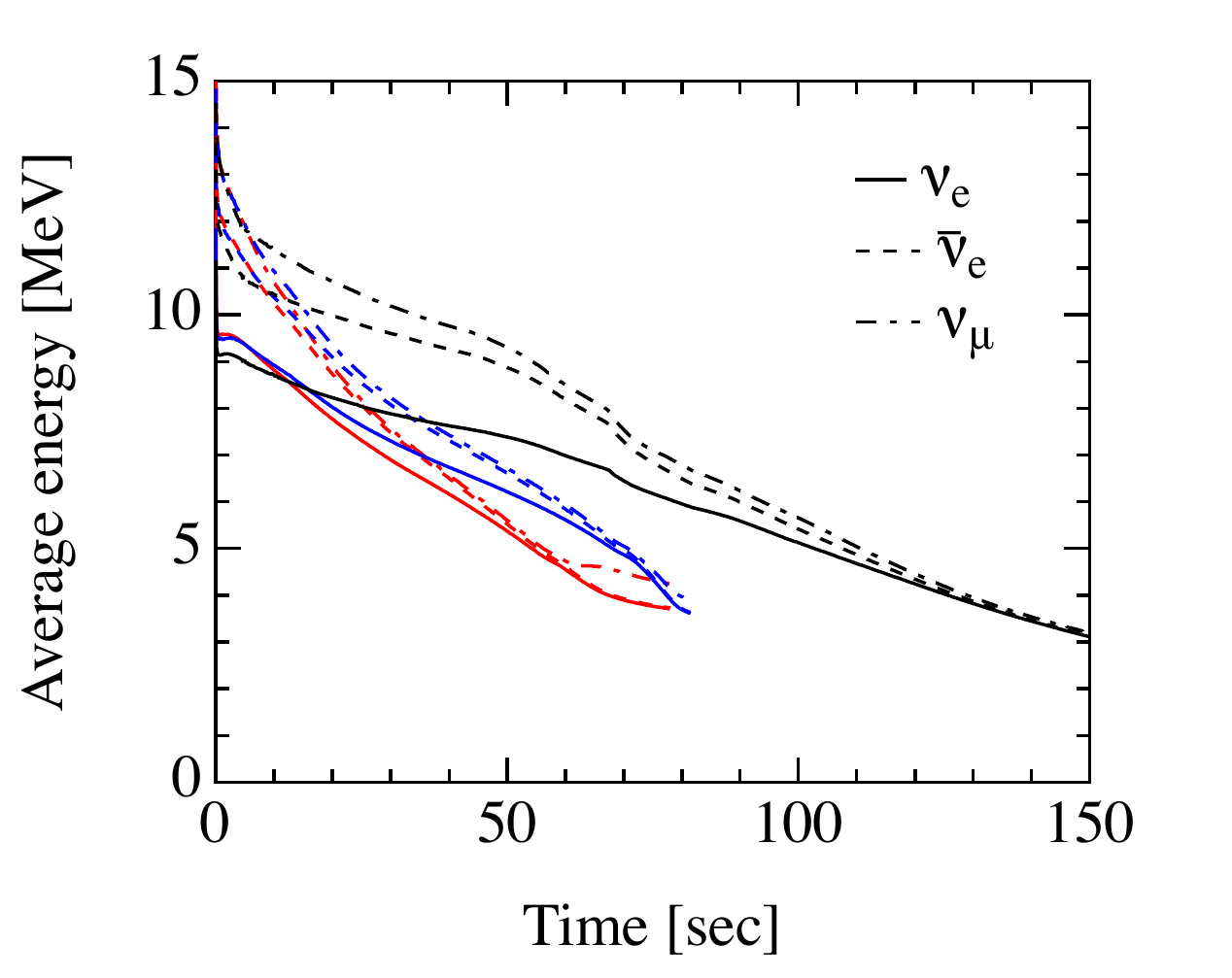}
\includegraphics[width=0.48\textwidth]{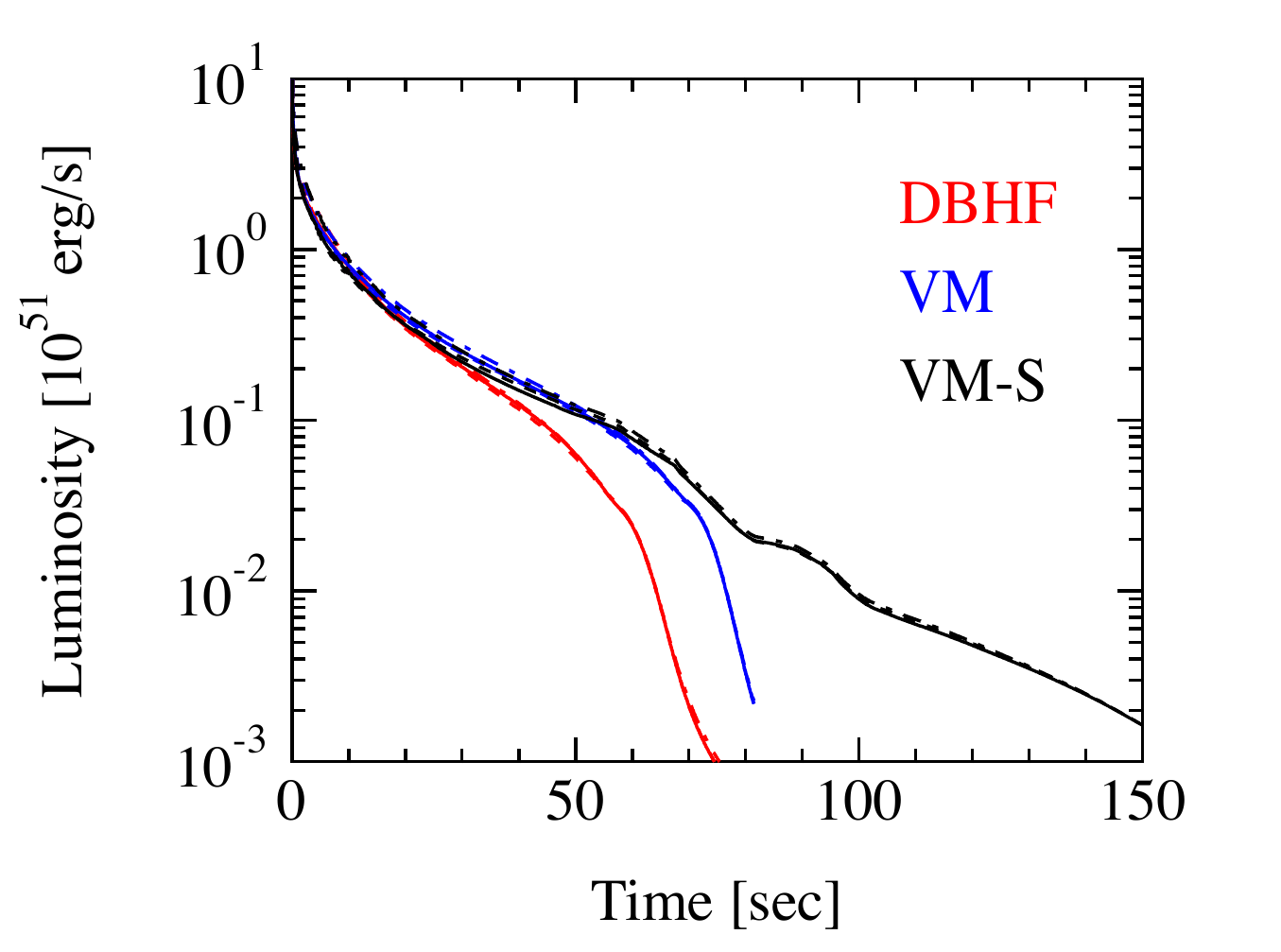}
\caption{Luminosities and average energies of neutrino emitted from the proto-neutron star evolution using the three sets of EOS.  
The DBHF, VM-S, and VM models are shown by red, black, and blue colors, respectively.  
The neutrino species, $\nu_{e}$, $\bar{\nu}_e$, and $\nu_{\mu}$ are denoted by solid, dashed, and dot-dashed lines, respectively.  
\label{fig:pnsc_Nu_3eos}}
\end{figure*}

We compare properties of neutrino emission during the cooling of proto-neutron stars with three sets of EOS in Fig. \ref{fig:pnsc_Nu_3eos}.  
The neutrino luminosities are similar among three models until 20 s and show differences at later stages in the right panel. 
This is because the difference of profiles becomes larger at late stages through the increase of density due to the contraction of proto-neutron stars.  

The luminosity of DBHF model decreases faster than the other models 
after $\sim$20 s.  
The density of the proto-neutron star in the DBHF model is getting lower than that of the VM model as seen in Fig. \ref{fig:pnsc_profile_DBHF}.  The diffusion time scale in the DBHF model is shorter than that in the VM model accordingly and the evolution via neutrino flow proceeds fast.  
Since the radii of cold neutron stars for DBHF EOS are larger than those for the VM EOS, 
the gravitational binding energy is smaller and the total energy of emitted neutrinos is smaller.  For reference, the initial values of the gravitational mass of proto-neutron star are 1.430M$_{\odot}$, 1.440M$_{\odot}$, and 1.439M$_{\odot}$ for DBHF, VM-S, and VM models, respectively.  The corresponding values at the end of each simulation are 1.341M$_{\odot}$, 1.332M$_{\odot}$, and 1.330M$_{\odot}$, respectively\footnote{Note that they have not yet reached the final values of cold neutron stars.  }.  Hence, the released energy is smaller in the DBHF model than the VM-S and VM models.  Note that the initial amount of neutrino fraction in the DBHF model is smaller than the others.  These facts are consistent with the shorter time scale of the evolution of proto-neutron star and luminosities in the DBHF model than the VM-S and VM models.  It is interesting to see that
the thermal evolution from the initial configuration to the low temperature state proceeds fast in the DBHF model.  This tendency is brought by a smaller amount of the thermal energy contained in the proto-neutron star even with lower neutrino luminosity at the late phase.  
The time scale of cooling also depends on the effective mass through the behavior of thermal energy and entropy \cite{nak19}.  It has been shown that large effective masses and small symmetry energies lead to the long time scale.  It is to be noted that the entropy of nucleons depends on the effective mass of nucleons through the energy levels in the distribution function.  The systematic study of cooling time scale can be applied to constrain the neutron star masses and radii \cite{nak20}.  

The duration of luminosity of VM-S model is longer than that of VM model due to the influence of heavy nuclei in the surface region (See profiles in Fig. \ref{fig:pnsc_composition_VMS}).  
The opacity due to the neutrino-nucleus scattering with a large mass number of nuclei hinders the transport of neutrinos.   Therefore, the energy transport is slow and the cooling proceeds slowly in the VM-S model. 
This phenomena is in accord with the finding of the thermal insulation by heavy nuclei \cite{nak18}.  
It is interesting to see that the luminosities for three species are almost similar due to the equipartition realized deep inside the energy neutrinosphere.  

The neutrino average energies show differences due to the influence of three sets of EOS after $\sim$10 s (left panel).  
The difference between the models of VM-S and VM is remarkable after 10 s although the two sets of EOS are based on the same input of uniform nuclear matter.  
This difference mainly comes from the composition obtained by the two methods for the EOS tables.  
In the VM-S model, the mass number of nuclei is large at the surface region of the proto-neutron star (Fig. \ref{fig:pnsc_composition_VMS}).  
Because of the large opacity, the outer layer becomes hot (Fig. \ref{fig:pnsc_profile_VMS}) and the average energy becomes higher as compared with the VM model.  
In contrast, the difference between the models of DBHF and VM arises in a different manner.  
As seen in the profiles in Fig. \ref{fig:pnsc_profile_DBHF}, 
the temperature of the DBHF model is lower than that of the VM model in the whole region from 10 s.  
As a result, the average energy of the DBHF model is lower than that of the VM model.  
The DBHF and VM models have similar composition of nuclei in the surface region and those medium mass nuclei do not have any large impact as seen in the VM-S model.  
Comparing three models, the influence on neutrino emission from the compositional treatment is stronger than that from the input of nuclear matter.  

The average energies for three species have hierarchy of values from the initial phase in the standard manner.  
This difference of average energies comes from the different position of neutrinospheres depending on the channels of neutrino reactions for opacities.  
The neutrinosphere for $\nu_{\mu}$ is located at the innermost region due to the lack of charged current reactions and hence the average energy of $\nu_{\mu}$ is high, reflecting high temperature inside.  
Those of $\bar{\nu}_e$ and $\nu_{e}$ are located at larger radii with lower temperatures, which provides lower average energies than $\nu_{\mu}$.  
The radial position of the neutrinosphere for $\nu_{e}$ is larger than that of $\bar{\nu}_e$ since the free proton are available even at low densities, leading to the lowest value among the three species.  
As the cooling of the proto-neutron star proceeds over 20 s, the density gradient at the outer layer becomes steep and the positions of neutrinosphere are similar to each other. Average energies become similar at the endpoint.  

\section{Summary}\label{section:Summary}

We study the influence of the hot and dense matter in supernova simulations by adopting the tables of equation of state (EOS) based on the microscopic nuclear many body frameworks.  
We perform the numerical simulations of the gravitational collapse and bounce of the central core of the massive stars and the long-term evolution of proto-neutron star cooling.  
We adopt the EOS tables based on the Dirac Br\"uckner Hartree-Fock theory and the variational method.  The former is recently applied to construct the EOS table by utilizing the functional formulae for the uniform matter evaluated by the Dirac Br\"uckner Hartree-Fock theory \cite{fur20} (DBHF EOS).  The latter is applied to the construction of the EOS tables by using the data of uniform matter evaluated by the variational method.  There are two versions of the EOS tables in the latter by treating the non-uniform matter to determine the composition in a single nucleus approximation with the Thomas-Fermi approximation \cite{tog17} (VM-S EOS) and in the nuclear statistical equilibrium with the compressible liquid drop model \cite{fur17b} (VM EOS).  

We examine the effects of the EOS in the two aspects: the uniform matter and the treatment of composition.  We investigate the numerical simulations with the DBHF EOS table and explore differences with respect to those with the VM EOS table.  These EOS tables adopt the same framework for the non-uniform matter but with different inputs of the uniform matter.  We explore the impact of the uniform matter by the Dirac Br\"uckner Hartree-Fock theory with respect to the variational method since they show characteristic differences of uniform matter inherent in relativistic and non-relativistic many body theories.  
We also make the comparison of the numerical simulations adopting the VM and VM-S EOS tables with a different treatment of the non-uniform matter with the same uniform matter by the variational method.  We examine the size of influence due to different composition in the non-uniform matter at low densities with the same uniform matter at high densities.  

We explore the EOS effects in the gravitational collapse, core bounce, and shock propagation in the central core of the massive stars in the numerical simulations by solving the general relativistic neutrino-radiation hydrodynamics under the spherical symmetry.  
When we compare the numerical results with the DBHF and VM EOS tables, we found a slightly larger size of the bounce core in the DBHF model due to a larger lepton fraction.  This is brought by a smaller amount of electron captures on free protons during the gravitational collapse due to a larger symmetry energy in DBHF EOS.  The differences in the profiles during the collapse and bounce are rather small and the composition of nuclei are found similar.  
In the comparison with the VM and VM-S tables, we found noticeable differences due to the effects of the composition even with the same input of the uniform matter.  We found a larger size of the bounce core with VM-S EOS than VM EOS.  The difference is actually larger than the one seen between DBHF EOS and VM EOS.  This is due to a larger lepton fraction mainly through stronger neutrino trapping via the neutrino-nucleus scattering with large mass nuclei in VM-S EOS.  The composition in the mixture of nuclei and nucleons is distinctively different in the two models throughout the evolution.  The nuclei with larger masses generally appear in the model with VM-S EOS.  
It may also affect the condition for the convection right after the core bounce.  

We examine the influence of EOS in the cooling of the proto-neutron stars in the general relativistic numerical simulation of the thermal evolution by solving the quasi-static configuration with the flux-limited neutrino diffusion under the spherical symmetry.  
In the comparison with DBHF EOS and VM EOS, the efficient cooling is seen in the DBHF model having the rapid decrease of temperature and neutrino fraction.  The decline of the neutrino luminosity is rapid reflecting a shorter diffusion time scale for lower density due to the stiffness of DBHF EOS.  
In the comparison with VM-S EOS and VM EOS, the thermal evolution for $\sim$10 s is similar to each other, however, the temperature profiles become different afterward.  The composition of nuclei is found different near the surface region.  As the nuclei with larger masses appear in the VM-S model, they hinder the neutrino transport as a source of the neutrino opacity and the thermal energy is stored inside the neutrinosphere.  This blocking of neutrinos as seen in the insulation of heat \cite{nak18} leads to the higher average energy of emitted neutrinos and longer duration of luminosity in the VM-S model.  

In summary, the supernova simulations with DBHF EOS show qualitative differences from those with the VM EOS due to the stiffness and the symmetry energy in the uniform matter.  The corresponding simulations adopting VM-S EOS and VM EOS reveal noticeable differences due to the composition.  The size of influence due to the composition is comparable or even more larger as compared with that due to the uniform matter.  
In order to extract the effects of the EOS from the supernova dynamics and supernova neutrinos, one has to carefully separate the effects due to composition and uniform matter.  
It is necessary to study further the effects due to the composition by careful evaluation of electron captures and neutrino scattering with mixture of nuclei.  
The effects of the compositions using the elaborated electron capture rates and the impact of composition on the convection in multi-dimensional calculations is under investigation and will be reported elsewhere \cite{nag22x,har22x}.  



\section*{Acknowledgment}

This work is supported by 
Grant-in-Aid for Scientific Research
(19K03837, 20H01905, 
20K03973, 21K13913, 21K13924, 21H01088) 
and 
Grant-in-Aid for Scientific Research on Innovative areas 
"Gravitational wave physics and astronomy:Genesis"
(17H06357, 17H06365)
and "Unraveling the history of the universe and matter evolution with underground physics" 
(19H05802, 19H05811) 
from the Ministry of Education, Culture, Sports, Science and Technology (MEXT), Japan.  
For providing high performance computing resources, 
Computing Research Center, KEK, 
JLDG on SINET of NII, 
Research Center for Nuclear Physics, Osaka University, 
Yukawa Institute of Theoretical Physics, Kyoto University, 
Nagoya University, 
and 
Information Technology Center, University of Tokyo are acknowledged.  
This work was supported by 
MEXT as "Program for Promoting Researches on the Supercomputer Fugaku" 
(Toward a unified view of the universe: from large scale structures to planets, JPMXP1020200109)
and
the Particle, Nuclear and Astro Physics Simulation Program (Nos. 2020-004, 2021-004, 2022-003) of Institute of Particle and Nuclear Studies, High Energy Accelerator Research Organization (KEK).


%



\let\doi\relax


\bibliographystyle{ptephy}
\bibliography{sumi}{}

\begin{thebibliography}{10}

\bibitem{bet90}
H.~A. {Bethe}, Reviews of Modern Physics, {\bf 62}(4), 801--866 (October 1990).

\bibitem{oer17}
M.~{Oertel}, M.~{Hempel}, T.~{Kl{\"a}hn}, and S.~{Typel}, Reviews of Modern
  Physics, {\bf 89}(1), 015007 (Jan 2017),  {{arXiv:1610.03361}}.

\bibitem{bar85}
E.~{Baron}, J.~{Cooperstein}, and S.~{Kahana}, \prl, {\bf 55}(1), 126--129 (Jul
  1985).

\bibitem{tak88}
Mariko {Takahara} and Katsuhiko {Sato}, \apj, {\bf 335}, 301 (Dec 1988).

\bibitem{swe94}
F.~Douglas {Swesty}, James~M. {Lattimer}, and Eric~S. {Myra}, \apj, {\bf 425},
  195 (Apr 1994).

\bibitem{sum05}
K.~{Sumiyoshi}, S.~{Yamada}, H.~{Suzuki}, H.~{Shen}, S.~{Chiba}, and H.~{Toki},
  \apj, {\bf 629}(2), 922--932 (August 2005),  {{arXiv:astro-ph/0506620}}.

\bibitem{hem12}
M.~{Hempel}, T.~{Fischer}, J.~{Schaffner-Bielich}, and M.~{Liebend{\"o}rfer},
  \apj, {\bf 748}(1), 70 (Mar 2012),  {{arXiv:1108.0848}}.

\bibitem{sum19}
Kohsuke {Sumiyoshi}, Ken'ichiro {Nakazato}, Hideyuki {Suzuki}, Jinniu {Hu}, and
  Hong {Shen}, \apj, {\bf 887}(2), 110 (December 2019),  {{arXiv:1908.02928}}.

\bibitem{mar09b}
A.~{Marek}, H.~T. {Janka}, and E.~{M{\"u}ller}, \aap, {\bf 496}(2), 475--494
  (Mar 2009),  {{arXiv:0808.4136}}.

\bibitem{cou13}
Sean~M. {Couch}, \apj, {\bf 765}(1), 29 (Mar 2013),  {{arXiv:1206.4724}}.

\bibitem{suw13}
Yudai {Suwa}, Tomoya {Takiwaki}, Kei {Kotake}, Tobias {Fischer}, Matthias
  {Liebend{\"o}rfer}, and Katsuhiko {Sato}, \apj, {\bf 764}(1), 99 (Feb 2013),
  {{arXiv:1206.6101}}.

\bibitem{fis14}
Tobias {Fischer}, Matthias {Hempel}, Irina {Sagert}, Yudai {Suwa}, and
  J{\"u}rgen {Schaffner-Bielich}, European Physical Journal A, {\bf 50}, 46
  (Feb 2014),  {{arXiv:1307.6190}}.

\bibitem{yas18}
H.~{Yasin}, S.~{Sch{\"a}fer}, A.~{Arcones}, and A.~{Schwenk}, \prl, {\bf
  124}(9), 092701 (March 2020).

\bibitem{sch19b}
A.~S. {Schneider}, L.~F. {Roberts}, C.~D. {Ott}, and E.~{O'Connor}, \prc, {\bf
  100}(5), 055802 (November 2019),  {{arXiv:1906.02009}}.

\bibitem{har20}
Akira {Harada}, Hiroki {Nagakura}, Wakana {Iwakami}, Hirotada {Okawa}, Shun
  {Furusawa}, Kohsuke {Sumiyoshi}, Hideo {Matsufuru}, and Shoichi {Yamada},
  arXiv e-prints, page arXiv:2003.08630 (March 2020),  {{arXiv:2003.08630}}.

\bibitem{jan12a}
Hans-Thomas {Janka}, Annual Review of Nuclear and Particle Science, {\bf
  62}(1), 407--451 (Nov 2012),  {{arXiv:1206.2503}}.

\bibitem{bur13}
A.~{Burrows}, Reviews of Modern Physics, {\bf 85}, 245--261 (January 2013),
  {{arXiv:1210.4921}}.

\bibitem{kot13}
K.~{Kotake}, Comptes Rendus Physique, {\bf 14}, 318--351 (April 2013),
  {{arXiv:1110.5107}}.

\bibitem{jan16}
Hans-Thomas {Janka}, Tobias {Melson}, and Alexander {Summa}, Annual Review of
  Nuclear and Particle Science, {\bf 66}(1), 341--375 (October 2016),
  {{arXiv:1602.05576}}.

\bibitem{jan17b}
Hans-Thomas {Janka},
\newblock {\em {Neutrino-Driven Explosions}}, page 1095,
\newblock Springer (2017).

\bibitem{bur88}
A.~Burrows, Astrophys.\ J., {\bf 334}, 891 (1988).

\bibitem{suz94}
H.~Suzuki,
\newblock In M.~Fukugita and A.~Suzuki, editors, {\em Physics and Astrophysics
  of Neutrinos}, page 763. Springer-Verlag, Tokyo (1994).

\bibitem{pon99}
J.~A. {Pons}, S.~{Reddy}, M.~{Prakash}, J.~M. {Lattimer}, and J.~A. {Miralles},
  \apj, {\bf 513}(2), 780--804 (March 1999),  {{arXiv:astro-ph/9807040}}.

\bibitem{rob12}
L.~F. {Roberts}, G.~{Shen}, V.~{Cirigliano}, J.~A. {Pons}, S.~{Reddy}, and
  S.~E. {Woosley}, \prl, {\bf 108}(6), 061103 (Feb 2012),  {{arXiv:1112.0335}}.

\bibitem{cam17}
Giovanni {Camelio}, Alessandro {Lovato}, Leonardo {Gualtieri}, Omar {Benhar},
  Jos{\'e}~A. {Pons}, and Valeria {Ferrari}, \prd, {\bf 96}(4), 043015 (August
  2017),  {{arXiv:1704.01923}}.

\bibitem{nak18}
Ken'ichiro {Nakazato}, Hideyuki {Suzuki}, and Hajime {Togashi}, \prc, {\bf
  97}(3), 035804 (Mar 2018),  {{arXiv:1710.10441}}.

\bibitem{nak19}
Ken{\textquoteright}ichiro {Nakazato} and Hideyuki {Suzuki}, \apj, {\bf
  878}(1), 25 (Jun 2019),  {{arXiv:1905.00014}}.

\bibitem{nak20}
Ken'ichiro {Nakazato} and Hideyuki {Suzuki}, \apj, {\bf 891}(2), 156 (March
  2020),  {{arXiv:2002.03300}}.

\bibitem{jan17a}
Hans-Thomas {Janka},
\newblock {\em {Neutrino Emission from Supernovae}}, page 1575,
\newblock Springer (2017).

\bibitem{mul19}
B.~{M{\"u}ller}, Annual Review of Nuclear and Particle Science, {\bf 69},
  253--278 (October 2019),  {{arXiv:1904.11067}}.

\bibitem{suw19}
Yudai {Suwa}, Kohsuke {Sumiyoshi}, Ken{\textquoteright}ichiro {Nakazato},
  Yasufumi {Takahira}, Yusuke {Koshio}, Masamitsu {Mori}, and Roger~A.
  {Wendell}, \apj, {\bf 881}(2), 139 (Aug 2019),  {{arXiv:1904.09996}}.

\bibitem{war20}
MacKenzie~L. {Warren}, Sean~M. {Couch}, Evan~P. {O'Connor}, and Viktoriya
  {Morozova}, \apj, {\bf 898}(2), 139 (August 2020),  {{arXiv:1912.03328}}.

\bibitem{bax21}
Amanda {Baxter}, Segev {BenZvi}, Joahan {Jaimes}, Alexis {Coleiro}, Marta
  {Molla}, Damien {Dornic}, Spencer {Griswold}, Tomer {Goldhagen}, Anne {Graf},
  Alec {Habig}, Remington {Hill}, Shunsaku {Horiuchi}, James {Kneller}, Mathieu
  {Lamoureux}, Rafael {Lang}, Massimiliano {Lincetto}, Jost {Migenda}, McKenzie
  {Myers}, Evan {O'Connor}, Andrew {Renshaw}, Kate {Scholberg}, Andrey
  {Sheshukov}, Jeff {Tseng}, Christopher {Tunnell}, Navya {Uberoi}, and Arkin
  {Worlikar}, The Journal of Open Source Software, {\bf 6}(67), 3772 (November
  2021),  {{arXiv:2109.08188}}.

\bibitem{nag21a}
Hiroki {Nagakura}, Adam {Burrows}, David {Vartanyan}, and David {Radice},
  \mnras, {\bf 500}(1), 696--717 (January 2021),  {{arXiv:2007.05000}}.

\bibitem{nag21b}
Hiroki {Nagakura}, Adam {Burrows}, and David {Vartanyan}, \mnras, {\bf 506}(1),
  1462--1479 (September 2021),  {{arXiv:2102.11283}}.

\bibitem{nag22a}
Hiroki {Nagakura} and David {Vartanyan}, \mnras, {\bf 512}(2), 2806--2816 (May
  2022),  {{arXiv:2111.05869}}.

\bibitem{nak22}
Ken'ichiro {Nakazato}, Fumi {Nakanishi}, Masayuki {Harada}, Yusuke {Koshio},
  Yudai {Suwa}, Kohsuke {Sumiyoshi}, Akira {Harada}, Masamitsu {Mori}, and
  Roger~A. {Wendell}, \apj, {\bf 925}(1), 98 (January 2022),
  {{arXiv:2108.03009}}.

\bibitem{tog17}
H.~{Togashi}, K.~{Nakazato}, Y.~{Takehara}, S.~{Yamamuro}, H.~{Suzuki}, and
  M.~{Takano}, \nphysa, {\bf 961}, 78--105 (May 2017),  {{arXiv:1702.05324}}.

\bibitem{fur17b}
S.~{Furusawa}, H.~{Togashi}, H.~{Nagakura}, K.~{Sumiyoshi}, S.~{Yamada},
  H.~{Suzuki}, and M.~{Takano}, Journal of Physics G Nuclear Physics, {\bf
  44}(9), 094001 (Sep 2017),  {{arXiv:1707.06410}}.

\bibitem{fur20}
Shun {Furusawa}, Hajime {Togashi}, Kohsuke {Sumiyoshi}, Koichi {Saito}, Shoichi
  {Yamada}, and Hideyuki {Suzuki}, Progress of Theoretical and Experimental
  Physics, {\bf 2020}(1), 013D05 (January 2020).

\bibitem{bro90}
R.~{Brockmann} and R.~{Machleidt}, \prc, {\bf 42}(5), 1965--1980 (November
  1990).

\bibitem{nag19c}
Hiroki {Nagakura}, Kohsuke {Sumiyoshi}, and Shoichi {Yamada}, \apjl, {\bf
  880}(2), L28 (August 2019),  {{arXiv:1907.04863}}.

\bibitem{iwa20}
Wakana {Iwakami}, Hirotada {Okawa}, Hiroki {Nagakura}, Akira {Harada}, Shun
  {Furusawa}, Kosuke {Sumiyoshi}, Hideo {Matsufuru}, and Shoichi {Yamada},
  arXiv e-prints, page arXiv:2004.02091 (April 2020),  {{arXiv:2004.02091}}.

\bibitem{nak21}
Ken'ichiro {Nakazato}, Kohsuke {Sumiyoshi}, and Hajime {Togashi}, \pasj, {\bf
  73}(3), 639--651 (June 2021),  {{arXiv:2103.14386}}.

\bibitem{sum21b}
Kohsuke {Sumiyoshi}, European Physical Journal A, {\bf 57}(12), 331 (December
  2021),  {{arXiv:2112.11159}}.

\bibitem{sum04}
K.~Sumiyoshi, H.~Suzuki, S.~Yamada, and H.~Toki, Nucl.\ Phys., {\bf A730}, 227
  (2004).

\bibitem{ste13}
A.~W. {Steiner}, M.~{Hempel}, and T.~{Fischer}, \apj, {\bf 774}(1), 17 (Sep
  2013),  {{arXiv:1207.2184}}.

\bibitem{ser86}
B.~D. Serot and J.~D. Walecka,
\newblock In J.~W. Negele and E.~Vogt, editors, {\em Advances in Nuclear
  Physics}, volume~16, page~1. Plenum Press, New York (1986).

\bibitem{sum94}
K.~{Sumiyoshi} and H.~{Toki}, \apj, {\bf 422}, 700 (Feb 1994).

\bibitem{hix03}
W.~R. {Hix}, O.~E. {Messer}, A.~{Mezzacappa}, M.~{Liebend{\"o}rfer},
  J.~{Sampaio}, K.~{Langanke}, D.~J. {Dean}, and G.~{Mart{\'\i}nez-Pinedo},
  \prl, {\bf 91}(20), 201102 (November 2003),  {{arXiv:astro-ph/0310883}}.

\bibitem{hem10}
Matthias {Hempel} and J{\"u}rgen {Schaffner-Bielich}, \nphysa, {\bf 837}(3-4),
  210--254 (Jun 2010),  {{arXiv:0911.4073}}.

\bibitem{fur17c}
Shun {Furusawa}, Hiroki {Nagakura}, Kohsuke {Sumiyoshi}, Chinami {Kato}, and
  Shoichi {Yamada}, \prc, {\bf 95}(2), 025809 (February 2017),
  {{arXiv:1701.08414}}.

\bibitem{fur13b}
Shun {Furusawa}, Hiroki {Nagakura}, Kohsuke {Sumiyoshi}, and Shoichi {Yamada},
  \apj, {\bf 774}(1), 78 (Sep 2013),  {{arXiv:1305.1510}}.

\bibitem{tog13}
H.~{Togashi} and M.~{Takano}, \nphysa, {\bf 902}, 53--73 (March 2013),
  {{arXiv:1302.4261}}.

\bibitem{kan07}
H.~{Kanzawa}, K.~{Oyamatsu}, K.~{Sumiyoshi}, and M.~{Takano}, \nphysa, {\bf
  791}(1-2), 232--250 (July 2007),  {{arXiv:nucl-th/0701069}}.

\bibitem{wir95}
R.~B. {Wiringa}, V.~G.~J. {Stoks}, and R.~{Schiavilla}, \prc, {\bf 51}(1),
  38--51 (January 1995),  {{arXiv:nucl-th/9408016}}.

\bibitem{car83}
J.~{Carlson}, V.~R. {Pandharipande}, and R.~B. {Wiringa}, \nphysa, {\bf
  401}(1), 59--85 (May 1983).

\bibitem{pud95}
B.~S. {Pudliner}, V.~R. {Pandharipande}, J.~{Carlson}, and R.~B. {Wiringa},
  \prl, {\bf 74}(22), 4396--4399 (May 1995),  {{arXiv:nucl-th/9502031}}.

\bibitem{akm98}
A.~{Akmal}, V.~R. {Pandharipande}, and D.~G. {Ravenhall}, \prc, {\bf 58}(3),
  1804--1828 (September 1998),  {{arXiv:nucl-th/9804027}}.

\bibitem{muk09}
Abhishek {Mukherjee}, \prc, {\bf 79}(4), 045811 (April 2009),
  {{arXiv:0811.3528}}.

\bibitem{sch79}
K.~E. {Schmidt} and V.~R. {Pandharipande}, Physics Letters B, {\bf 87}(1-2),
  11--14 (October 1979).

\bibitem{muk07}
Abhishek {Mukherjee} and V.~R. {Pandharipande}, \prc, {\bf 75}(3), 035802
  (March 2007),  {{arXiv:nucl-th/0609058}}.

\bibitem{kat13}
Tetsuya {Katayama} and Koichi {Saito}, \prc, {\bf 88}(3), 035805 (September
  2013),  {{arXiv:1307.2067}}.

\bibitem{gro99}
T.~{Gross-Boelting}, C.~{Fuchs}, and Amand {Faessler}, \nphysa, {\bf 648}(1),
  105--137 (March 1999),  {{arXiv:nucl-th/9810071}}.

\bibitem{dem10}
P.~B. {Demorest}, T.~{Pennucci}, S.~M. {Ransom}, M.~S.~E. {Roberts}, and
  J.~W.~T. {Hessels}, \nat, {\bf 467}(7319), 1081--1083 (October 2010),
  {{arXiv:1010.5788}}.

\bibitem{ant13}
John {Antoniadis}, Paulo C.~C. {Freire}, Norbert {Wex}, Thomas~M. {Tauris},
  Ryan~S. {Lynch}, Marten~H. {van Kerkwijk}, Michael {Kramer}, Cees {Bassa},
  Vik~S. {Dhillon}, and Thomas {Driebe}, Science, {\bf 340}(6131), 448 (Apr
  2013),  {{arXiv:1304.6875}}.

\bibitem{cro20}
H.~T. {Cromartie}, E.~{Fonseca}, S.~M. {Ransom}, P.~B. {Demorest},
  Z.~{Arzoumanian}, H.~{Blumer}, P.~R. {Brook}, M.~E. {DeCesar}, T.~{Dolch},
  J.~A. {Ellis}, R.~D. {Ferdman}, E.~C. {Ferrara}, N.~{Garver-Daniels}, P.~A.
  {Gentile}, M.~L. {Jones}, M.~T. {Lam}, D.~R. {Lorimer}, R.~S. {Lynch}, M.~A.
  {McLaughlin}, C.~{Ng}, D.~J. {Nice}, T.~T. {Pennucci}, R.~{Spiewak}, I.~H.
  {Stairs}, K.~{Stovall}, J.~K. {Swiggum}, and W.~W. {Zhu}, Nature Astronomy,
  {\bf 4}, 72--76 (January 2020),  {{arXiv:1904.06759}}.

\bibitem{fon21}
E.~{Fonseca}, H.~T. {Cromartie}, T.~T. {Pennucci}, P.~S. {Ray}, A.~Yu.
  {Kirichenko}, S.~M. {Ransom}, P.~B. {Demorest}, I.~H. {Stairs},
  Z.~{Arzoumanian}, L.~{Guillemot}, A.~{Parthasarathy}, M.~{Kerr},
  I.~{Cognard}, P.~T. {Baker}, H.~{Blumer}, P.~R. {Brook}, M.~{DeCesar},
  T.~{Dolch}, F.~A. {Dong}, E.~C. {Ferrara}, W.~{Fiore}, N.~{Garver-Daniels},
  D.~C. {Good}, R.~{Jennings}, M.~L. {Jones}, V.~M. {Kaspi}, M.~T. {Lam}, D.~R.
  {Lorimer}, J.~{Luo}, A.~{McEwen}, J.~W. {McKee}, M.~A. {McLaughlin},
  N.~{McMann}, B.~W. {Meyers}, A.~{Naidu}, C.~{Ng}, D.~J. {Nice}, N.~{Pol},
  H.~A. {Radovan}, B.~{Shapiro-Albert}, C.~M. {Tan}, S.~P. {Tendulkar}, J.~K.
  {Swiggum}, H.~M. {Wahl}, and W.~W. {Zhu}, \apjl, {\bf 915}(1), L12 (July
  2021),  {{arXiv:2104.00880}}.

\bibitem{abb18}
B.~P. {Abbott}, R.~{Abbott}, T.~D. {Abbott}, F.~{Acernese}, K.~{Ackley},
  C.~{Adams}, T.~{Adams}, P.~{Addesso}, R.~X. {Adhikari}, and V.~B. {Adya},
  \prl, {\bf 121}(16), 161101 (Oct 2018),  {{arXiv:1805.11581}}.

\bibitem{mcmil19}
M.~C. {Miller}, F.~K. {Lamb}, A.~J. {Dittmann}, S.~{Bogdanov},
  Z.~{Arzoumanian}, K.~C. {Gendreau}, S.~{Guillot}, A.~K. {Harding}, W.~C.~G.
  {Ho}, J.~M. {Lattimer}, R.~M. {Ludlam}, S.~{Mahmoodifar}, S.~M. {Morsink},
  P.~S. {Ray}, T.~E. {Strohmayer}, K.~S. {Wood}, T.~{Enoto}, R.~{Foster},
  T.~{Okajima}, G.~{Prigozhin}, and Y.~{Soong}, \apjl, {\bf 887}(1), L24
  (December 2019),  {{arXiv:1912.05705}}.

\bibitem{ril19}
T.~E. {Riley}, A.~L. {Watts}, S.~{Bogdanov}, P.~S. {Ray}, R.~M. {Ludlam},
  S.~{Guillot}, Z.~{Arzoumanian}, C.~L. {Baker}, A.~V. {Bilous},
  D.~{Chakrabarty}, K.~C. {Gendreau}, A.~K. {Harding}, W.~C.~G. {Ho}, J.~M.
  {Lattimer}, S.~M. {Morsink}, and T.~E. {Strohmayer}, \apjl, {\bf 887}(1), L21
  (December 2019),  {{arXiv:1912.05702}}.

\bibitem{mcmil21}
M.~C. {Miller}, F.~K. {Lamb}, A.~J. {Dittmann}, S.~{Bogdanov},
  Z.~{Arzoumanian}, K.~C. {Gendreau}, S.~{Guillot}, W.~C.~G. {Ho}, J.~M.
  {Lattimer}, M.~{Loewenstein}, S.~M. {Morsink}, P.~S. {Ray}, M.~T. {Wolff},
  C.~L. {Baker}, T.~{Cazeau}, S.~{Manthripragada}, C.~B. {Markwardt},
  T.~{Okajima}, S.~{Pollard}, I.~{Cognard}, H.~T. {Cromartie}, E.~{Fonseca},
  L.~{Guillemot}, M.~{Kerr}, A.~{Parthasarathy}, T.~T. {Pennucci}, S.~{Ransom},
  and I.~{Stairs}, arXiv e-prints, page arXiv:2105.06979 (May 2021),
  {{arXiv:2105.06979}}.

\bibitem{ril21}
Thomas~E. {Riley}, Anna~L. {Watts}, Paul~S. {Ray}, Slavko {Bogdanov}, Sebastien
  {Guillot}, Sharon~M. {Morsink}, Anna~V. {Bilous}, Zaven {Arzoumanian},
  Devarshi {Choudhury}, Julia~S. {Deneva}, Keith~C. {Gendreau}, Alice~K.
  {Harding}, Wynn C.~G. {Ho}, James~M. {Lattimer}, Michael {Loewenstein},
  Renee~M. {Ludlam}, Craig~B. {Markwardt}, Takashi {Okajima}, Chanda
  {Prescod-Weinstein}, Ronald~A. {Remillard}, Michael~T. {Wolff}, Emmanuel
  {Fonseca}, H.~Thankful {Cromartie}, Matthew {Kerr}, Timothy~T. {Pennucci},
  Aditya {Parthasarathy}, Scott {Ransom}, Ingrid {Stairs}, Lucas {Guillemot},
  and Ismael {Cognard}, \apjl, {\bf 918}(2), L27 (September 2021),
  {{arXiv:2105.06980}}.

\bibitem{sum93}
K.~{Sumiyoshi}, D.~{Hirata}, H.~{Toki}, and H.~{Sagawa}, \nphysa, {\bf 552}(4),
  437--450 (February 1993).

\bibitem{she98a}
H.~{Shen}, H.~{Toki}, K.~{Oyamatsu}, and K.~{Sumiyoshi}, Nuclear Physics A,
  {\bf 637}, 435--450 (July 1998),  {{arXiv:nucl-th/9}}.

\bibitem{she98b}
H.~{Shen}, H.~{Toki}, K.~{Oyamatsu}, and K.~{Sumiyoshi}, Progress of
  Theoretical Physics, {\bf 100}, 1013--1031 (November 1998),
  {{arXiv:nucl-th/9}}.

\bibitem{she11}
H.~{Shen}, H.~{Toki}, K.~{Oyamatsu}, and K.~{Sumiyoshi}, \apjs, {\bf 197}, 20
  (December 2011),  {{arXiv:1105.1666}}.

\bibitem{she20}
Hong {Shen}, Fan {Ji}, Jinniu {Hu}, and Kohsuke {Sumiyoshi}, \apj, {\bf
  891}(2), 148 (March 2020),  {{arXiv:2001.10143}}.

\bibitem{fur11}
Shun {Furusawa}, Shoichi {Yamada}, Kohsuke {Sumiyoshi}, and Hideyuki {Suzuki},
  \apj, {\bf 738}(2), 178 (Sep 2011),  {{arXiv:1103.6129}}.

\bibitem{fur13a}
Shun {Furusawa}, Kohsuke {Sumiyoshi}, Shoichi {Yamada}, and Hideyuki {Suzuki},
  \apj, {\bf 772}(2), 95 (Aug 2013),  {{arXiv:1305.1508}}.

\bibitem{fur17a}
Shun {Furusawa}, Kohsuke {Sumiyoshi}, Shoichi {Yamada}, and Hideyuki {Suzuki},
  \nphysa, {\bf 957}, 188--207 (Jan 2017),  {{arXiv:1612.01852}}.

\bibitem{sum06}
K.~{Sumiyoshi}, S.~{Yamada}, H.~{Suzuki}, and S.~{Chiba}, \prl, {\bf 97}(9),
  091101 (September 2006),  {{arXiv:astro-ph/0608509}}.

\bibitem{sum07}
K.~{Sumiyoshi}, S.~{Yamada}, and H.~{Suzuki}, \apj, {\bf 667}(1), 382--394
  (September 2007),  {{arXiv:0706.3762}}.

\bibitem{woo02}
S.~E. {Woosley}, A.~{Heger}, and T.~A. {Weaver}, Reviews of Modern Physics,
  {\bf 74}, 1015--1071 (November 2002).

\bibitem{woo95}
S.~E. {Woosley} and Thomas~A. {Weaver}, \apjs, {\bf 101}, 181 (November 1995).

\bibitem{nak13a}
K.~{Nakazato}, K.~{Sumiyoshi}, H.~{Suzuki}, T.~{Totani}, H.~{Umeda}, and
  S.~{Yamada}, \apjs, {\bf 205}, 2 (March 2013),  {{arXiv:1210.6841}}.

\bibitem{sum95c}
K.~{Sumiyoshi}, H.~{Suzuki}, and H.~{Toki}, \aap, {\bf 303}, 475 (Nov 1995),
  {{arXiv:astro-ph/9506024}}.

\bibitem{bru89a}
Stephen~W. {Bruenn}, \apj, {\bf 340}, 955 (May 1989).

\bibitem{bur84}
A.~{Burrows} and J.~M. {Lattimer}, \apj, {\bf 285}, 294--303 (October 1984).

\bibitem{lan03}
K.~Langanke and G.~Mart\'inez-Pinedo, Rev.\ Mod.\ Phys., {\bf 75}, 819 (2003).

\bibitem{lan03prl}
K.~{Langanke}, G.~{Mart{\'\i}nez-Pinedo}, J.~M. {Sampaio}, D.~J. {Dean}, W.~R.
  {Hix}, O.~E. {Messer}, A.~{Mezzacappa}, M.~{Liebend{\"o}rfer}, H.~Th.
  {Janka}, and M.~{Rampp}, \prl, {\bf 90}(24), 241102 (June 2003),
  {{arXiv:astro-ph/0302459}}.

\bibitem{joh22}
Zac {Johnston}, Sheldon {Wasik}, Rachel {Titus}, MacKenzie~L. {Warren}, Evan~P.
  {O'Connor}, Remco {Zegers}, and Sean~M. {Couch}, \apj, {\bf 939}(1), 15
  (November 2022),  {{arXiv:2202.09370}}.

\bibitem{nag18}
Hiroki {Nagakura}, Wakana {Iwakami}, Shun {Furusawa}, Hirotada {Okawa}, Akira
  {Harada}, Kohsuke {Sumiyoshi}, Shoichi {Yamada}, Hideo {Matsufuru}, and Akira
  {Imakura}, \apj, {\bf 854}(2), 136 (Feb 2018),  {{arXiv:1702.01752}}.

\bibitem{nag19a}
Hiroki {Nagakura}, Shun {Furusawa}, Hajime {Togashi}, Sherwood {Richers},
  Kohsuke {Sumiyoshi}, and Shoichi {Yamada}, \apjs, {\bf 240}(2), 38 (Feb
  2019),  {{arXiv:1812.09811}}.

\bibitem{nag22x}
H.~{Nagakura}, S.~{Furusawa}, and et~al. (2022),
\newblock in preparation.

\bibitem{bar90}
E.~{Baron} and J.~{Cooperstein}, \apj, {\bf 353}, 597 (April 1990).

\bibitem{tim96}
F.~X. {Timmes}, S.~E. {Woosley}, and Thomas~A. {Weaver}, \apj, {\bf 457}, 834
  (February 1996),  {{arXiv:astro-ph/9510136}}.

\bibitem{lat91}
James~M. {Lattimer} and Douglas~F. {Swesty}, \nphysa, {\bf 535}(2), 331--376
  (December 1991).

\bibitem{har22x}
H.~{Harada}, S.~{Furusawa}, K~{Sumiyoshi}, and et~al. (2022),
\newblock in preparation.

\end{thebibliography}

\appendix

\section{Appendix}\label{section:Appendix}
We examine here influence of different treatments of the translational motion of nuclei in the tables of EOS.  We show that it is negligible in the dynamics of core bounce by performing numerical simulations using sets of the EOS with and without the translational contribution.  

In the evaluation of the free energy of dense matter for the construction of table of EOS, there are some differences in the contributions of free energy of nuclei.  While the contribution from the translational motion of nuclei is included in the free energy of nuclei in the models such as \cite{lat91,fur11}, it is not included in the models such as a series of the Shen EOS \cite{she98a,she98b,she11,she20}.  The translational energy of nuclei has a contribution only at the low temperature regime where the nuclei are dominant composition and its contribution is usually small.  However, the contribution to the entropy is not negligible, especially in the initial model of massive stars, and may have influence on the following evolution.  

In order to examine this point, we additionally prepare the table of VM-S EOS with the translational contribution.  In the framework of VM-S EOS, the contribution of the translational motion is not included in the free energy following the convention of the Shen EOS.  We add contributions of the translational motion of nuclei to the original entries of energy, pressure and entropy in VM-S EOS by using the expression of the ideal Boltzmann gas.  This is a reasonable approximation to test this influence.  It is to be noted, in principle, that the minimization of the free energy with the translational energy is necessary to obtain the correct state of dense matter.   However, reconstruction of the EOS table is a formidable task and out of the scope here.  

\begin{figure*}[ht]
\centering
\includegraphics[width=0.45\textwidth]{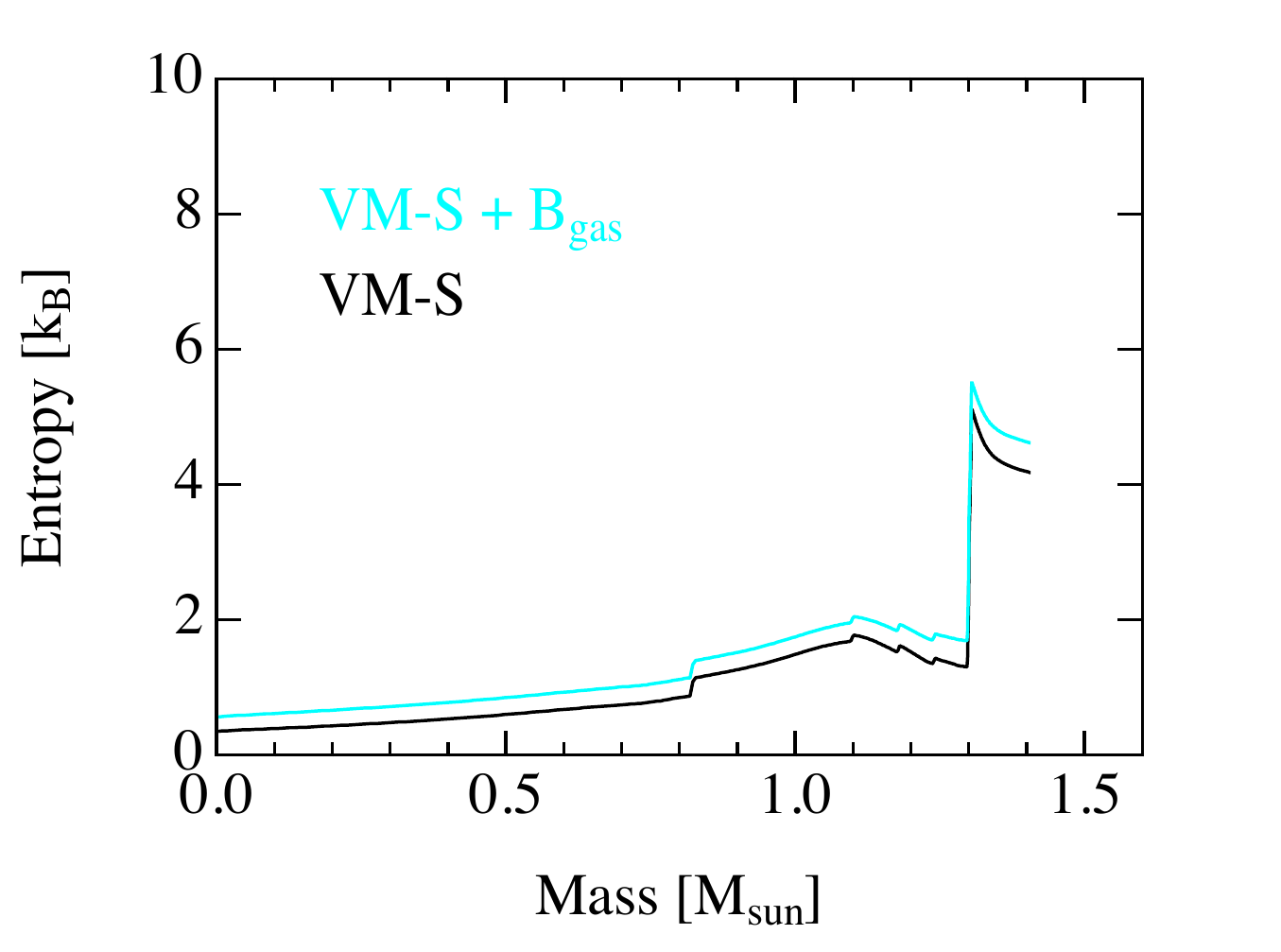}
\includegraphics[width=0.45\textwidth]{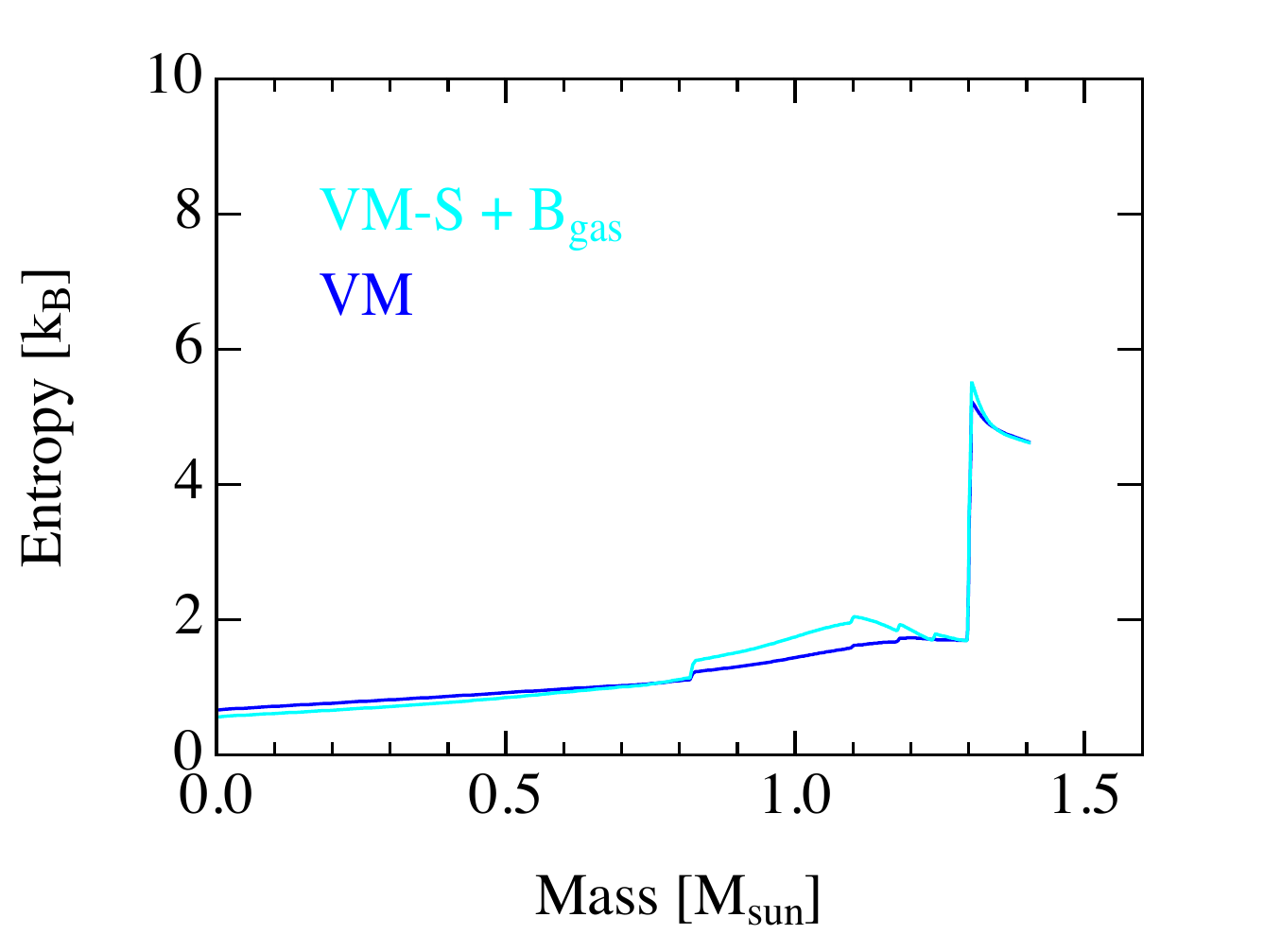}
\includegraphics[width=0.45\textwidth]{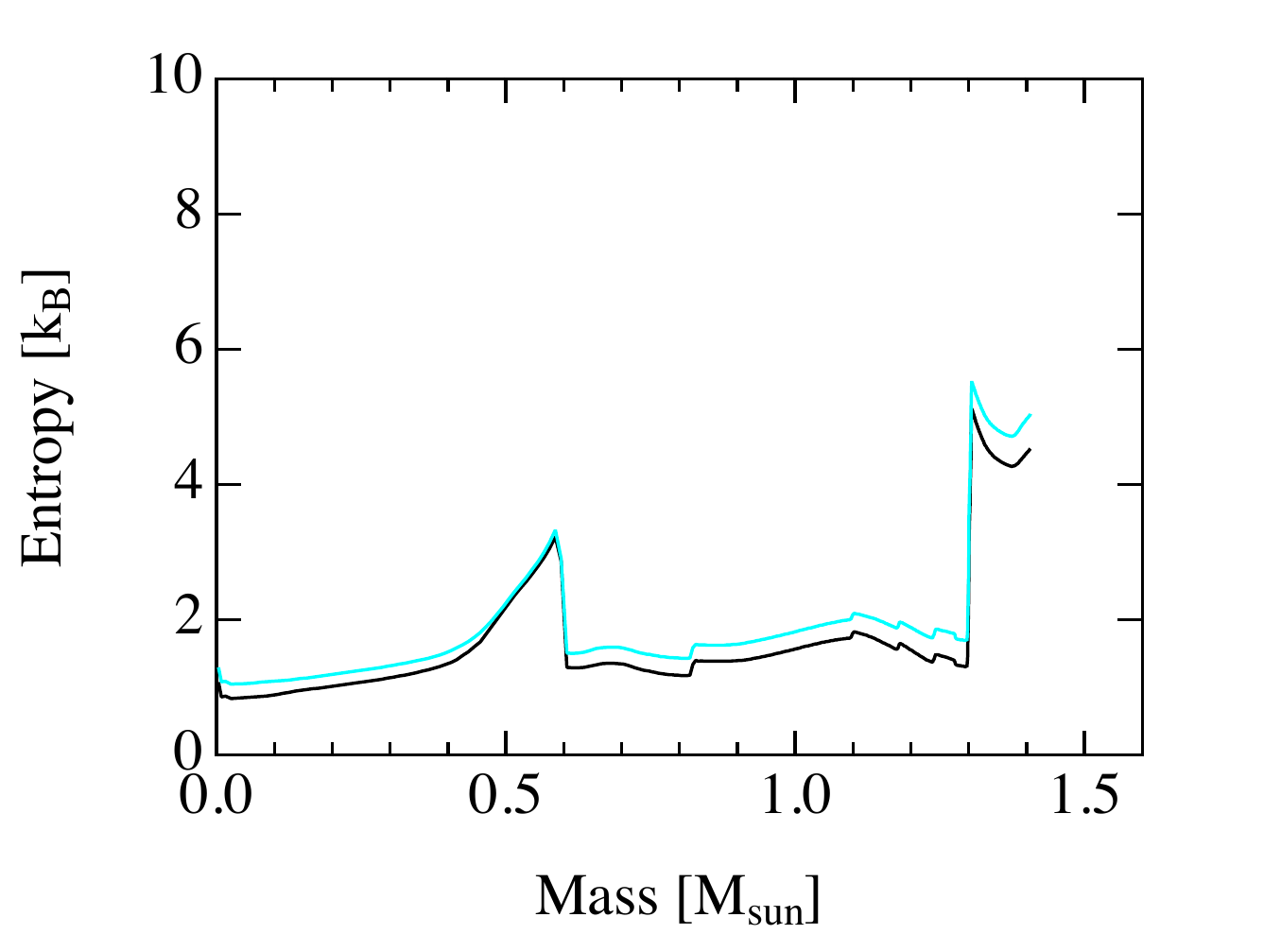}
\includegraphics[width=0.45\textwidth]{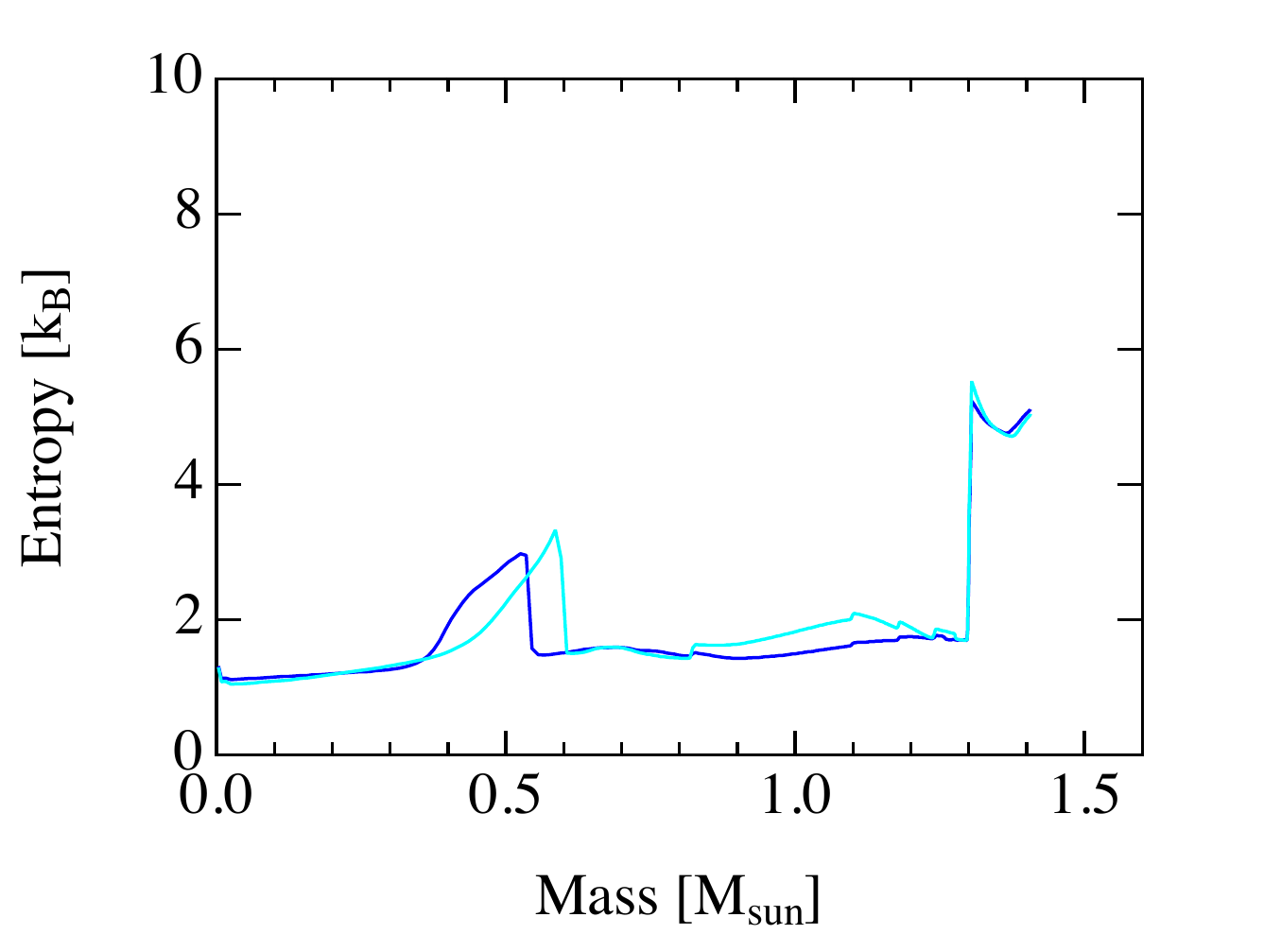}
\caption{Radial profiles of entropy per baryon at the initial condition (upper panels) and the core bounce (lower panels) are shown as a function of the baryon mass coordinate for the models of 11.2M$_{\odot}$.  
The comparison of the model using VM-S EOS with the Boltzmann gas contribution with respect to the model using VM-S EOS and VM EOS is shown in the left and right panels, respectively.  
The VM-S+B$_{\mathrm{gas}}$, VM-S, and VM models are shown by sky blue, black, and blue colors, respectively.  
\label{fig:radhyd_tpb0_entropy_Boltz}}
\end{figure*}

\begin{figure*}[ht]
\centering
\includegraphics[width=0.9\textwidth]{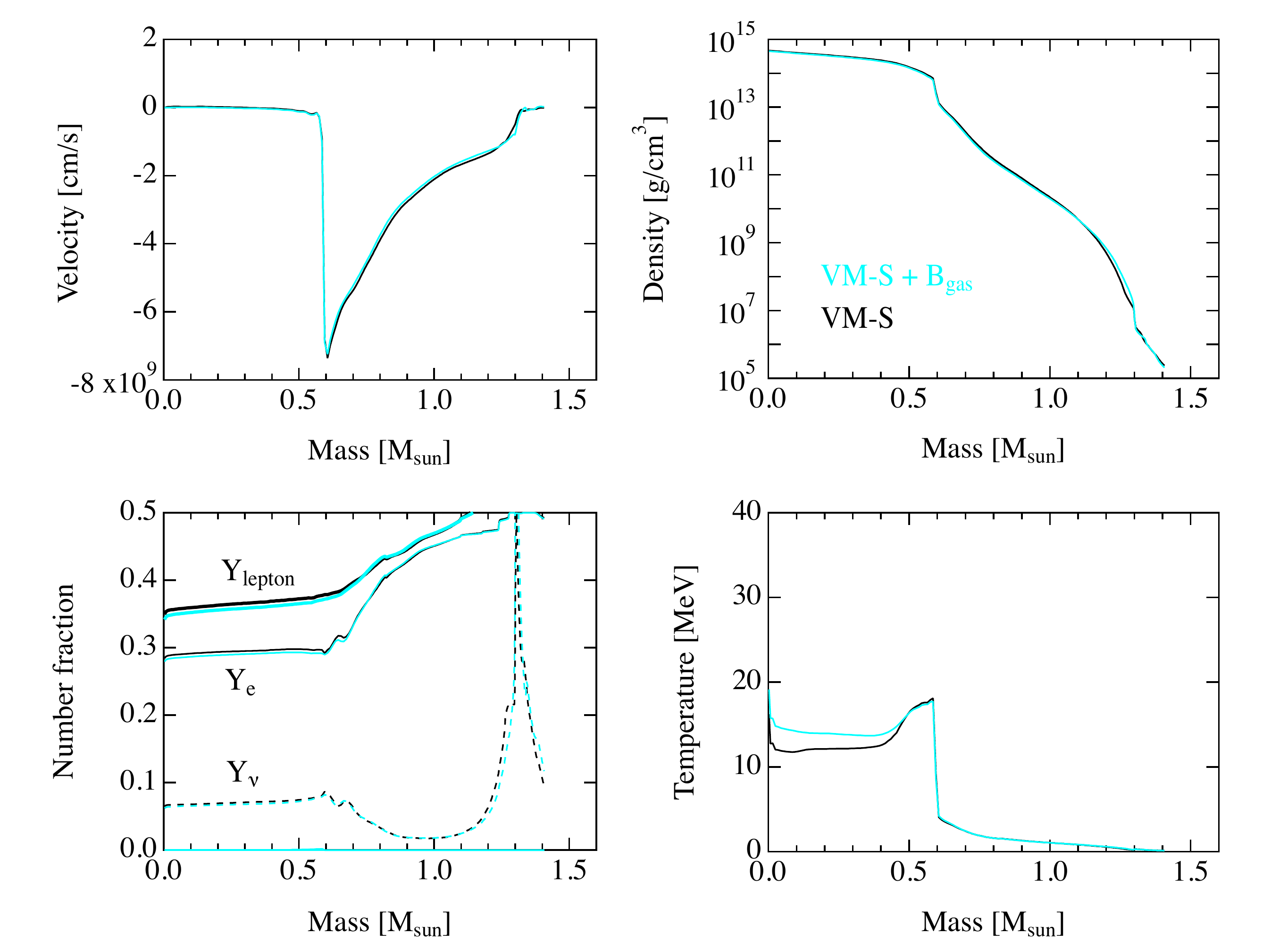}
\caption{The profiles at the core bounce are compared for the two models of 11.2M$_{\odot}$ star using VM-S EOS with the Boltzmann gas contribution and VM-S EOS.  
The velocity, density, lepton fractions, and temperature are shown as functions of the baryon mass coordinate in the upper-left, upper-right, lower-left, and lower-right panels, respectively.  
The lepton, electron, and neutrino fractions are shown by thick-solid, solid, and dashed lines, respectively, in the lower-left panel.  
The VM-S+B$_{\mathrm{gas}}$ and VM-S models are shown by sky blue and black colors, respectively.  
\label{fig:radhyd_tpb0_Boltz}}
\end{figure*}

We performed a numerical simulation of core collapse of the 11.2M$_{\odot}$ star using the modified VM-S EOS (VM-S+B$_{\mathrm{gas}}$ EOS) in the same way as the VM-S EOS model.  
We show in Fig. \ref{fig:radhyd_tpb0_entropy_Boltz} the radial profiles of the entropy per baryon at the initial condition and the core bounce.  Note that we set the temperature from the progenitor model as input to determine the entropy per baryon in each set of the EOS.  At the initial condition (upper panels), the entropy per baryon for VM-S+B$_{\mathrm{gas}}$ EOS is upward shifted from the original values of the entropy per baryon of VM-S EOS.  When we compare with VM EOS, which contains the contribution of the translational motion, the entropy per baryon of VM-S+B$_{\mathrm{gas}}$ EOS is roughly accord with that of VM EOS, indicating that the ad hoc addition of the contribution of translational motion works well.  At the core bounce (lower panels), the profiles of the entropy per baryon in the VM-S+B$_{\mathrm{gas}}$ and VM-S models are similar to each other except for an upward shift.  The position of the shock wave remains the same.  When we compare the VM-S+B$_{\mathrm{gas}}$ and VM models, the difference in the profiles of the entropy per baryon (including the position of shock wave) arises essentially from the difference of VM-S EOS and VM EOS.  
We additionally show in Fig. \ref{fig:radhyd_tpb0_Boltz} the profiles at the core bounce for VM-S+B$_{\mathrm{gas}}$ and VM-S models.  
The positions of shock waves are similar to each other as well as other quantities of density and lepton fractions.  The temperature profiles look similar except for a difference of inner temperature due to the entropy shift.  

In short, the effect on the dynamics due to the different treatments of the translational motion of nuclei is small except for overall shifts of the entropy per baryon.  
Initial setting of the central core needs some cautions about the values of the entropy per baryon although choosing the temperature as input may provide reasonable setting.  
The lack of translational energy in the series of the Shen EOS (and other EOSs in the same formalism) has been a minor but persistent concern in core-collapse simulations.  However, the current study implies that there are no major impacts in the core bounce and after.  
It is, of course, the better treatment of the translational motion is recommended in new developments of the EOS tables unless one sets the same setting without it to keep the consistency with the conventional EOS tables.  


\end{document}